\documentclass[11pt,a4paper]{article}
\usepackage[utf8]{inputenc}

\usepackage[sort&compress,numbers]{natbib}
\usepackage[colorlinks=true,citecolor=blue,urlcolor=blue]{hyperref}
\usepackage{braket}
\usepackage{tikz}
\usepackage{wrapfig}
\usepackage[rightcaption]{sidecap}
\usepackage{floatrow}
\usepackage{pgfplots}
\usepackage{float}
\usepackage{xcolor}
\usepackage{colortbl}
\usepackage{multirow}
\usepackage{appendix}

\usepackage{esvect}
\usepackage{multicol}
\usepackage{svg}
\usepackage{comment}
\usepackage{indentfirst}
\usepackage{physics}

\usepackage{siunitx}
\usepackage{subcaption}
\usepackage{graphicx} 
\usepackage{amsmath} 
\usepackage{amssymb} 
\usepackage[T1]{fontenc} 
\usepackage[english]{babel} 
\usepackage{csquotes}
\usepackage{tabularx}
\usepackage{psfrag} 
\usepackage{eurosym} 
\def\�{\euro{}}
\usepackage{color} 
\definecolor{linkcolor}{rgb}{0,0,0.6} 
\usepackage{fancyhdr} 

\makeatletter
\@ifpackageloaded{babel}%
{}
{}
\makeatother

\begin{document}

	\setlength{\parindent}{0pt}
	\thispagestyle{empty}
	
	\includegraphics[height=2cm]{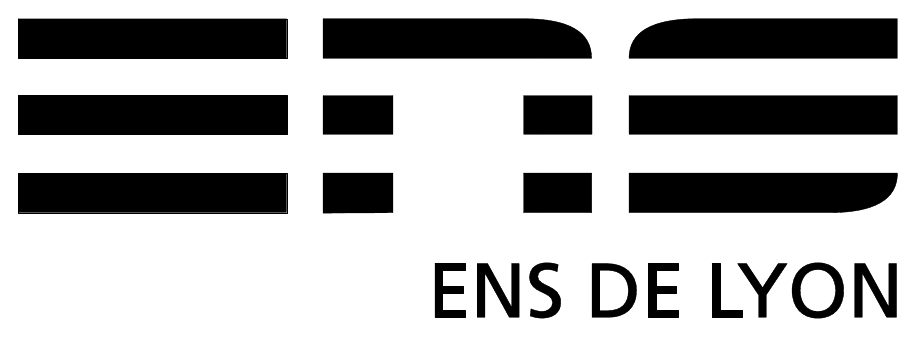} \hfill \includegraphics[height=2cm]{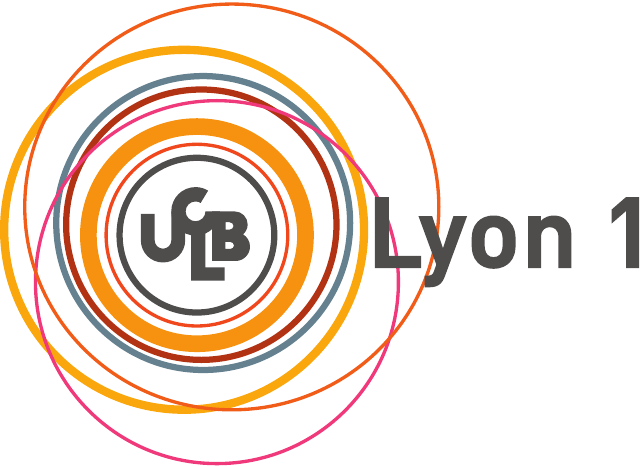} \hfill \includegraphics[height=2cm]{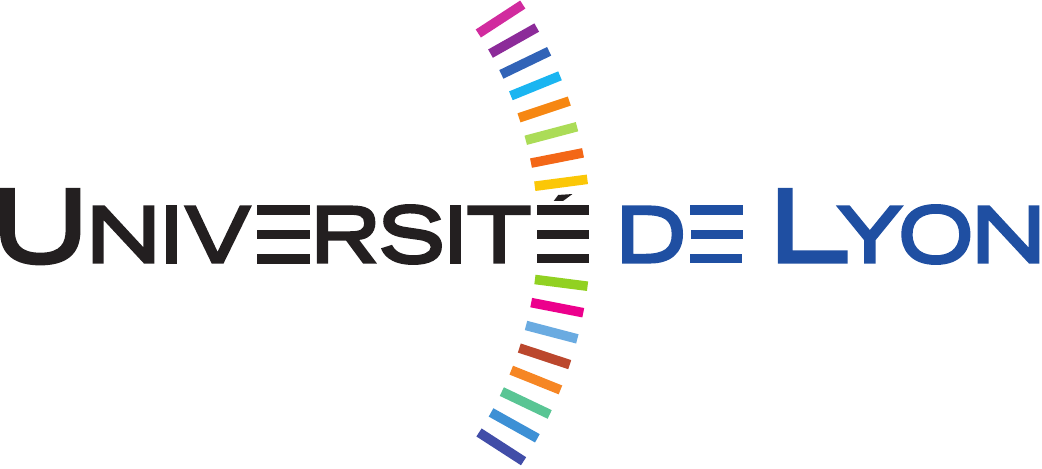}
	
	\vspace{0.5cm}
	
	\begin{tabularx}{\textwidth}{@{} l X l @{} }
		{\sc Master Science de la mati\`ere} & & Internship 2024 \\
		{\it Ecole Normale Sup\'erieure de Lyon} & & Brichet Lyse \\
		{\it Universit\'e Claude Bernard Lyon I} & & M2 Physics
	\end{tabularx}
	
	\begin{center}
		
		\vspace{1.5cm}
		
		\rule[11pt]{5cm}{0.5pt}
		
		\textbf{\huge Experimental investigation of vortex shedding in superfluid $^4$He.}
		
		\rule{5cm}{0.5pt}
		
		\vspace{1.5cm}
		
		\parbox{15cm}{\small
			\textbf{Abstract} : \it During this internship, the starting vortex and its shedding from the trailing edge of an airfoil moving at constant acceleration in $^4$He has been studied. The airfoil has an angle of attack of 49.1° and corresponds to the NACA 0012, in contrast to most of the simulations using an angle of attack of 90° with a flat plate. An optical access combined with force and torque sensors enabled the study of this phenomenon. Particles are added to the fluid andfrom their position and velocity we obtained an estimate of the vorticity field. A tracking algorithm could then detect and track the vortex. From this tracking, we found no difference in the behaviour of the vortices in HeI and HeII. Furthermore, the trajectories do not correspond to the theoretical expectations, exhibiting straight trajectories until the vortex is independant of the airfoil, when a deviation in the trajectory happens. The vortex also moves faster than predicted. The differences are attributed to viscous and 3D effects.}

		\vspace{0.5cm}
		
		\parbox{15cm}{
		}
		
		\vspace{0.5cm}
		
		\parbox{15cm}{
			Internship supervised by :
			
			{\bf Marco La Mantia}
			
			\href{mailto:marco.la-mantia@mff.cuni.cz}{\tt marco.la-mantia@mff.cuni.cz} / t\`el. +420 95155 2894

			\vspace{0.5cm}
			
			Departement of low-temperature physics
			
			{\it Charles University \\
				121 16 Prague\\
				Czech Republic}
			
			\url{https://www.mff.cuni.cz/en/faculty/organizational-structure/department?code=107}
		}
		
		\vspace{0.4cm}

		\includegraphics[height=4.3cm]{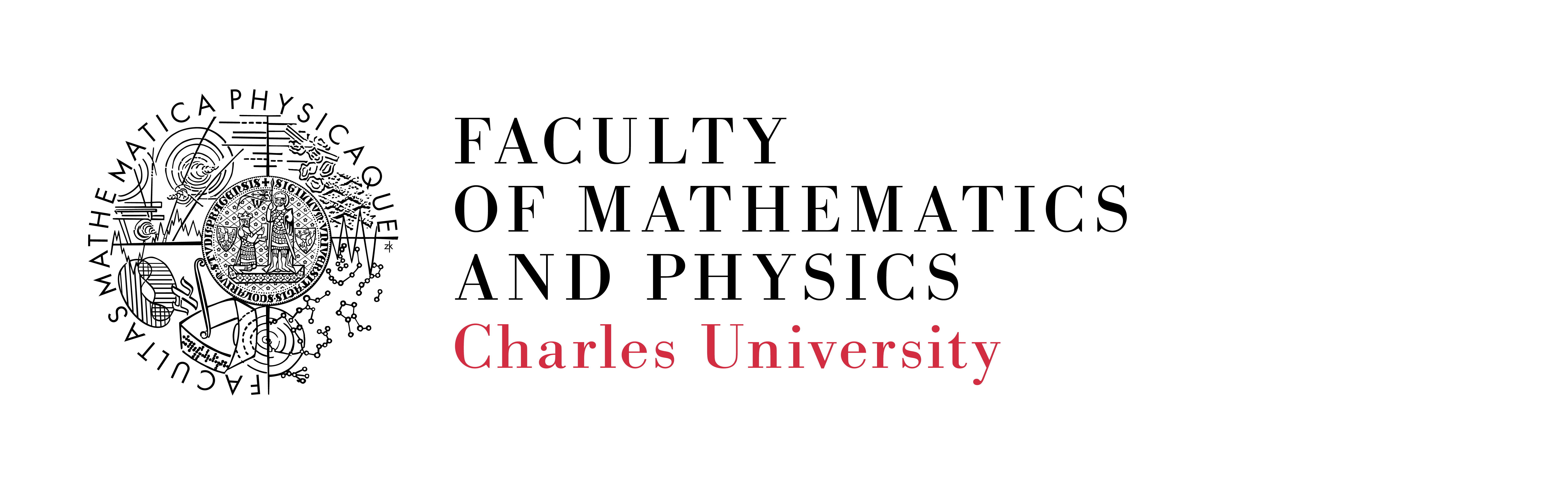}

	\end{center}
	
	\vfill
	\hfill 
	\newpage
	\thispagestyle{empty}
	\setlength{\parindent}{16pt}
	\section*{Acknowledgments}
	I would like to thank Marco La Mantia for accepting me for this internship and for his throughrough supervising.
	
	I am also grateful to Jirka who helped me find my feet and to Miloš for his indications during the internship and all the little discussions.
	
	I would also like to thank all of my friends and especially Alex and Ella for their support, in particular in last moments with the political situation complicating the state of mind.
	
	I would finally like to thank the university for welcoming me during the intership.
	
	\tableofcontents
	\newpage

	\setcounter{page}{1}

	\section{Introduction}

\subsection{Properties of Helium.}

Low temperature physics began in the late 19$^{\textrm{th}}$ century with the liquefaction of O$_2$ in 1877 by Cailletet and Pictet. However, helium has only been liquefied for the first time in 1908 by Onnes. This achievement lead to many other discoveries such as superfluidity in $^4$He\cite{Kapitza1938}. But the research was limited to some laboratories until the 40' when the commercial development of Helium liquefaction began, allowing helium cryogenics to be more broadly studied\cite{HeliumCryogenics}.

Two isotopes of helium exist, $^4_2$He which is a boson following the Bose-Einstein distribution\cite{Tisza1938}, and $^3_2$He a fermion following the Fermi-Dirac statistics, thus having different properties at low temperature. In the following we will only be considering the isotope $^4_2$He  referring to it as He. Because $^4$He has the lowest critical point, it has a weak intermolecular potential, and hence its behaviour is nearly the one of an ideal fluid. Because of this, much of its behaviour in the gaseous or liquid above the superfluid transition states can be treated with classical models as the Navier-Stokes equation. HeI, which is the normal liquid, acts as a weakly interacting gas due to its weak intermolecular interaction and low viscosity ($\nu\approx 2.5\times 10^{-8}$m$^2\cdot$s$^{-1}$ at 3 K\cite{TheObservedPropertiesofLiquidHeliumattheSaturatedVaporPressure}). On the contrary, the behaviour of the liquid below the transition (HeII) and the solide states are described by quantum mechanics. Helium has other several particularities that can be highlighted with its phase diagram (see figure \ref{fig:PT}). First, the solid state is never obtained, even at 0K unless pressures higher than 2.5 MPa are applied due to its quantum nature. It has a large zero point energy so its lowest energy state corresponds to liquid helium. Because the solid state cannot exist without applying external pressure, the triple point does not exist. The transition from He I to He II is a second order phase transition called, in the case of helium, $\lambda$-transition, because the specific heat has the shape of the Greek letter near the transition. This transition occurs at $T_\lambda = 2.17$ K at saturated vapor pressure. The weak intermolecular interaction of helium is also the reason why helium liquefies at very low temperature, its normal boiling point being at 4.2K. Normal liquid helium is, after hydrogen, the lowest density condensed fluid.\newline

\begin{figure}[h!]
	\centering
	\begin{subfigure}[b]{0.45\textwidth}
		\centering
\begin{tikzpicture}[scale=0.9]

\definecolor{darkgray176}{RGB}{176,176,176}
\definecolor{navy}{RGB}{0,0,128}

\begin{axis}[
log basis y={10},
tick align=outside,
tick pos=left,
x grid style={darkgray176},
xlabel={Temperature (K)},
xmajorgrids,
xmin=0.57993, xmax=5.42147,
xtick style={color=black},
y grid style={darkgray176},
ylabel={Pressure (Pa)},
ymajorgrids,
ymin=0.650551149008877, ymax=43101429.3335881,
ymode=log,
ytick style={color=black}
]
\addplot [semithick, navy]
table {%
1.28 2567575.5
1.32 2578721.25
1.36 2591893.5
1.4 2608105.5
1.44 2628370.5
1.48 2648635.5
1.52 2672953.5
1.56 2704364.25
1.6 2741854.5
1.64 2786437.5
1.68 2843179.5
1.72 2914107
1.76 3006312.75
1.8 3117770.25
1.84 3241386.75
1.88 3371082.75
1.92 3503818.5
1.96 3640607.25
2 3779422.5
};
\addplot [semithick, navy]
table {%
2 3829000
2.5 5778000
3 7970000
3.5 10400000
4 13050000
4.5 15920000
5 19010000
};
\addplot [semithick, navy]
table {%
1.7678 3013000
1.8 2829000
1.85 2530000
1.9 2214000
1.95 1879000
2 1523000
2.05 1145000
2.1 737200
2.15 287300
2.1773 5040
};
\addplot [semithick, navy]
table {%
2.1773 5000
2.2 5300
2.3 6700
2.35 7500
2.4 8300
2.45 9200
2.5 10200
2.55 11200
2.6 12400
2.65 13500
2.7 14800
2.75 16100
2.8 17500
2.85 19000
2.9 20600
2.95 22300
3 24000
3.05 25900
3.1 27800
3.15 29800
3.2 32000
3.25 34200
3.3 36500
3.35 39000
3.4 41500
3.45 44200
3.5 47000
3.55 49900
3.6 52900
3.65 56000
3.7 59300
3.75 62600
3.8 66100
3.85 69800
3.9 73500
3.95 77400
4 81500
4.05 85600
4.1 90000
4.15 94400
4.2 99000
4.224 101300
4.25 103800
4.3 108700
4.35 113800
4.4 119000
4.45 124400
4.5 129900
4.55 135700
4.6 141600
4.65 147600
4.7 153900
4.75 160300
4.8 167000
4.85 173800
4.9 180800
4.95 188000
5 195400
5.05 203100
5.2014 227500
};
\addplot [semithick, navy]
table {%
0.8 1.474996547969
0.85 2.91399318019096
0.9 5.37898741120356
0.95 9.37897804976356
1 15.5699635605948
1.05 24.7799420058792
1.1 37.99991106632
1.15 56.449867886678
1.2 81.4798093074672
1.25 114.699731560708
1.3 157.899630457156
1.35 212.899501737356
1.4 281.99934001848
1.45 367.399140151736
1.5 471.49889652026
1.55 596.99860280508
1.6 746.398253155296
1.65 922.397841251936
1.7 1127.99736007392
1.75 1365.99680306824
1.8 1637.99616649032
1.85 1948.99543863836
1.9 2289.9946405756
1.95 2691.99369975088
2 3128.99267701356
2.05 3611.99154661968
2.1 4140.99030856924
2.15 4714.9889652026
2.2 5334.9875141794
2.25 6004.9859461382
2.3 6729.9842493772
2.35 7511.98241921568
2.4 8353.98044863256
2.45 9257.97833294712
2.5 10229.9760581172
2.55 11259.9736475464
2.6 12369.9710497468
2.65 13549.968288122
2.7 14809.9653392684
2.75 16139.9622265896
2.8 17549.958926682
2.85 19049.955416142
2.9 20629.9517183732
2.95 22289.9478333756
3 24049.943714342
3.05 25889.9394080796
3.1 27839.9348443776
3.15 29869.9300934468
3.2 32009.9250850764
3.25 34249.91984267
3.3 36589.9143662276
3.35 39039.9086323456
3.4 41589.9026644276
3.45 44259.8964156664
3.5 47039.8899094656
3.55 49939.8831224216
3.6 52959.8760545344
3.65 56089.8687292076
3.7 59349.861099634
3.75 62739.8531658136
3.8 66249.84495115
3.85 69889.8364322396
3.9 73659.8276090824
3.95 77569.8184582748
4 81619.8089798168
4.05 85809.7991737084
4.1 90139.7890399496
};
\draw (100,5) node{HeII} ;
\draw (370,6) node{HeI} ;
\draw (300,3) node{Gas} ;
\draw (70,7) node{Solid} ;
\end{axis}

\end{tikzpicture}
		\caption{Equilibrium phase diagram of $^4He$ \cite{bookDensity}\cite{GRILLY1962250}\cite{TheObservedPropertiesofLiquidHeliumattheSaturatedVaporPressure}. The critical point is at $T_c\approx 5.2$ K and $P_c\approx 226$ kPa.}
		\label{fig:PT}
	\end{subfigure}
	\hfill
	\begin{subfigure}[b]{0.45\textwidth}
		\centering
\begin{tikzpicture}[scale=0.79]

\definecolor{darkgray176}{RGB}{176,176,176}
\definecolor{darkorange25512714}{RGB}{255,127,14}
\definecolor{steelblue31119180}{RGB}{31,119,180}
\definecolor{deepskyblue}{RGB}{0,191,255}
\definecolor{navy}{RGB}{0,0,128}

\begin{axis}[
tick align=outside,
tick pos=left,
x grid style={darkgray176},
xlabel={Temperature (K)},
xmajorgrids,
xmin=-0.10884, xmax=2.28564,
xtick style={color=black},
y grid style={darkgray176},
every axis y label/.style={
            at={(ticklabel cs:0.5)},rotate=90,anchor=near ticklabel,
        },
ylabel={Density (g/cm\(\displaystyle ^3\))},
ymajorgrids,
ymin=-0.0073055, ymax=0.1534155,
ytick style={color=black},
yticklabels={,0,0.05,0.1,0.15}
]
\addplot [semithick, navy, dashed]
table {%
0 0.14513
0.05 0.14513
0.1 0.14512
0.15 0.14511
0.2 0.1451
0.25 0.14509
0.3 0.14509
0.35 0.14508
0.4 0.14508
0.45 0.14509
0.5 0.14511
0.55 0.14512
0.6 0.14513
0.65 0.14513
0.7 0.14511
0.75 0.14506
0.8 0.14498
0.85 0.14486
0.9 0.14469
0.95 0.14446
1 0.14414
1.05 0.14371
1.1 0.14313
1.15 0.14235
1.2 0.14137
1.25 0.14012
1.3 0.1386
1.35 0.13676
1.4 0.13457
1.45 0.13199
1.5 0.129
1.55 0.12556
1.6 0.12163
1.65 0.11715
1.7 0.11206
1.75 0.1063
1.8 0.09982
1.85 0.09254
1.9 0.08444
1.95 0.07542
2 0.06507
2.05 0.05275
2.1 0.03779
2.11 0.03443
2.12 0.03091
2.13 0.02718
2.14 0.02311
2.15 0.01862
2.16 0.01359
2.17 0.00729
2.171 0.00652
2.172 0.00572
2.173 0.00489
2.174 0.00401
2.175 0.003
2.176 0.00175
2.1768 0
};
\addplot [semithick, deepskyblue]
table {%
0 0
0.05 2.95e-09
0.1 5.9e-09
0.15 8.85e-09
0.2 2.771e-08
0.25 6.687e-08
0.3 1.368e-07
0.35 2.502e-07
0.4 4.242e-07
0.45 7.03e-07
0.5 1.249e-06
0.55 2.588e-06
0.6 6.069e-06
0.65 1.452e-05
0.7 3.295e-05
0.75 6.923e-05
0.8 0.0001345
0.85 0.0002431
0.9 0.0004127
0.95 0.0006636
1 0.00102
1.05 0.00141
1.1 0.00199
1.15 0.00276
1.2 0.00375
1.25 0.00499
1.3 0.00652
1.35 0.00837
1.4 0.01057
1.45 0.01316
1.5 0.01617
1.55 0.01963
1.6 0.02358
1.65 0.02809
1.7 0.03321
1.75 0.039
1.8 0.04554
1.85 0.05286
1.9 0.06103
1.95 0.07012
2 0.08055
2.05 0.09297
2.1 0.10804
2.11 0.11143
2.12 0.11497
2.13 0.11874
2.14 0.12284
2.15 0.12737
2.16 0.13243
2.17 0.13878
2.171 0.13955
2.172 0.14036
2.173 0.1412
2.174 0.14208
2.175 0.1431
2.176 0.14436
2.1768 0.14611
};
\end{axis}

\end{tikzpicture}
		\caption{The thermal dependence of the density of liquid helium for the normal (in light blue) and superfluid (dark blue) components as described in the two fluid model \cite{TheObservedPropertiesofLiquidHeliumattheSaturatedVaporPressure}.}
		\label{fig:Density}
	\end{subfigure}
	\label{fig:fluide}
\end{figure}

To describe the heat and mass transport in HeII Tisza and Landau\cite{Kapitza1938}\cite{Landau1941} developed the two fluid model theory. In this model He II is made of two interpenetrating fluid components, the superfluid and the normal fluid which contains the excitation in the liquid. The latter is expected to behave as an ordinary fluid with a viscosity $\mu_n$ and a density $\rho_n$. On the other hand, the superfluid of density $\rho_s$ has no viscosity. The properties of He II is thus a linear combination of these two fluids. Therefore, the density is $\rho = \rho_s + \rho_n$ and the momentum density $\vb{j}$ is $\vb{j} = \rho_n\vb{v_n} + \rho_s\vb{v_s}$ with $\vb{v_n}$ and $\vb{v_s}$ the velocity of the normal and superfluid fluid respectively. Despite the fact that the densities of the normal and superfluid phases strongly vary with the temperature (see figure \ref{fig:Density}), the total density remains quite constant ($\rho\sim146$ kg$\cdot$m$^{-3}$). A more detailed mathematical description of the two fluid model can be found in the appendix \ref{App:2fluid}. In the absence of viscosity, the dissipation is only due to the normal phase. The superfluid does not experience resistance to flow, and therefore $\grad\times\vb{v_s}=0$\cite{HeliumCryogenics}. However, this description ceases to be correct above a critical velocity $\vb{v_c}$ where the superfluid transition to a turbulence state. Furthermore, some effects due to quantum behaviour of the HeII phase, such as being frictionless, cannot be understood with this classical approach. We can therefore define the wave function $\psi (\vb{r},t) = \psi_0e^{i\phi(\vb{r},t)}$ with $\psi_0$ the amplitude and $\phi$ the phase, $\vb{r}$ the position and $t$ the time. The number of atoms in the superfluid state is given by $\psi^*(\vb{r},t)\psi (\vb{r},t) = |\psi |^2 = \rho_s/m_4$ with $m_4$ the mass of $^4$He atom. The amplitude is thus $\psi_0 = \sqrt{\rho_s/m_4}$. This quantum description enables us to obtain the analytical expression $\vb{v}_s$.

\begin{equation}
	\vb{v}_s = \frac{\vb{p}}{m_4} = \frac{\hbar}{m_4}\grad\phi.
\end{equation}

The flow is potential, and therefore the vorticity is null in a simple connected region, $\vb{\omega} = \grad\times\vb{v}_s =0$. If we assume the flow to be incompressible and inviscid, we can according to Kelvin-Helmotz  theory define the circulation $\Gamma$. However, the vortices, as the fluid, is described by quantum theory. Therefore, it is quantized\cite{IntroductionToVortexTheory} and can be non-zero\cite{ANDRONIKASHVILI1966}.

\begin{equation}
	\Gamma = \oint_S\vb{v}_s\cdot d\vb{l} = \frac{\hbar}{m_4}\oint\grad\phi\cdot d\vb{l} = n\kappa,
\end{equation}

with $l$ a vector tangent to $S$, $\kappa=h/m_4 = 9.997\times10^{-8}$m$^2\cdot$s$^{-1}$ a quantum of circulation. The quantum vortices are of the same strength and the vorticity is constrained to their cores. The size of such vortices is of the order of ångströms. \newline

\subsection{Starting vortex}

The starting vortex is a subject of study since the work of Prandtl in 1904\cite{PrandtlsBondaryLayer}. Since then, starting vortex has mainly been studied throug the prism of forces because of its essential aspect in aeronautics. Its mathematical formulation was written by Rott \cite{Rott_1956} with similarity laws. The equations from this work were later solved in 1969\cite{Blendermann} and 1978\cite{Pullin1978}. Since then, various experiments\cite{Taneda1971}\cite{Kriegseis_Kinzel_Rival_2013}\cite{RINGUETTE_MILANO_GHARIB_2007}\cite{Patrik2015} and simulations\cite{Koumoutsakos_Shiels_1996} have been done with constant velocity or acceleration of the moving object.

The generation of vorticity is related to the interaction between the fluid and the solid boundary. This create a shear layer which is a product of viscous effects. Instead of following the contour of the body, the viscous stress lead to the formation of a vortex. However in the framework of ideal fluid, another theory using vortex sheets was developed\cite{Xia2017}\cite{Saffman_1993}. In this other model, the boundary layer becomes the vortex sheet where the vorticity and the vortex lines are confined. When the vortex sheet separates itself from the body that generated it, it remains sheet-like with the vorticity contained inside. It rolls up into a spiral shape. The separation is assumed to be irrotational\cite{VorticityAndVortexDynamics}. In real fluid, the shear layers merge into a rigid core, where the vorticity is uniformly distributed and the outer shear layer remains discrete with great velocity gradients\cite{Lian1989}.\newline

The caracteristic parameters choosen to study the starting vortex are the kinematic velocity of the fluid $\nu$, the caracteristic velocity of the flow $V$ and $L$ the caracteristic dimension of the airfoil. According to the Buckingham $\pi$-theorem, the starting vortex only has one control parameter, so we decided it to be the Reynolds number defined as 
$$
\textrm{Re} = \frac{VL}{\nu}.
$$
Xu\cite{Xu2014} defines a caracteristic timescale $T = (L/a)^{1/2}$ with $a$ the acceleration of the moving object. The Reynolds number thus becomes
\begin{equation}
	\textrm{Re} = \frac{L^2}{T\nu} = \frac{a^{1/2}L^{3/2}}{\nu}.
\end{equation}

Defining the kinematic viscosity in the case of the superfluid helium can be ambigus as several possibilities can be used\cite{Babuin_2014}. However, to be able to compare with the work of\cite{MasterThesis} the kinematic viscosity is similarly defined as in classical theory $\nu = \mu_n/\rho$. Its featuring values come from\cite{TheObservedPropertiesofLiquidHeliumattheSaturatedVaporPressure}. Still following the work of Xu\cite{Xu2014}, one can define dimensionless quantities to study the phenomenon using the caracteristic timescale to anayse the results. We can then write dimensionless coordinates and time as:

\begin{equation}\label{eq:adimensionnement}
	\tilde{x}=\frac{x}{L},\quad\tilde{y} = \frac{y}{L}, \quad\tilde{t} = \frac{t}{T}
\end{equation}
with $(x,y)$ the cartesian coordinates  in the laboratory reference and $t$ the time.

\subsection{Motivation}

During this internship, I focused on the vortex shedding created by a moving airfoil with an angle of attack of 49.1° (see section \ref{SecAirfoil}), when most studies concentrate on angles of attack of 90° (flow perpendicular to the wing). Furthermore, computing simulations with an arbitrary airfoil instead of a flat plate with the current models is challenging\cite{Xia2017}. So our experiments will provide new data for the vortex shedding with an airfoil that does not match the flat plate and with a non-negligible thickness. It also confirms or points out the limits of the theories developed on the subject. Furthermore, the previous studies mainly focused on the forces, the circulation\cite{Nitsche_2014}, the size of the vortex\cite{Taneda1971} or its stability\cite{Schneider2014}. The trajectories, however, have been less studied. The internship will thus contribute to extend on this subject. The study will also provide new data on physical phenomena in liquid helium, thereby extending our knowledge of this subject, and add to our knowledge of the similarities between quantum and classical behaviour.

\section{The set-up}

\begin{table}[h!]
	\centering
	\begin{tabular}{c|c|c|c|c|c|c}
		Experiment & Temperature & f (Hz) & Number of & Number of file & Medium inside & Re\\
		& (K) &  & movies & for the sensors & the cryostat & \\  \hline 
		Aa & 2.26(0.01) & 1 & 100 & 102 & liquid & 9.2$\times 10^{4}$\\ \hline
		Ba & 1.95(0.001) & 1 & 147 & 300 & liquid & 1.9$\times 10^{5}$\\ \hline
		Ca & 1.75(0.002) & 1 & 104 & 255 & liquid & 2.0$\times 10^{5}$\\
		Cb & 1.75(0.001) & 2 & 50 & 112 & liquid & 4.0$\times 10^{5}$\\
		Cc & 1.75(0.001) & 0.5 & 50 & 113 & liquid & 1.0$\times 10^{5}$\\ \hline
		Da & 1.5(0.004) & 1 & 50 & 150 & liquid & 1.9$\times 10^{5}$\\
		Db & 1.5(0.005) & 2 & 50 & 150 & liquid & 3.9$\times 10^{5}$\\        
		Dc & 1.5(0.004) & 0.5 & 50 & 103 & liquid & 9.7$\times 10^{4}$\\ \hline
		Ea & 1.3(0.007) & 1 & 50 & 100 & liquid & 1.7$\times 10^{5}$\\
		Eb & 1.3(0.007) & 2 & 50 & 100 & liquid & 3.4$\times 10^{5}$\\
		Ec & 1.3(0.006) & 0.5 & 50 & 100 & liquid & 8.6$\times 10^{4}$\\ \hline
		Fa,b,c & 4.22 & 1, 2, 0.5 & 0 & 100 & liquid & \\ \hline
		Ga,b,c & <100 & 1, 2, 0.5 & 0 & 100 & cold gas & \\ \hline
		Ha,b,c & <100 & 1, 2, 0.5 & 0 & 100 & cold vacuum & \\
	\end{tabular}
	\caption{Parameters of the different experiments. The capital letter of the name of the experiment is related to the temperature (in parenthesis is the standard deviation) at which it was run. The small letter indicates the frequency $f$ of the motor (see \ref{SecLinMot}). The number of movies recorded for each situation is given as the number of experiment recorded by the sensors.}
	\label{tab:Param}
\end{table}

In order to use liquid helium, the experiments need to happen in a cryostat (see subsection \ref{SecApparatus}) to reach low temperature. The experiments follow a specific procedure. First, deuterium particles are injected in the system (see subsection \ref{SecSeeding}) the first day. During the experiments, the particles need to be mixed with the fluid, so we are doing injections of helium gas before each experiment. Those injections consist of pulses (between 3 and 6 depending on the pressure of the gas injected, how well the particles are mixed and how much it perturbs the temperature of liquid helium) of 50 $\mu$s every 500 ms. We wait for several seconds for the fluid to rest. We then turn on the sensors (see subsection \ref{SecSensors}) and run a calibration before taking any data. This calibration consist of measuring the signal from the sensors for 1s which will be averaged and then subtracted to the data of the experiment. After that we start the recording with the camera (see subsection \ref{SecOptical}) and then the airfoil (see subsection \ref{SecAirfoil}) oscillates driven by a linear motor (see subsection \ref{SecLinMot}). The camera films longer than the theoretical time needed for buffer. Once the oscillation is done, a second calibration is run to potentially remove a drift that occurred during the experiment, and then the sensors are turned off. Finally, the images are saved during several seconds. This time enables the fluid to rest, and the experiment is run again. Particularly during the last days of the experiments, some particles sedimented on the wing before the motor could start. To avoid this problem, we run one (or two if needed) additional movement of the wing between the injections of helium gas and the filmed movement. All the devices are run with labview codes. The parameters of the experiments are given table \ref{tab:Param}.

\subsection{The apparatus}\label{SecApparatus}

\begin{figure}[H]
	\centering
	\begin{subfigure}[b]{0.45\textwidth}
		\centering
		\includegraphics[scale=0.28]{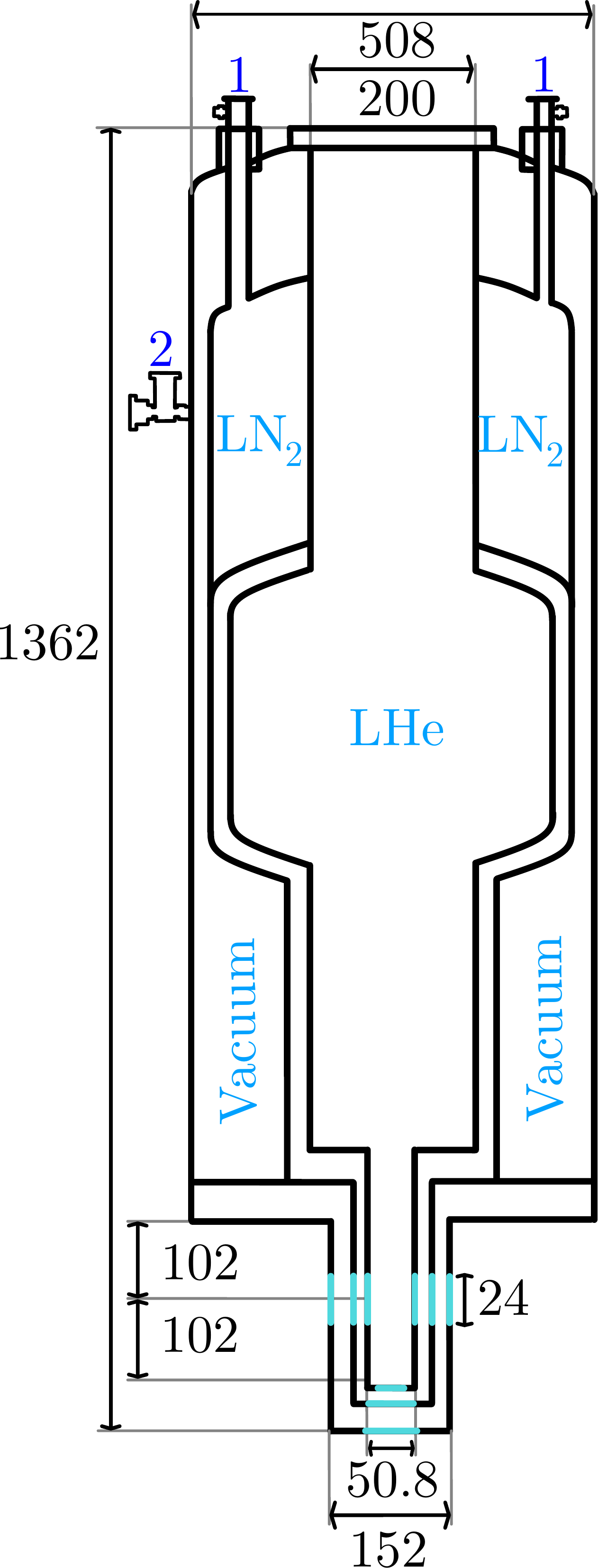}
		\caption{Drawing of the cryostat. The distances are in mm. 1: Access to the vessel filled with liquid nitrogen; 2: Access to the vacuum space.}
		\label{fig:cryostat}
	\end{subfigure}
	\hfill
	\begin{subfigure}[b]{0.45\textwidth}
		\centering
		\includegraphics[scale=0.5]{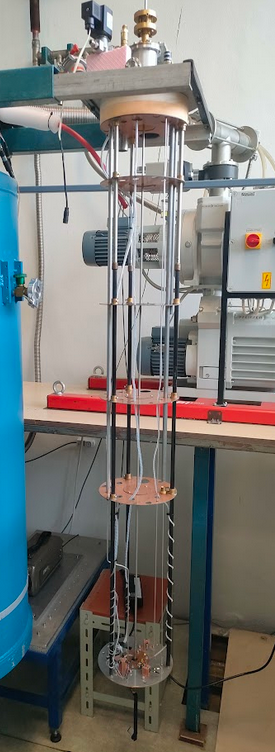}
		\caption{Photograph of the insert.}
		\label{fig:innerCryostat}
	\end{subfigure}
\end{figure}

The cryostat used during this study was designed in Prague and manufactured by Precision Cryogenic Systems\footnote{\href{https://precisioncryo.com/}{website}}. The inner volume of the cryostat can contain 60L of liquid helium. Another vessel with a volume of 35L is filled with liquid nitrogen at 77K. This vessel is in contact with a part of the inner cryostat (see figure \ref{fig:cryostat}). It is used to pre-cool the walls of the crystat. This pre-cooling take a day and a half and helps to avoid as much as possible to lose helium, which is costly. Both vessels are separated from the outer cryostat by a vacuum cavity. This cavity is pumped prior to the experiment to reach a pressure of $\sim 5 \cdot 10^{-5}$ Pa at room temperature. The process is stopped before cooling down the cryostat. The liquid nitrogen bath and the vacuum cavity create isolation, minimizing heat transfer. At saturated vapour, the relation between the pressure and the temperature (see appendix \ref{App:SatVap}), enables us to control the temperature of liquid helium by pumping the vapours with a Pfeiffer pump connected to the cryostat (see appendix \ref{App:SatVap}). At the lowest pressure (and therefore at the lowest temperature) a butterfly valve, is used instead of the by-pass. This valve has been cleaned prior to the experiments. It has a control system particularly useful to keep as much as possible the wished pressure that can be easily disturbed by the injections of helium gas used to mix the particles with the fluid. The pressure is monitored by a Barotron gauge with a sensitivity of 0.1 Torr \footnote{1 Torr = 0.75 hPa}.

Visualization is allowed thanks to windows on each wall. Every window has a diameter of 24 mm and is made of three glasses. The inner one is made of sapphire to reduce the heat input, and the other ones are made of quartz. There are five optical accesses. One is at the bottom, the other four are on each side of the apparatus.

\begin{figure}[h!]
	\centering
	\includegraphics[scale=0.25]{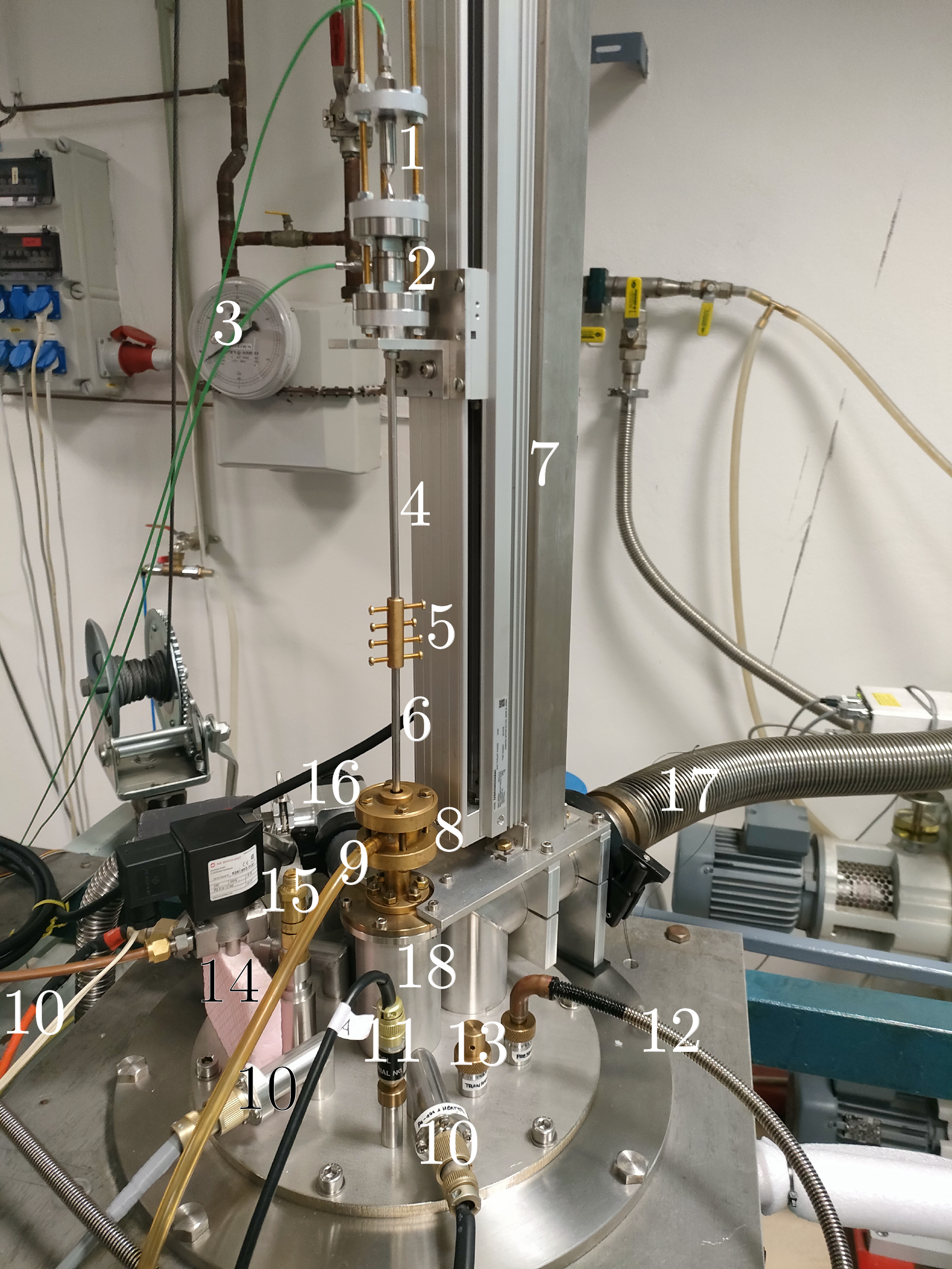}
	\caption{Photograph of the top of the cryostat. 1: force sensor, 2: torque sensor, 3: high insulation cable, 4: shaft connected to the motor and the sensors, 5: connector of the shafts, 6: shaft going in the cryostat and connected to the airfoil, 7: cavity over-pressurised with helium, 8: tube filled with helium to keep the cavity over-pressurised, 9: access to the inner cryostat, 10: cables of the cernex thermometers and the PT-100, 11: level meter to measure the level of liquid helium in the cryostat, 12: access fo the Barotron gauge, 13: access to do the helium transfer, 14: valve used for the injections of helium and deuterium gas, 15: security valve which opens if the pressure inside the crystat is above 900 Torr, 16: access to the retur line for the helium gas, 17: connection to the pump.}
	\label{fig:topCryostat}
\end{figure}

Through the top of the cryostat (see figure \ref{fig:topCryostat}) are the various access point for the Barotron gauge, the wires to the PT-100 and the cernox thermometers (see subsection \ref{SecTemp}), the valve through which are injected the deuterium particles, the shaft connected at both ends to the airfoil motor/sensors assembly and the connections to the pump and for the helium. Various of the connection points are made leak tight O-rings that have been cleaned and greased before the experiments. The access through which the shaft goes through is a cavity overpressured with helium, so if there is a leak it is helium that goes in and not air. This shaft is also regularly cleaned and greased between the experiments.

\subsection{The seeding system}\label{SecSeeding}

The movement of the fluid is known through the presence of tracers. Due to the low density of liquid helium ($\rho_{He} = 146$ kg$\cdot$m$^{-2}$ at 2 K), it is hard to obtain tracers with matching densities. A possible solution is to use solidified hydrogen ($\rho_{H_2} = 88$ kg$\cdot$m$^{-2}$) and deuterium ($\rho_{D_2} = 200$ kg$\cdot$m$^{-2}$) with a specific ratio. However, their melting temperature being different, respectively 14 K and 19 K, $\textbf{D}_2$ solidifies before the hydrogen resulting in pure particles of $\textbf{H}_2$ and $\textbf{D}_2$. Therefore, only pure $\textbf{D}_2$ particles were used. To creates these particles, $\textbf{D}_2$ is first mixed with helium at room temperature for the future particle to all be of similar sizes. The ratio was 1 $\textbf{D}_2$ for 100 He. The mixture ratio between the gases controls the size of the particles \cite{MarcoTracers}. The temperature of injection is slightly higher than that of superfluid transition to delay the aggregation of $\textbf{D}_2$ into the cryostat.

\subsection{The sensors}\label{SecSensors}

\begin{wrapfigure}{r}{0.5\textwidth}
	\centering
	\includegraphics[scale=0.3]{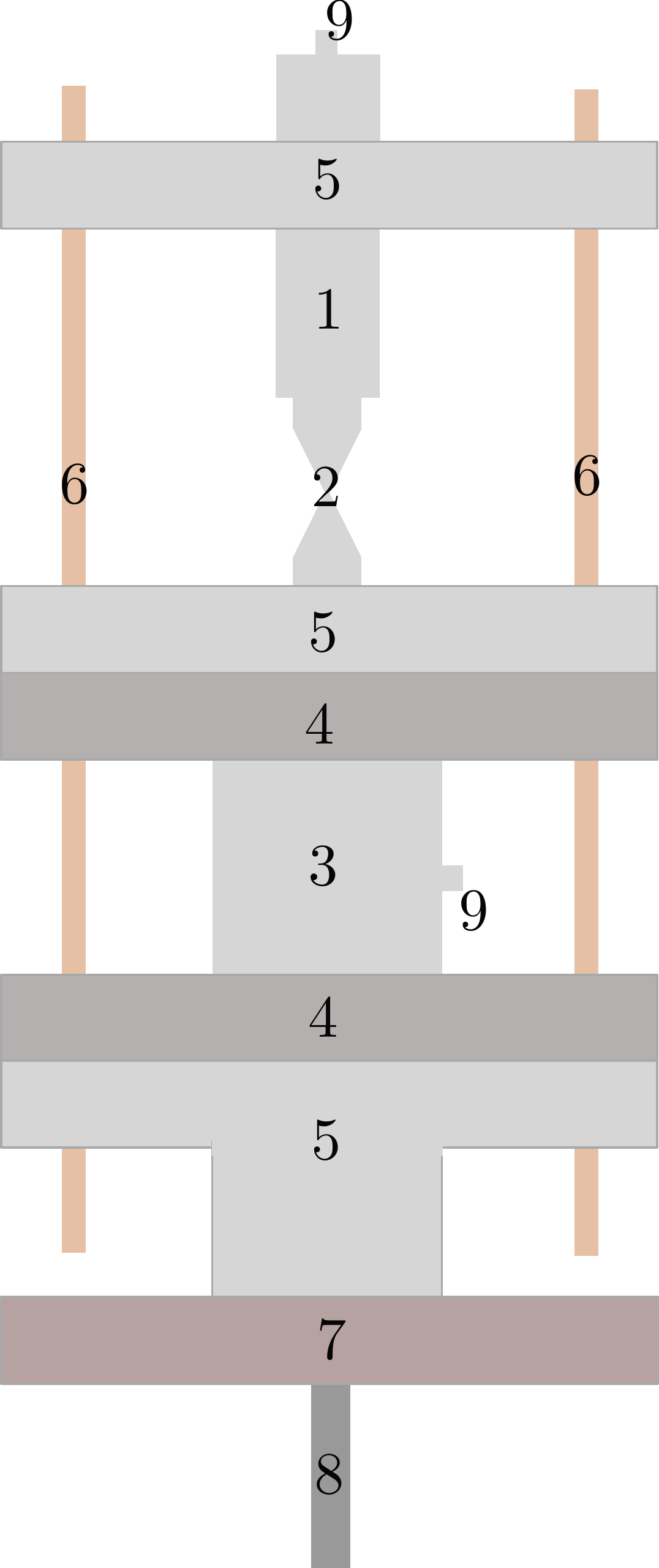}
	\caption{Drawing of the set-up of the sensors. 1: Force sensor; 2: Coupling element (type 9405) ;3: Torque sensor; 4: Mounting flanges; 5: Plates to maintain and use the sensors; 6: (threaded shafts) to fix the set-up; 7: Plate of the motor; 8: Shaft with the airfoil at the end; 9: Connection to the high insulation cable. Drawing not to scale.}
	\label{fig:DrawSensors}
\end{wrapfigure}

To complete the study of the starting vortex, we used a force (9217A1 from Kistler company\footnote{\href{https://www.kistler.com/INT/fr/cp/capteurs-de-force-piezoelectrique-pe-low-force-sensor/P0000647}{Product reference}}) and a torque (9329A from Kistler company\footnote{\href{https://www.kistler.com/INT/fr/cp/couplemetres-de-reaction-93x9a/P0001300}{Product reference}}) sensor. These sensors are positioned above the plate of the motor (n°7 on the figure \ref{fig:DrawSensors}) through which the sensors (n°3 and 1) and the shaft (n°8) controlling the movement of the airfoil are fixed together to the motor. A drawing of the set-up can be seen figure \ref{fig:DrawSensors}. The sensors produce a charge that is transformed into a voltage by a charge amplifier. They are connected to this charge amplifier with a high insulation cable. The charge amplifier is driven by a labview code that I wrote. We then collect the data from the charge amplifier to access the signal of the force and the torque. During the experiments, the range of the torque sensor is set to -2191 pC/N$\cdot$m and the sensitivity to 0.1 N$\cdot$m. A clockwise torque produces a negative charge signal at the sensor and a positive voltage at the analog output of the charge amplifier and vice versa. The range of the sensor is set to -0.5 N and the sensitivity to -103.3 pC/N. A compressive force produces a negative charge signal at the sensor and a positive voltage at the analog output of the charge amplifier and vice versa for a tensile force. The parameters (sensitivity and range) have be chosen to have the less noise possible. Some additional information about the installation of the force and torque sensors can be found in the appendix \ref{App:ForceTorque}.

\subsection{The optical system}\label{SecOptical}

\begin{figure}[h!]

	\centering
	\includegraphics[scale=0.3, angle=-90]{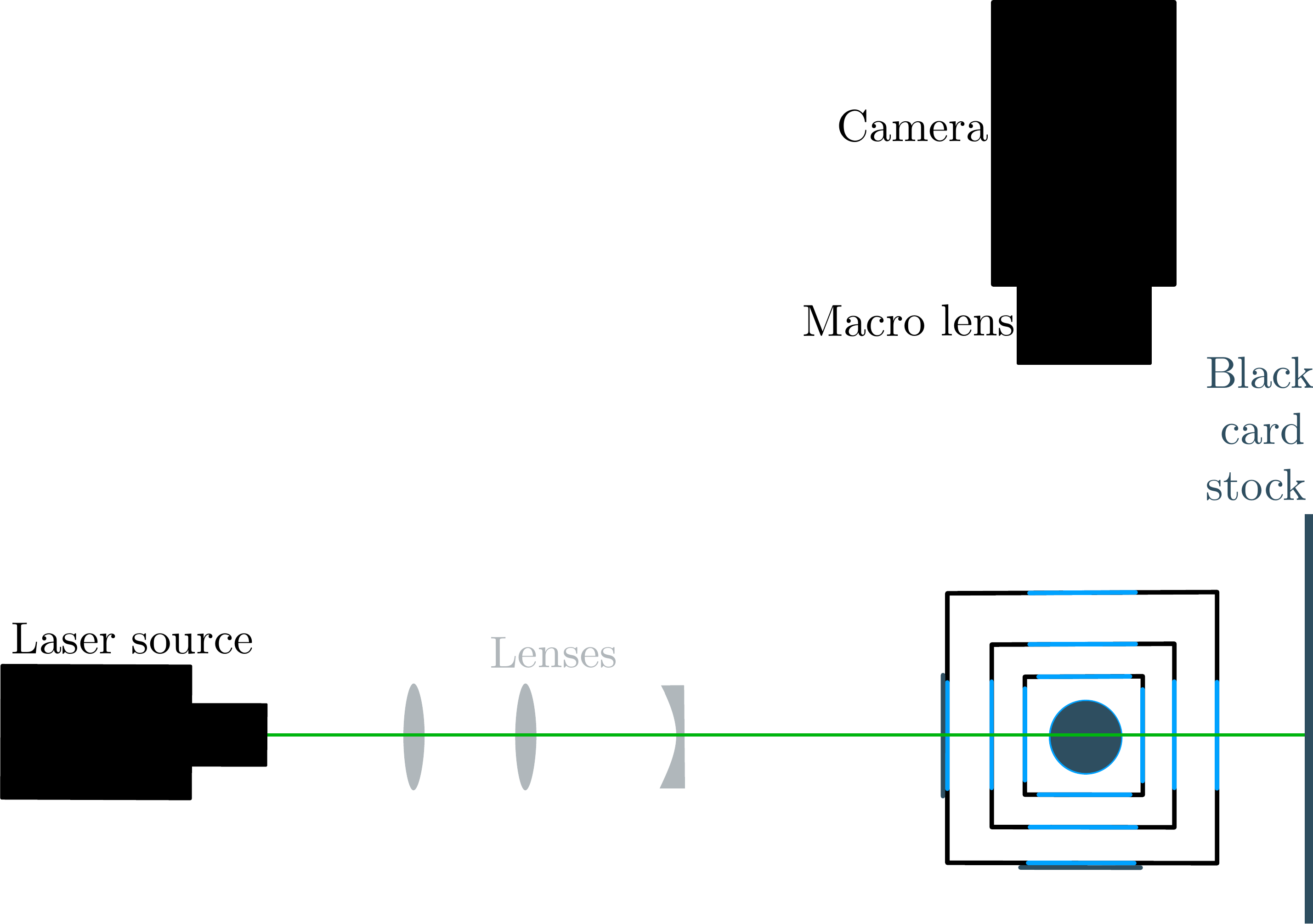}
	\caption{Drawing of the axial cross-section of the experimental apparatus. In blue the windows, in dark gray the black card stocks, in light gray the lenses, in green the light from the laser.}
	\label{fig:optical_access}

\end{figure}

To obtain videos of the experiment, the cryostat is illuminated by a laser sheet through one of the windows. This laser sheet is created with a laser (RayPower 5000 by Dantec Dynamics\footnote{\href{https://www.dantecdynamics.com/}{website}}) emitting at $\lambda = 532$ nm and is dispersed by an optical setup of 3 lenses. The sheet is about 5 cm  high and 1 mm thick. It intersects the airfoil approximately in its middle. Two of the windows that are not used, the one at the bottom of the cryostat and the one at the opposite of the camera, are covered with black card stock to insulate as much as possible the inner cryostat from the light of the room. The window through which the laser sheet enters is also covered with black card stock except from the part through which the laser sheet goes through. The opposite window is not covered to avoid the heating of the window due to the laser, but some black card stock is put further away to stop the laser light.

The high-speed CMOS camera Phantom v12 used to record the data is placed perpendicularly to the laser sheet and equipped with a Canon 180 mm macro lens to focus. The resolution of the camera is $1280\times 800$ px and the framerate during the experiment was $f_c=1$ kHz. In order to have a relation between the physical values and the pixels, a calibration must be performed. As the refractive index of liquid helium is close to 1, the calibration can be done with the cryostat filled with air at room temperature. To do that, a ruler is positioned in the middle of the area of interest and an image of it is taken. The scaling factor obtained is thus 39.1 px/mm.

\subsection{The airfoil}\label{SecAirfoil}

\begin{wrapfigure}{l}{0.5\textwidth}
	\centering
	\includegraphics[scale=0.2]{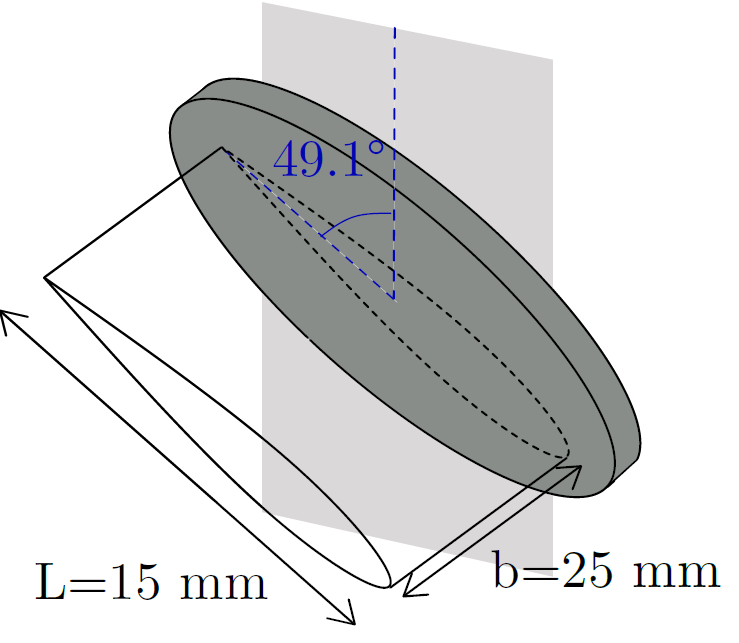}
	\caption{Drawing of the airfoil used during the experiments. In light dray the shaft and in dark gray the support of the airfoil.}
	\label{fig:airfoil}
\end{wrapfigure}

The airfoil used corresponds to the NACA 0012 \cite{NACA0012} airfoil and is made of PMMA as it doesn't break at low temperature and is transparent. The transparency is essential for the airfoil not to be heated by the laser and to avoid thermally driven flows that could be induced by this. It is connected to the main shaft through a metallic plate. The ends of the airfoil and the metallic plate are painted in black to reduce reflection into the camera. The experiment starts with the cross-section of the airfoil fully visible before going down according to the parameters set at the motor. The angle between the airfoil and the vertical is of 49.1°. The angle was determined with the software \texttt{ImageJ}.

\subsection{The linear motor}\label{SecLinMot}

\begin{table}[h!]
	\centering
	\begin{tabular}{c|c|c|c|c|c}
		Experiment  & a (mm$\cdot$s$^{-2}$) &  $v_\textrm{max}$ (mm$\cdot$s$^{-1}$) & D (mm) & $f$ (Hz) & $\delta$ (mm)\\ \hline
		c & 160 & 80 & 40 & 0.5 & 0.16\\
		a & 640 & 160 & 40 & 1 & 0.32\\
		b & 2560 & 320 & 40 & 2 & 0.64\\
	\end{tabular}
	\caption{Parameters used for the motor for each experiment in this study and the corresponding probe scale $\delta$.}
	\label{tab:ParamMotor}
\end{table}

The oscillating airfoil is driven by a linear motor. It is attached to a sliding platform controlled by a rotating screw. The motion chosen consist of segments of constant acceleration. The motor starts from rest and accelerates at $\tilde{a}$ until it has travelled the distance $\Delta D/2$. At the distance it reached the maximum velocity $v_{\textrm{max}}$. Then the motor decelerates at $-\tilde{a}$ until it reaches the final displacement $D$ with the velocity $v=0$. This motion is repeated back and forth to create the oscillations of amplitude $D$. During the experiments, we vary the frequency $f$ of the cycle through the value of the acceleration and the maximum speed that can be reached. The different type of motion used for this study are listed in the table \ref{tab:ParamMotor}.

Even though HeII can exhibit some quantum behaviour, our ability to observe it or not depend on the size of the probe scale $\delta$\cite{LaMantia_2014}. We estimate $\delta$ as being the minimal displacement of the tracers. As their velocity will be calculated with a linear differentiation, the time needed for the particle to cover the distance $\delta$ will be $2/f_c$. And thus

\begin{equation}
	\delta \approx\frac{2v_{\textrm{max}}}{f_c},
\end{equation}
with $v_{\textrm{max}}$ the maximum velocity reached by the airfoil.

\subsection{Monitor the temperature}\label{SecTemp}

To monitor the temperature, we used one PT-100 (platinium thermometer) during the cooling down until approximately the transition of the nitrogen where it ceases to work, and two cernox thermometer previously calibrated. One cernox thermometer have been calibrated during the cooling down of the cryostat, a second one not calibrated ceased to function during the cooling down. The cernox temperature sensors are then used at low temperature. The temperature for each experiment is the mean of the temperature for each each experiment of the set.

\section{Result from the sensors}\label{SecResSensors}

Prior experiments have shown that the state of opening of the butterfly valve does not influence the results. The data processing of the sensors is quite simple. For each signal we first remove the data points measured before the motor started, thanks to the knowledge of its starting time that is saved in .txt files through the experiments. This is necessary as the starting time after the beginning of the data recording vary from one experiment to another. However, this starting time is close to the real one, but some doubt can be raised as some randomness has been observed for the starting time of the second part of the movement. We then average all the signals from one set of experiments. However, the signals remain quite dispersed from one experiment to another. So I decided to remove the signals above (resp below) the mean signal plus (minus) the standard deviation as it reduces shifts. So respectively 81 and 94$\%$ of the signals were kept for the torque and the force.\newline

\begin{figure}[h!]
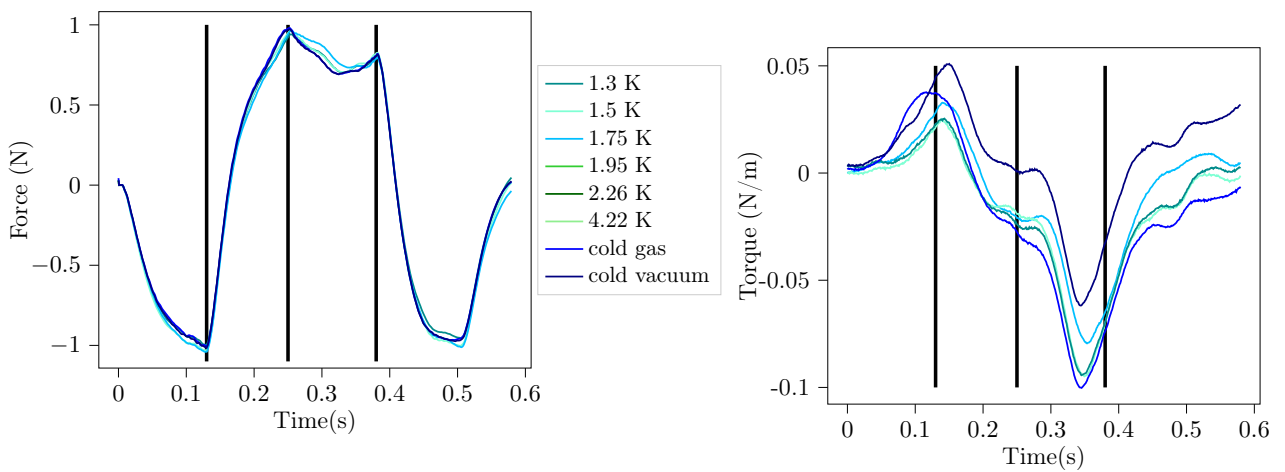

	\centering
	\begin{subfigure}[b]{0.54\textwidth}
		\centering
		\input{Images2Hz_sensors_force}
		\caption{Signal of the force at 2 Hz.}
		\label{fig:F2Hz}
	\end{subfigure}
	\hfill
	\begin{subfigure}[b]{0.44\textwidth}
		\centering
		\input{Images2Hz_sensors_torque}
		\caption{Signals of the torque at $f$=2 Hz.}
		\label{fig:T2Hz}
	\end{subfigure}
	\caption{Signal of the force and torque for the different set of experiments at $f$=2 Hz. The black lines symbolizes the times when the acceleration of the airfoil changes its sign. The legend will be the same for all the figures in \ref{SecSensors}.}
	\label{fig:FT2Hz}
\end{figure}

\begin{figure}[h!]
	\centering
	\includegraphics[scale=0.3]{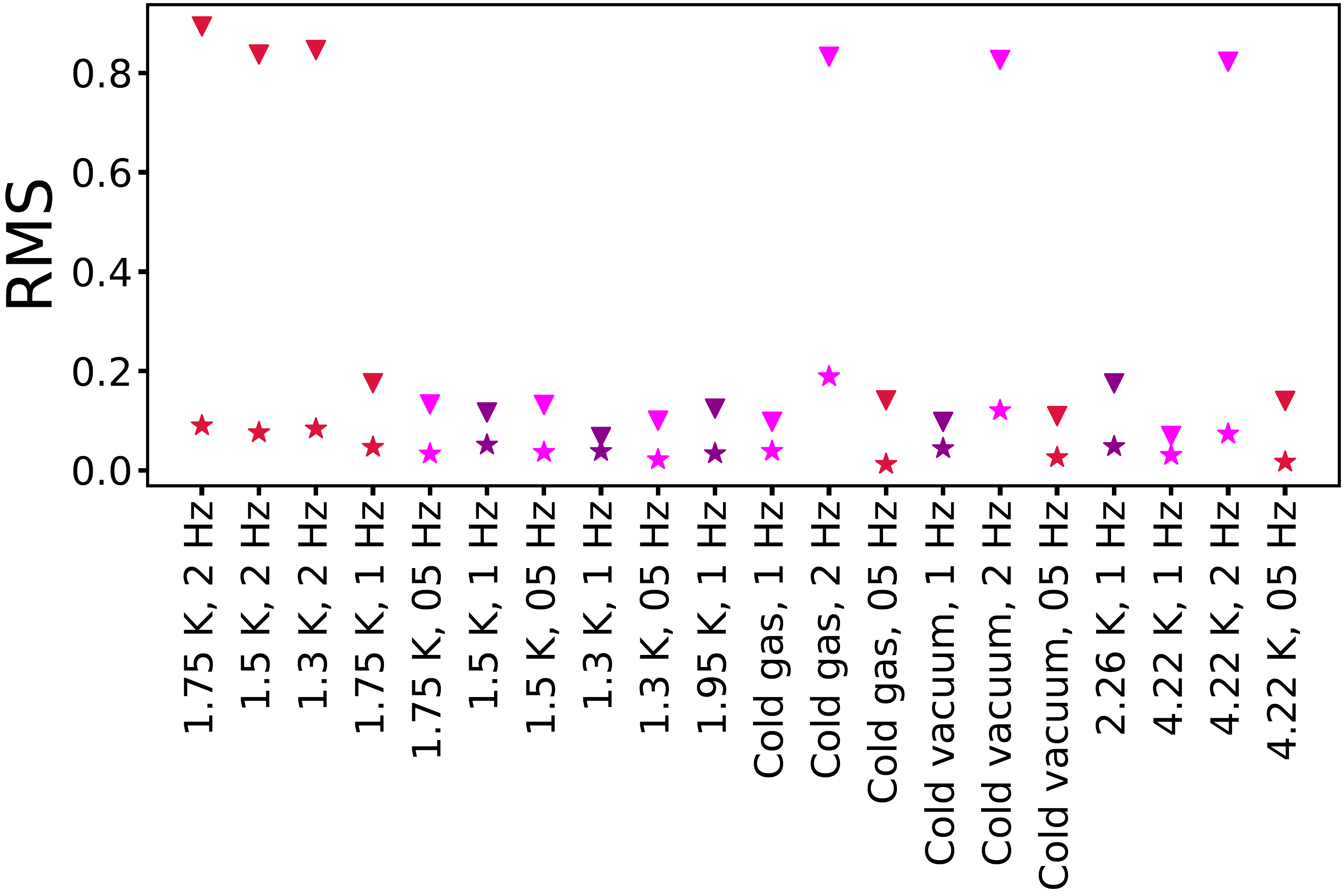}
	\caption{RMS of the signals for each set of experiment. In red: $f$=2 Hz; in purple: $f$=1 Hz; in pink: $f$=0.5 Hz; $\star$ correspond to the RMS for the torque and $\blacktriangledown$ correspond to the RMS for the force.}
	\label{fig:RMS}
\end{figure}

The pattern of the signal of the force is coherent with what we would expect. Indeed, it is consistent with the simple formula of the equation of motion of a particle from \cite{PhysRevB.90.014519}
\begin{equation}
	\frac{du_p}{dt} = \underbrace{K_p\frac{Du_f}{Dt}}_{\textrm{Added mass}} + \underbrace{Bg}_{\textrm{Buoyancy force}} + \underbrace{S(u_f-u_p)}_{\textrm{Stokes drag}} = \sum \vb{F},
\end{equation}
with the subscripts $p$ denoting the particle and $f$ denoting the fluid. As the movement of the airfoil consists of 4 phases of constant acceleration, this explains the plateaus and the symmetry between the slowest and the largest values. In the figure \ref{fig:F2Hz}, the only features we could pinpoint are the plateaus, the other parts being remarkably similar between the different experiments. Any difference in the signals that could stand out are not found at the other frequencies, so it would be hazardous to draw any conclusion from it. Furthermore, as it was previously expressed, the obtained signals are quite dispersed around the mean values, as a result, the RMS of the filtered set of signals (i.e. the RMS in the figure \ref{fig:RMS} was calculated after the signals out of the range of work were removed) is close to the extreme values for the force. That is also why any observation made on the values of the plateau could be criticized. At this frequency, we can observe that the torque is asymmetric. This asymmetry is due to the fact that the metallic plate holding the wing is not centered with the main shaft connected to the sensors. The torque in cold vacuum seems to be higher than the one in helium. This feature was already observed at room temperature between the gas and the vacuum (see appendix \ref{App:sensors}). On the contrary, we do not observe any shift or difference in the strength of the force between the data in the different mediums as expected (see appendix \ref{App:sensors}).\newline

\begin{figure}[h!]
	\centering
	\begin{subfigure}[b]{0.48\textwidth}
		\centering
		\input{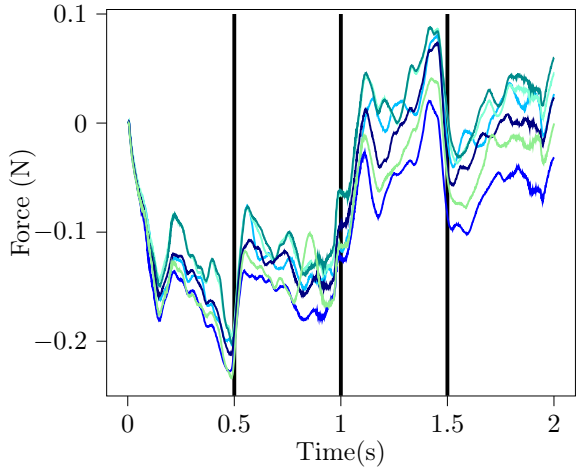}
		\caption{Signals of the force at $f$=0.5 Hz.}
		\label{fig:F05Hz}
	\end{subfigure}
	\hfill
	\begin{subfigure}[b]{0.48\textwidth}
		\centering
		\input{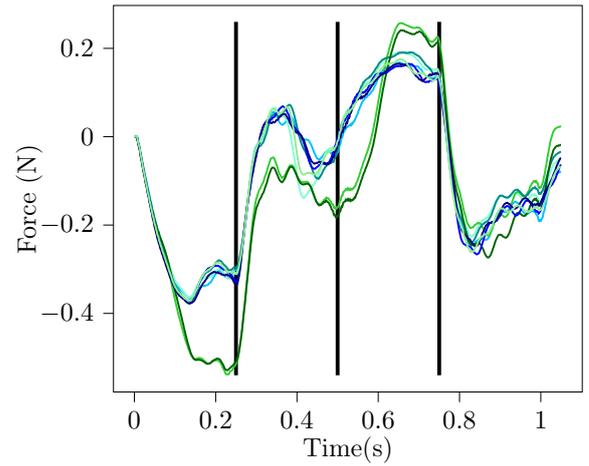}
		\caption{Signals of the force at $f$=1 Hz.}
		\label{fig:F1Hz}
	\end{subfigure}
	\caption{Signal of the force for the different set of experiments at $f$=1 and 0.5 Hz. The black lines symbolizes the times when the acceleration of the airfoil changes its sign.}
	\label{fig:F1and05Hz}
\end{figure}
\begin{figure}[h!]
	\centering
	\begin{subfigure}[b]{0.48\textwidth}
		\centering
		\input{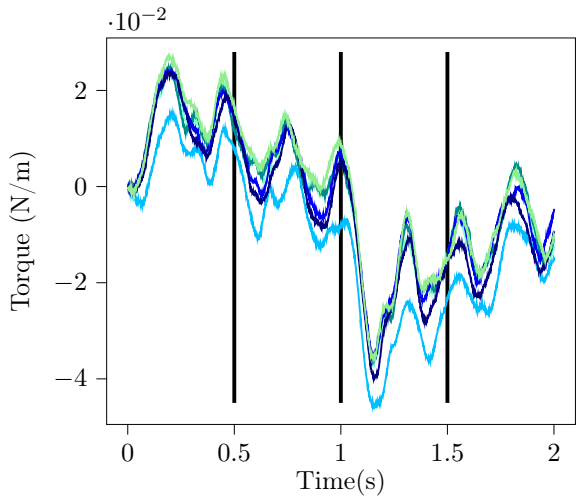}
		\caption{Signals of the torque at $f$=0.5 Hz.}
		\label{fig:T05Hz}
	\end{subfigure}
	\hfill
	\begin{subfigure}[b]{0.48\textwidth}
		\centering
		\input{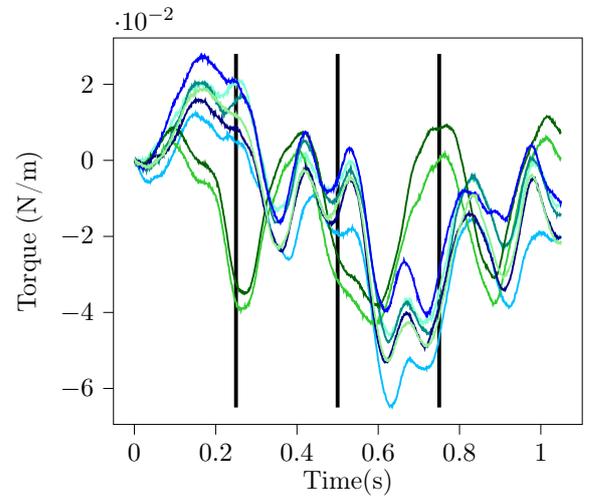}
		\caption{Signals of the torque at $f$=1 Hz.}
		\label{fig:T1Hz}
	\end{subfigure}
	\caption{Signal of the torque for the different set of experiments at $f$=1 and 0.5 Hz. The color are the same as in figure \ref{fig:F1and05Hz} and \ref{fig:FT2Hz}. The black lines symbolizes the times when the acceleration of the airfoil changes its sign.}
	\label{fig:T1and2and05Hz}
\end{figure}

The pattern of the signals at 0.5 and 1 Hz (see figures \ref{fig:F1and05Hz} and \ref{fig:T1and2and05Hz}) are quite different from the ones at 2Hz. Previous experiments in air and in water have shown the same behaviour at 0.5 Hz for the force when using the same sensor and the same apparatus. However, with a different sensor (less sensitive and with a different set-up), we observed the expected pattern as the one at 2 Hz. An explanation of these differences is that the force is not strong enough to be accurately sensed by the sensor and is more influenced by exterior parameters. However this was not the case at 1Hz. For the torque, some oscillations appear. Prior experiments with the same set-up in the cryostat in vacuum and helium gas at room temperature have shown no oscillations, drawing out the possibility of the wing being light or the shaft not being rigid enough. The differences of behaviour with the data at 1.95 and 2.26K hold doubt on any physical interpretation for these oscillations in the torque and the pattern of the signals of the force, and underlines the role of the high insulated cables in obtaining data. Indeed, the high insulated cables have been touched to try to obtain better results before doing the experiments at other temperatures. An interesting feature is that the higher the acceleration (and though the force and the torque), the less impacted the signals seem to be compared to the preliminary experiments. We can assume that the lower the signal, the more impacted it will be by noise of external influences through the cables. The position of the cables is thus the major criticism as it is extremely sensitive to their environment, another cable, how it touches the table... So drawing any real conclusion about the data obtained from these sensors would require a guide for the cable to reduce or even eliminate any influence from the environment. A last remark we can make is that we do not see any particular behaviour related to the beginning of the movement that could give us some information to the shedding of the starting vortex.

\section{Processing of the movies}

In order to process the data, we use several computing steps that each takes several hours. The first step is to determine the trajectory of the airfoil and the temperature during the experiments. Then, the airfoil is masked and the reflections from the support is reduced. Only after this pre-processing, the particle tracking velocimetry is done. The data processing, except for the Particle Tracking Velocimetry (see section \ref{PTV}), is done with \texttt{python}.

\subsection{Pre-processing of the images}

Before processing the data we wish to remove the airfoil and the reflections from its elliptical support, as it could be interpreted as particles. First, the airfoil must be located\cite{BachelorThesis} on each frame of the movie. To do so, we do a discrete convolution with an image of the airfoil created with its mathematical expression, rotated with the right angle and blurred with a Gaussian filter, as a kernel. From this convolution we obtain the position of the airfoil by taking the coordinates of the point where the convolution is the highest with the function \texttt{find}$\_$\texttt{peaks} from scipy. Therefore we can obtain the times at which the airfoil starts to move, escape from the image and re-enter the region we are filming. These times are crucial to determine the temperature of the fluid during the movement of the airfoil. To recover the temperature we average the temperature recorded between the start of the airfoil and its re-entering in the frame. We also cut the movie to only keep the images between the starting time and the time the airfoil re-enters the region of interest. We can then mask the airfoil. Thanks to the knowledge of its position, we apply a black mask (obtained with its mathematical formula and rotated accurately) on the airfoil. The mask is slightly larger than the airfoil itself because of the reflection all around that is detected by the particle detection code as particles. There are also a lot of reflections due to the support of the wing, even though it was painted to avoid this issue as much as possible. To reduce the reflections, for each set of data, one picture was taken, its particles were removed and it was cropped to only keep the support. Then for each pictures of the experiments this image of reference was subtracted. Some tests were done using several pictures and taking the mean instead. However, with this method some interference pattern appeared, probably due to the fact that the position of the wing might be known with a precision of some pixels, which is good enough to mask the airfoil but causes the interferences. An example of the data processing can be found in the appendix \ref{AppDataProcessing}.

\subsection{Particle Tracking Velocimetry (PTV)}\label{PTV}

The behaviour of the fluid is known through the deuterium particles that act as tracers. Their trajectory is determined with a Particle Tracking Velocimetry (PTV). The methods used to seed He does not enable a large enough amount of particles to run a Particle Imaging Velocimetry (PIV) instead. The PTV is done using the open software \texttt{ImageJ} with the plugin \texttt{mosaic} created by Sbalzarini et al. \cite{PTV} for biological purposes. It detects the particles as local intensity maxima in the upper $n$th percentile of intensity values of the frame. The parameter $n$ varies from 0.12 to 0.15 in our cases. The approximate radius of the particles used to run the code was 4 pixels. It was slightly larger than the particles but smaller than the inter-particle separation. Once the particles are detected, they are then linked into trajectories. The linking range $r$ , i.e. the particle can disappear between the frames $t$ and $t+r$, was set to 2. The maximum displacement between two succeeding frames was $N=10$ pixels. This algorithm can be used with a Brownian movement. In other words, contrary to the usual algorithms used to compute a PTV in the context of fluid dynamics, no assumption based on the previous direction, velocity or acceleration of the particle is used. Indeed, in superfluids, the particles can display some Brownian behaviour. Despite it being unlikely in our experiment, an other algorithm would miss such behaviour. The velocity of the particle in each frame is calculated using a second linear differentiation 
\begin{equation}
	\vec{v}_n = \frac{\vec{x}_{n+1}-\vec{x}_{n-1}}{2\tau},
\end{equation}
with $\vec{x}$ the position of the particle, $\tau$ the interval between two frames and the subscript indicate the number of the frame. For the start and the end of the movie, the velocity is calculated using the first and the last frame and divided by $\tau$. As the linking range is 2, a linear extrapolation between the frames $n-1$ and $n+1$ is used to know the position of the particle in the frame $n$. \newline

\begin{figure}[h!]
	\centering
	\begin{subfigure}[b]{0.45\textwidth}
		\centering
		\includegraphics[scale=0.3]{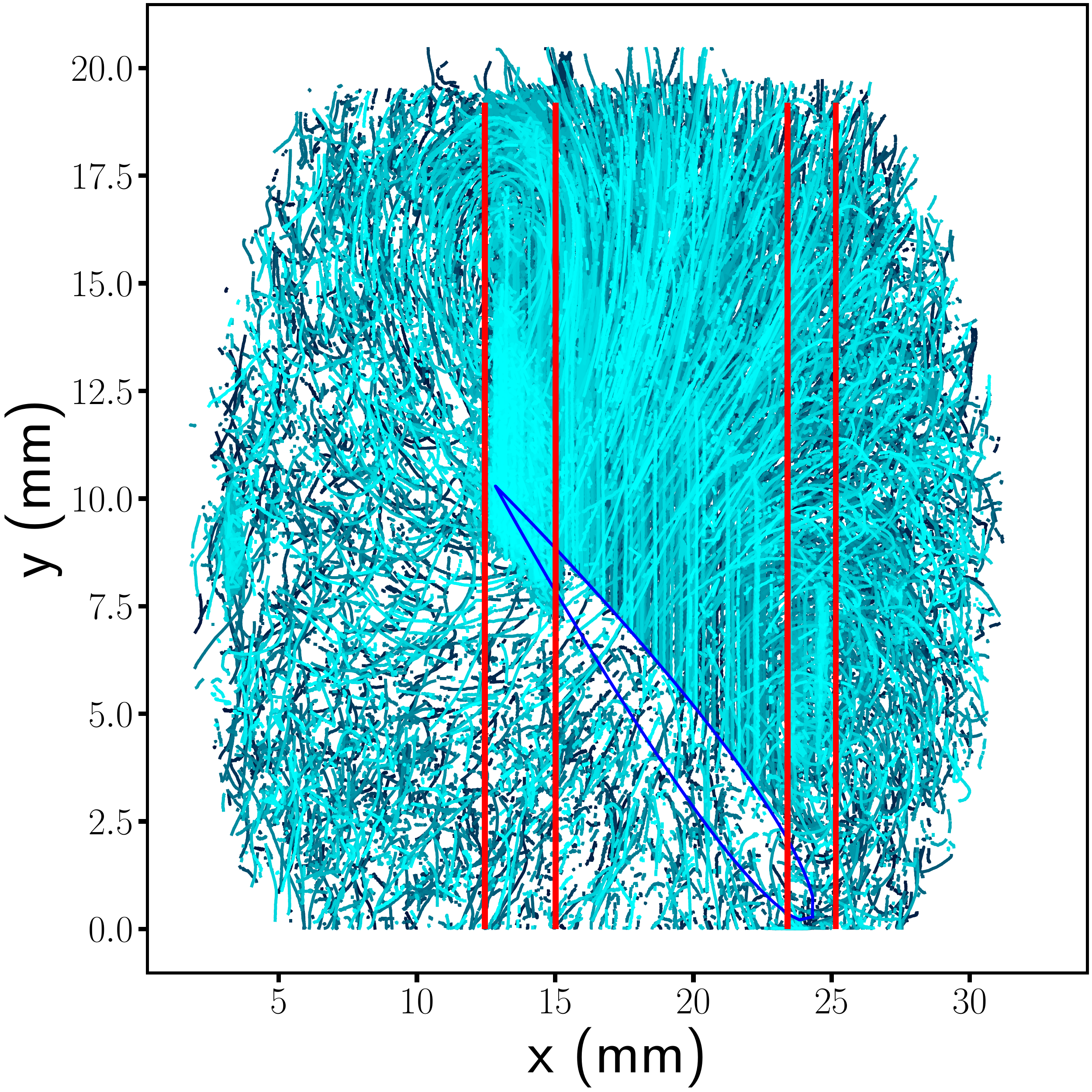}
		\caption{Trajectories before the filtering. Between the red vertical lines are the special areas where the filtering of the tracks has larger conditions.}
		\label{fig:TrajRaw}
	\end{subfigure}
	\hfill
	\begin{subfigure}[b]{0.45\textwidth}
		\centering
		\includegraphics[scale=0.3]{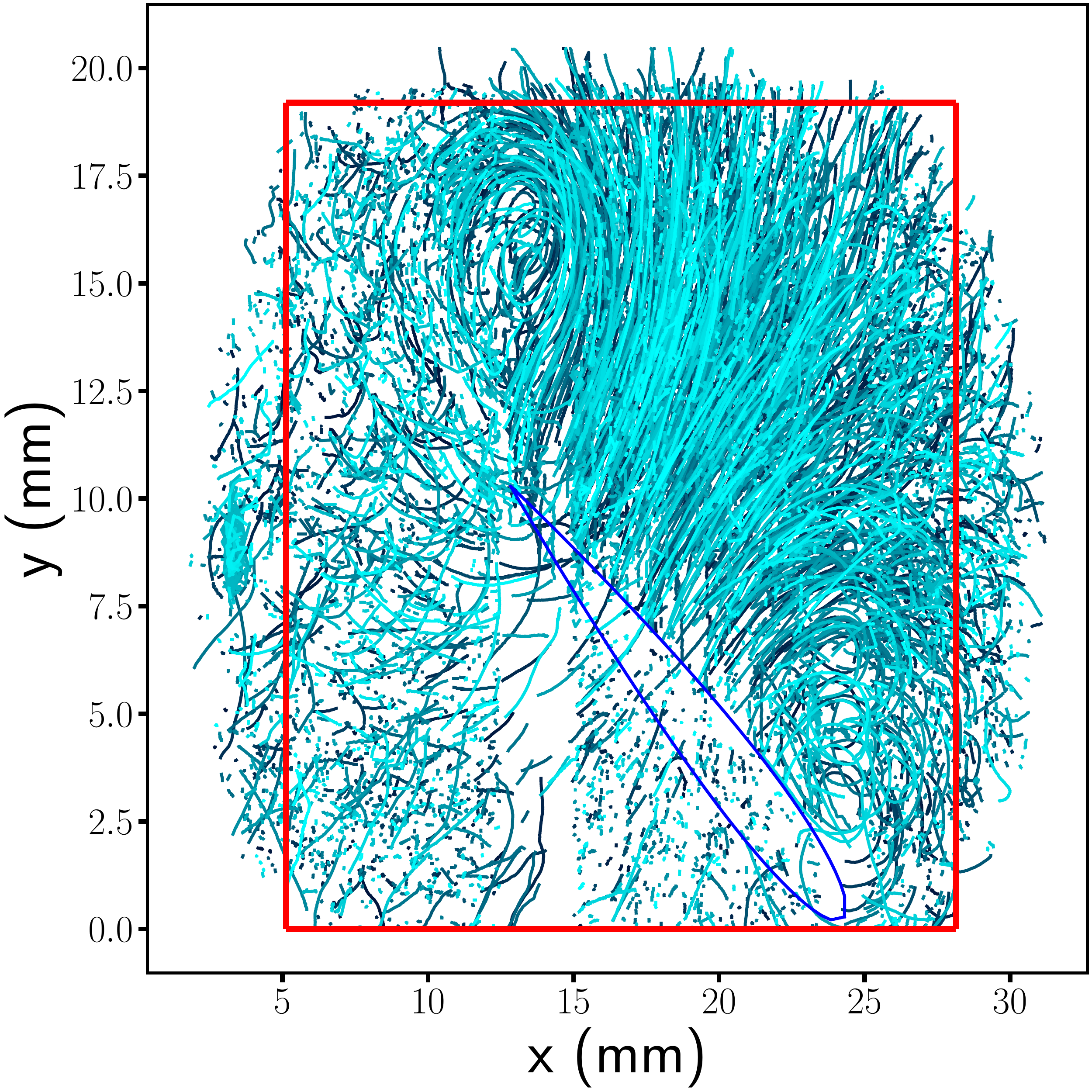}
		\caption{Trajectories after the filtering. Inside the red square is the region that will be considered for the following processing.}
		\label{fig:TrajFilt}
	\end{subfigure}
	\caption{Trajectories from the experiment Ec up to 0.3s before and after the filtering of the trajectories. In dark blue the outline of the airfoil.}
	\label{fig:Traj}
\end{figure}

Once we have the results from the PTV we can have a look at the tracks themselves. First, we remove the tracks with less than 5 frames. Then, despite the masking, we still have tracks related to the airfoil. This is due to reflections that have not been removed with the preprocessing, and to some particles that happen to have settled on the wing before the start of the movement. To reduce the number of those tracks, we try to remove as much as possible the straight trajectories. To be removed, several conditions must be verified. There is condition on their length, as the majority of the tracks due to the wing are long, and a condition on their straightness. 
\begin{itemize}
	\item For the tracks lasting between 15 and 50 frames, if less than 25$\%$ of the $x$ coordinate is out of the range [$x_{\textrm{mean}}$-5; $x_{\textrm{mean}}$+5] pixel, the trajectory is removed.
	\item For trajectories longer than 50 frames, the condition is broader. The range is [$x_{\textrm{mean}}$-10; $x_{\textrm{mean}}$+10] and the percentage is 45$\%$ as their coordinates can more vary through the frames.
	\item At the edges of the airfoil we have more trajectories related to the airfoil, so in the $x$ coordinate between the red boxes (cf figure \ref{fig:TrajRaw}), the range is [$x_{\textrm{mean}}$-17; $x_{\textrm{mean}}$+17] and the percentage is 50 $\%$ whatever the length of the trajectory.
\end{itemize}
Considering a percentage of the trajectory being out of the range enable us to take into account straight trajectories with an abrupt change in their $x$ values, as it is often the case for the tracks related to the wing. This filter does not remove all the trajectories from the airfoil, but most of it. A result of this filter can be found figure \ref{fig:Traj}. At the end of this process, 54.2$\%$ of the trajectories have been removed after the filtering of the short tracks. After all the filters, 27.1$\%$ of the initial tracks are remaining. 

In the following, we consider the data in the red square visible figure \ref{fig:TrajFilt}.

\subsection{Pseudovorticity and vortex tracking}

\begin{figure}[h!]
	\centering
	\includegraphics[scale=0.4]{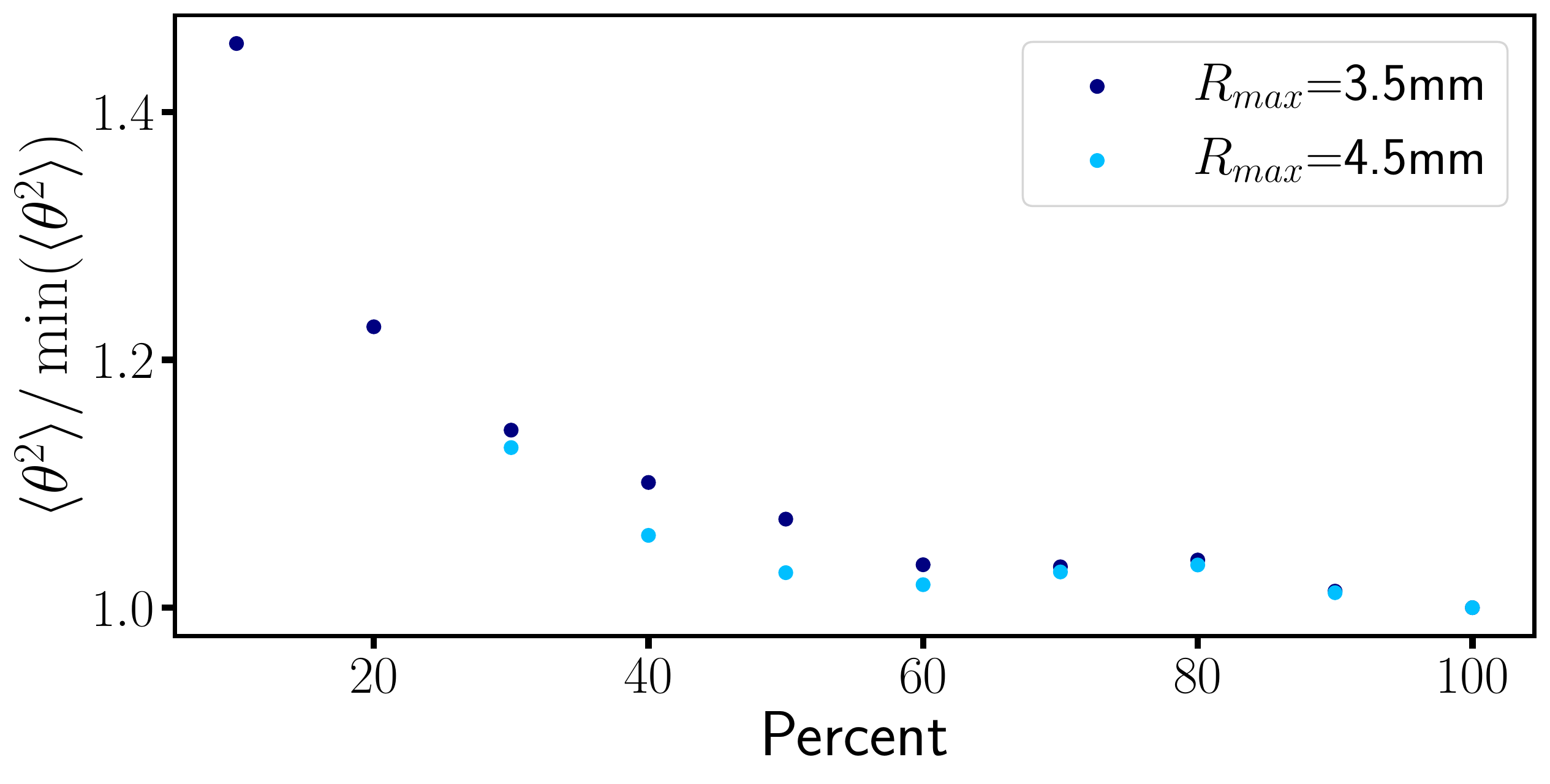}
	\caption{Value of $\langle\theta^2\rangle/ \min(\langle \theta^2 \rangle)$ for different  $R_{\textrm{max}}$ at different percentage of the of available particle positions used to compute the pseudovorticity.}
	\label{fig:convergence}
\end{figure}

As we wish to study the vortex shedding past an airfoil, an interesting parameter is the vorticity. The vorticity is defined as $\vb*{\Omega} (\vb{r}) = \grad\times \vb{v}(\vb{r})$ with $\vb{v}(\vb{r})$ the velocity field of the fluid. However, this formula cannot be computed when the fluid is studied by Lagrangian methods as it doesn't provide a full knowledge of the velocity field. That's why the pseudovorticity was introduced as

\begin{equation}\label{eq:pseudovorticity}
	\vb*{\theta} (\vb{r},t) = \langle \frac{[(\vb{r}_i-\vb{r})\times \vb{v}_i]_z}{(\vb{r}_i-\vb{r})^2}\rangle_{\mathcal{S}},
\end{equation}
with $i$ the label of the considered particle, $z$ the axis perpendicular to the observation plane (so we only consider particles in the observable plane) and the brackets stand for the average over all the particles in the set $\mathcal{S}$ of lagrangian particles \cite{Outrata_Pavelka_Hron_LaMantia_Polanco_Krstulovic_2021}. This set corresponds to a certain time window centered at time $t$ and within a annulus region of radius $R_{\textrm{max}}$ and $R_{\textrm{min}}$ centered at the point $r$.  If $R\rightarrow 0$, the pseudovorticity tends towars half the vorticity ($\theta\rightarrow\frac{\omega}{2}$). On the other hand, if $R\rightarrow\infty$, the pseudovorticity vanishes. A relation exists between the pseudovorticity and the vorticity (see appendix \ref{App:pseudovorticity}), however, we are not able to quantify it with our available data. Thus, this is the quantity that we will be computing in our data processing. The time window taken was $\Delta t$ = 5 and $R_{\textrm{min}}$=0.25 mm following the work of \cite{BachelorThesis}. I computed the pseudovorticity in an evenly spaced grid with  $\Delta x = \Delta y$ = 6 px = 0.15 mm and with $x\in$ [5.1;28.1] mm and $y\in$ [0;19.2] mm as shown in the figure \ref{fig:TrajFilt}. Even though we removed as much as possible the detected particles related to the airfoil, we still wish to avoid effects from it so we enforce the pseudovorticity inside a region slightly larger of the airfoil to zero. To decide which $R_{\textrm{max}}$ to choose, we compute a convergence study with the objective of the data to converge. To do so, we compute the pseudovorticity using 10 to 100 $\%$ of the available particle positions. Once it is done we can calculate the quantity $\langle\theta^2\rangle$ which is the mean over all the frames of the pseudovorticity. The results are given figure \ref{fig:convergence}.

The squared pseudovorticity is for each set, divided by the minimum value of the set of  $\langle\theta^2\rangle$ (which corresponds here to the value at 100$\%$. As we explained $\theta \xrightarrow[R\rightarrow 0]~\omega / 2$ and  $\theta \xrightarrow[R\rightarrow \infty]~0$, so we do not expect the values to become perfectly constant once it has converge. With the larger value of $R_{\textrm{max}}$ $\langle\theta^2\rangle$ seems to decrease to a local minima before reaching the real one after a small increase. For$R_{\textrm{max}}$=3.5mm, the values decrease more slowly. We could argue that the minimum in the largest $R_{\textrm{max}}$ reflect the fact that the data has converged. What is remarkable is that the value are quite similar between the different values of $R_{\textrm{max}}$ after 60$\%$. So we decide to use $R_{\textrm{max}}$=3.5mm in the following.\newline

Once we have the pseudovorticity field $\mathcal{F}$ we can locate and track the vortices. This is done using a code developed by Jiri Blaha and Sebastian Busch. First the code selects the points with the highest value of pseudovorticity, the points selected are the highest 40$\%$. Once it has the points with the highest values, it creates clusters with group of points separated by less than $\epsilon=2$ grid points. It then determines the center of the vortex computing the mean of the coordinates of the points in a cluster weighted by their pseudovorticity value. Finally, only the clusters with more than 40 points are kept as a real vortex. Then the different vortices in each frame must be connected to create a trajectory. If the distance between them is less than $r_v=10$ and if they are separated by less than 10 frames they are considered as part of the same trajectory. All trajectories with shorter than 20 points are removed.

\section{Results}

The results presented in the following section are only preliminary results as the experiment could have only been done at the end of May because there were delays in the arrival of materials and a lot of issues with the set-up occured.

\subsection{The Pseudovorticity}

\begin{figure}[h!]
	\centering
	\begin{subfigure}[b]{0.42\textwidth}
		\centering
		\includegraphics[scale=0.3]{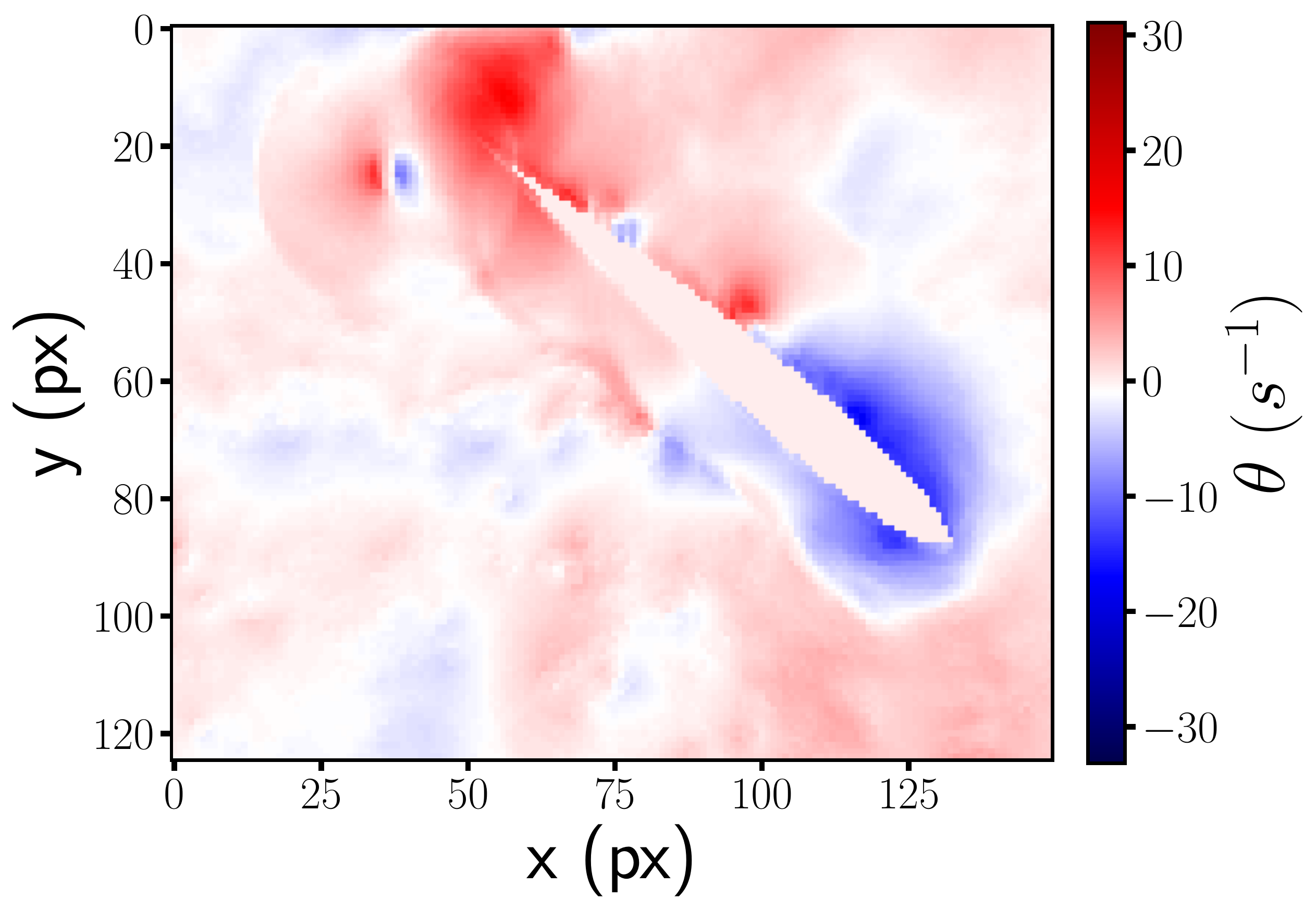}
		\caption{Pseudovorticity map at $\tilde{t}$=0.45.}
		\label{fig:psv1}
	\end{subfigure}
	\hfill
	\begin{subfigure}[b]{0.42\textwidth}
		\centering
		\includegraphics[scale=0.3]{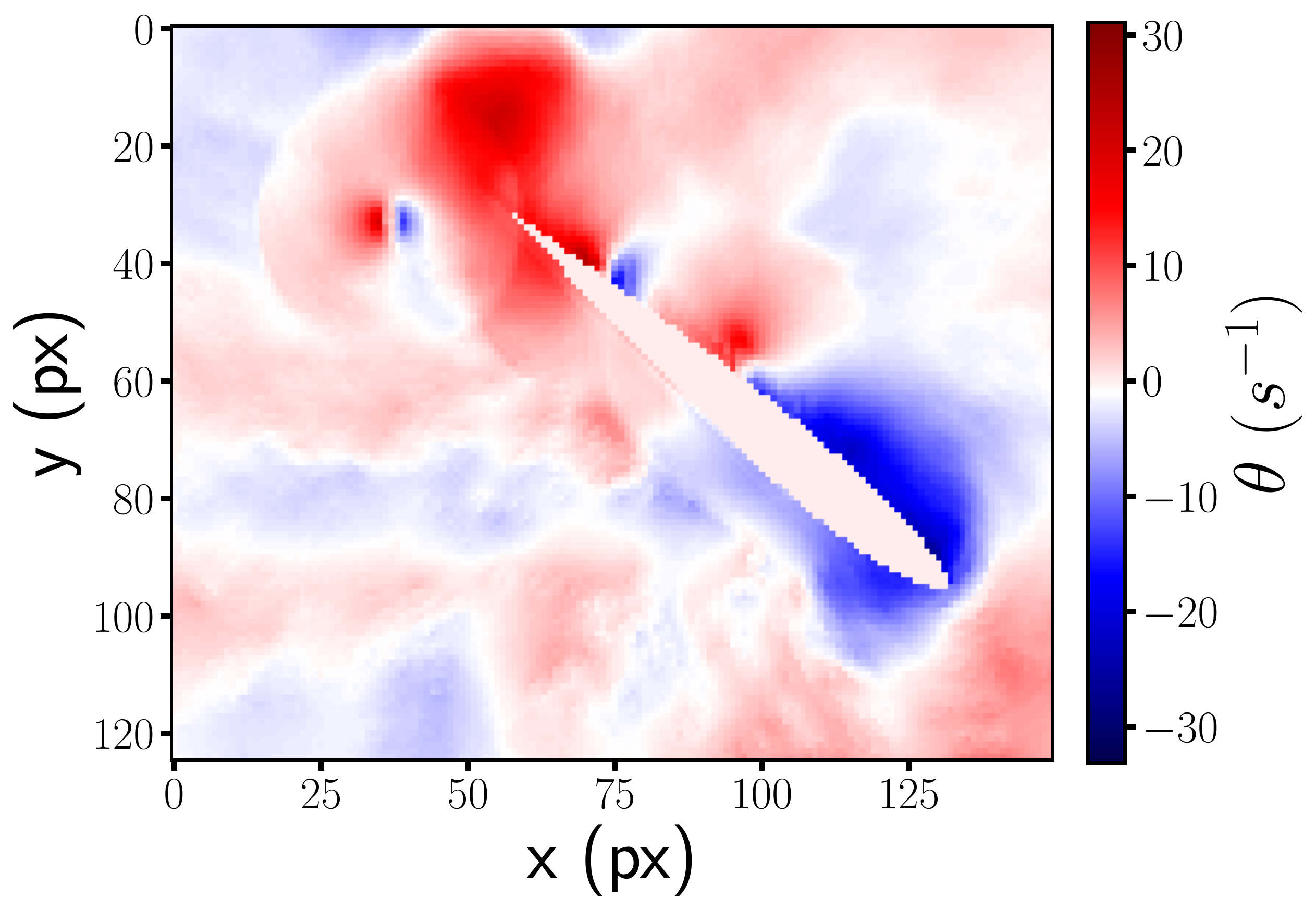}
		\caption{Pseudovorticity map at $\tilde{t}$=0.65.}
		\label{fig:psv2}
	\end{subfigure}
	\hfill
	\begin{subfigure}[b]{0.42\textwidth}
		\centering
		\includegraphics[scale=0.3]{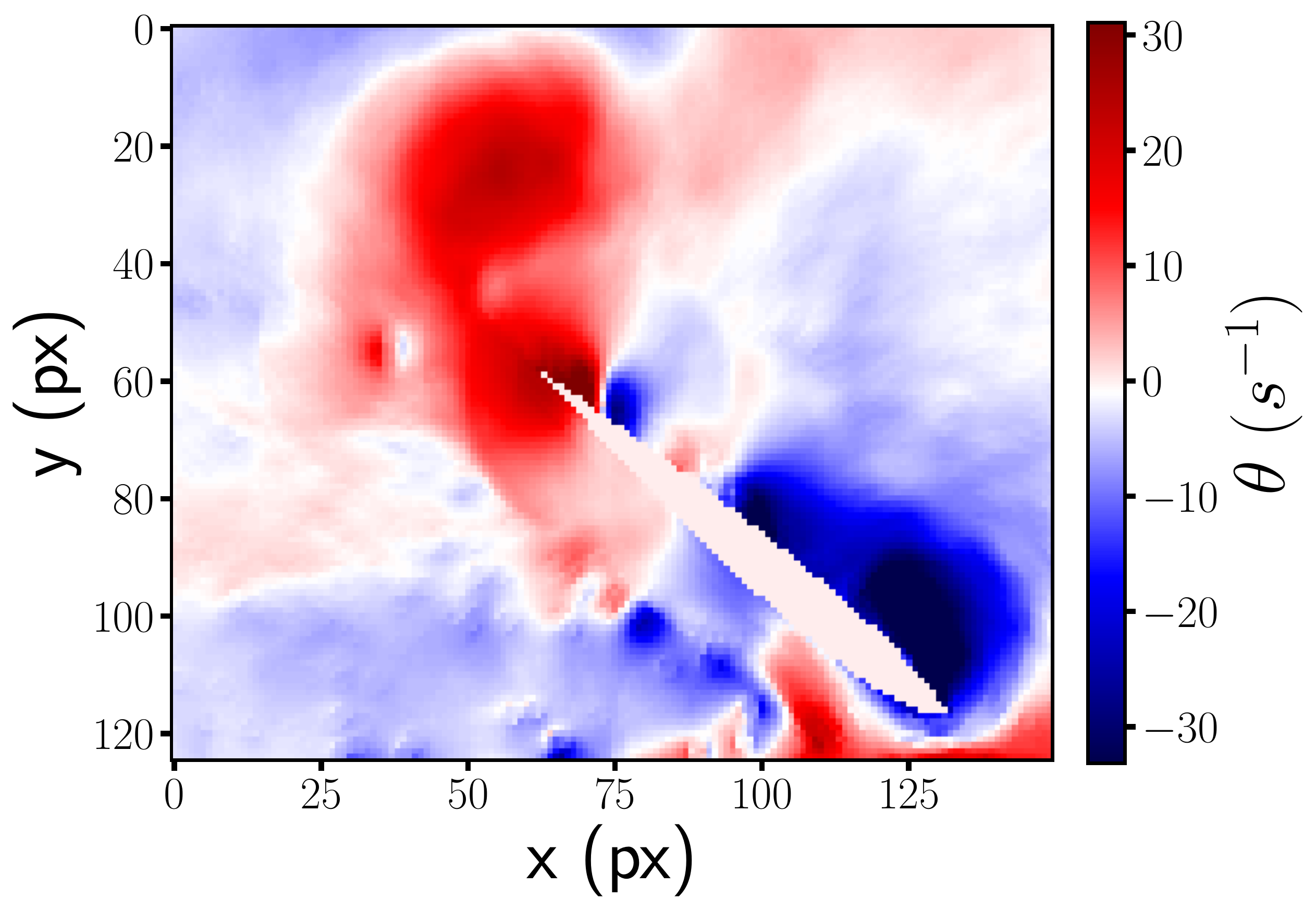}
		\caption{Pseudovorticity map at $\tilde{t}$=0.91.}
		\label{fig:psv3}
	\end{subfigure}
	\hfill
	\begin{subfigure}[b]{0.42\textwidth}
		\centering
		\includegraphics[scale=0.3]{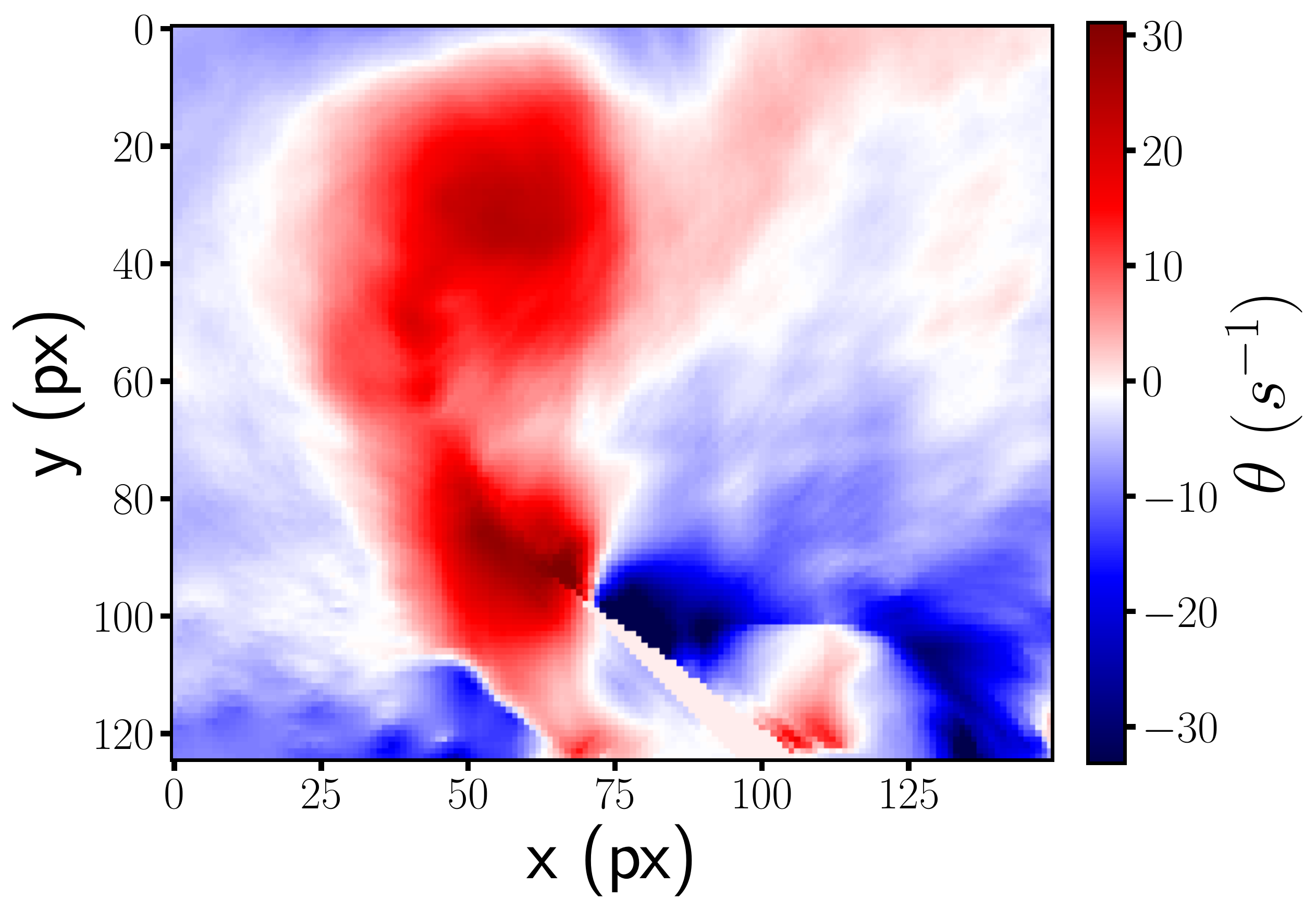}
		\caption{Pseudovorticity map at $\tilde{t}$=1.23.}
		\label{fig:psv4}
	\end{subfigure}
	\hfill
	\begin{subfigure}[b]{0.42\textwidth}
		\centering
		\includegraphics[scale=0.3]{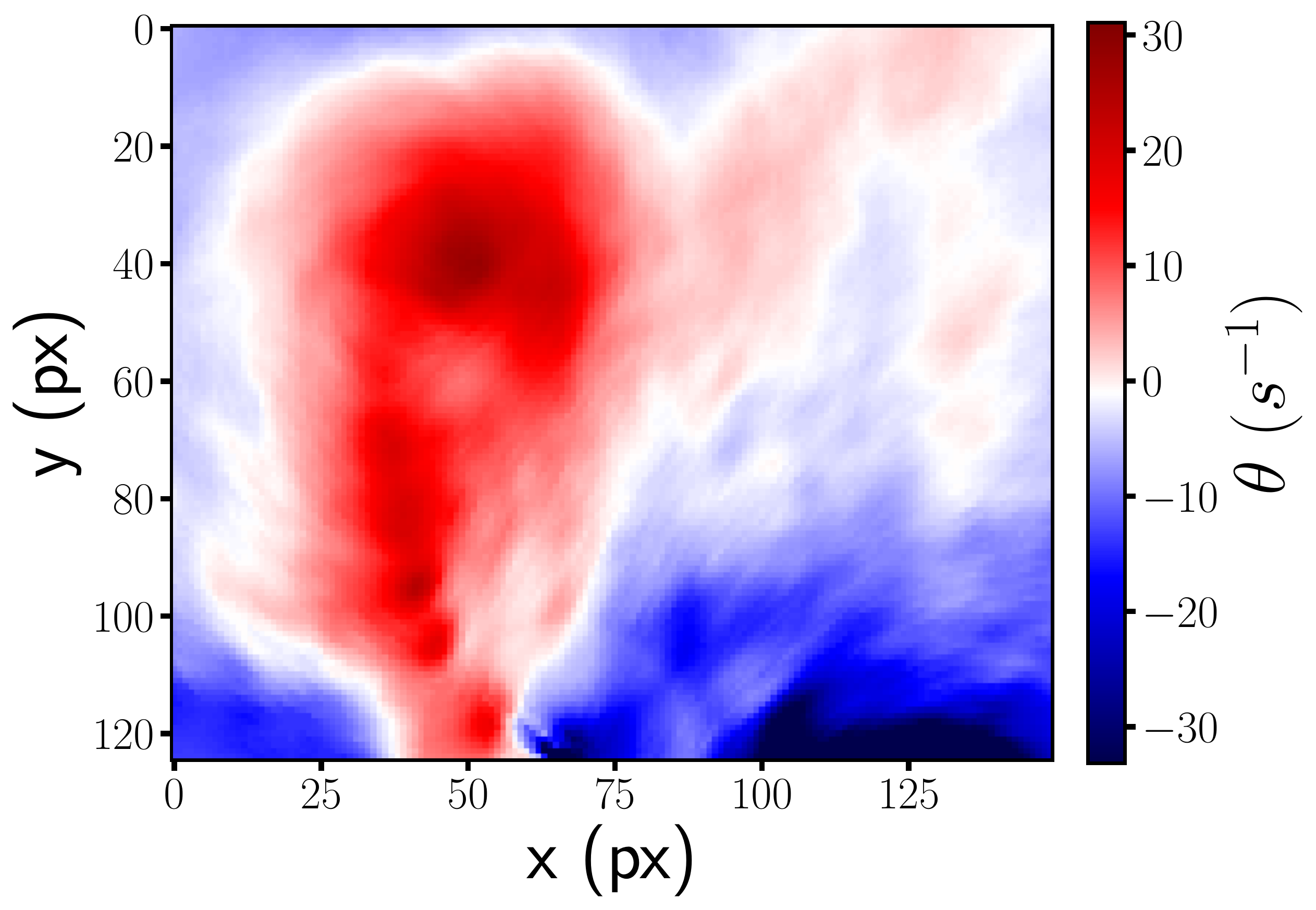}
		\caption{Pseudovorticity map at $\tilde{t}$=1.49.}        
		\label{fig:psv5}
	\end{subfigure}
	\hfill
	\begin{subfigure}[b]{0.42\textwidth}
		\centering
		\includegraphics[scale=0.3]{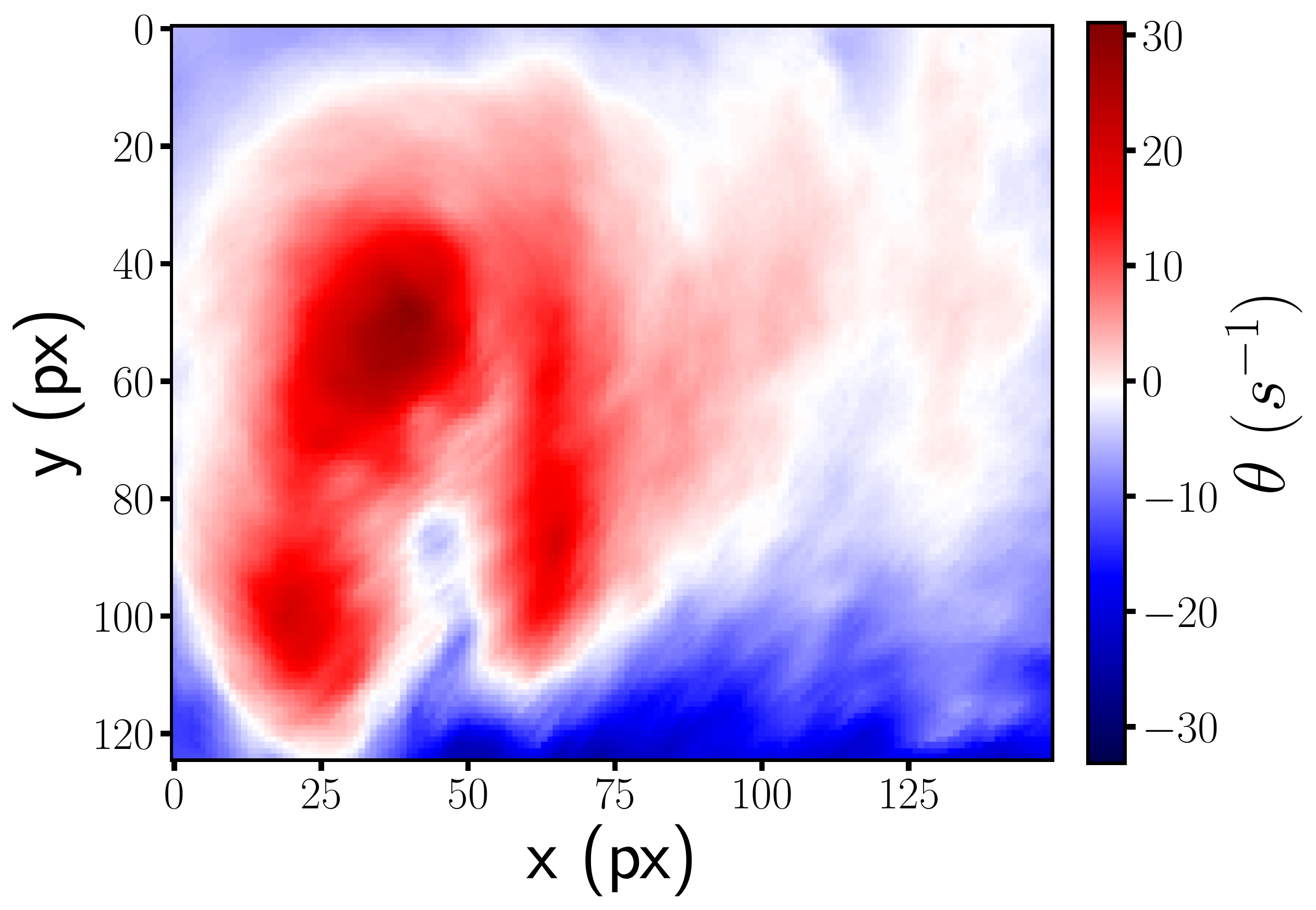}
		\caption{Pseudovorticity map at $\tilde{t}$=2.01.}
		\label{fig:psv6}
	\end{subfigure}
	\caption{Pseudovorticity maps at different time $\tilde{t}$ for the experiment Aa.}
	\label{fig:psv}
\end{figure}

Once we have the pseudovorticity map $\mathcal{F}$, one can calculate the mean intervortex distance $l$. It can be estimated from the vorticity of the superfluid component throught $\omega_s = \kappa/l^2$. We will assume that the flow is a coflow\footnote{$v_n=v_s$}. By considering $\omega \propto\sqrt{\langle\theta^2\rangle}$ we obtain \cite{BachelorThesisPatrik}\cite{Patrik2015}
\begin{equation}
	l\propto\frac{\sqrt{\kappa\rho/\rho_s}}{\langle\theta^2\rangle^{1/4}}.
\end{equation}
The values of $l$ and $\langle\theta^2\rangle$ are stored in the table \ref{tab:IntervotexDistance}. The mean intervortex distance is lower than the probe scale, indeed $\delta/l\in$[1.3;10], so we do not expect to observe any quantum behaviour. However, depending on the particle density and the chosen $R_{max}$, the pseudovorticity is usually between [$\omega/10$ ; $\omega/100$] and not $\omega/2$\cite{BachelorThesis}. So the mean intervortex distance is twice to seven times bigger than calculated. Those differences are linked to to the assumptions made in the definition of the mean intervortex distance, such as considering that all the vortices being a product of quanticized vortices.

\begin{table}[h!]
	\centering
	\begin{tabular}{c|c|c|c|c|c|c|c|c|c|c|c}
		Experiment & Aa & Ba & Ca & Cb & Cc & Da & Db & Dc & Ea & Eb & Ec \\ \hline
		$\langle\theta^2\rangle$ (s$^{-2}$) & 41 & 41 & 51 & 125 & 13 & 61 & 147 & 11  & 39 & 142  & 10 \\
		$l$ ($\mu$m) & 88 & 88 & 84 & 67 & 118 & 80 & 64 & 123 & 89 & 65 & 125
	\end{tabular}
	\caption{Estimated mean intervortex distance for each case studied.}
	\label{tab:IntervotexDistance}
\end{table}

\begin{figure}[h!]
	\centering
	\begin{subfigure}[b]{0.42\textwidth}
		\centering
		\includegraphics[scale=0.3]{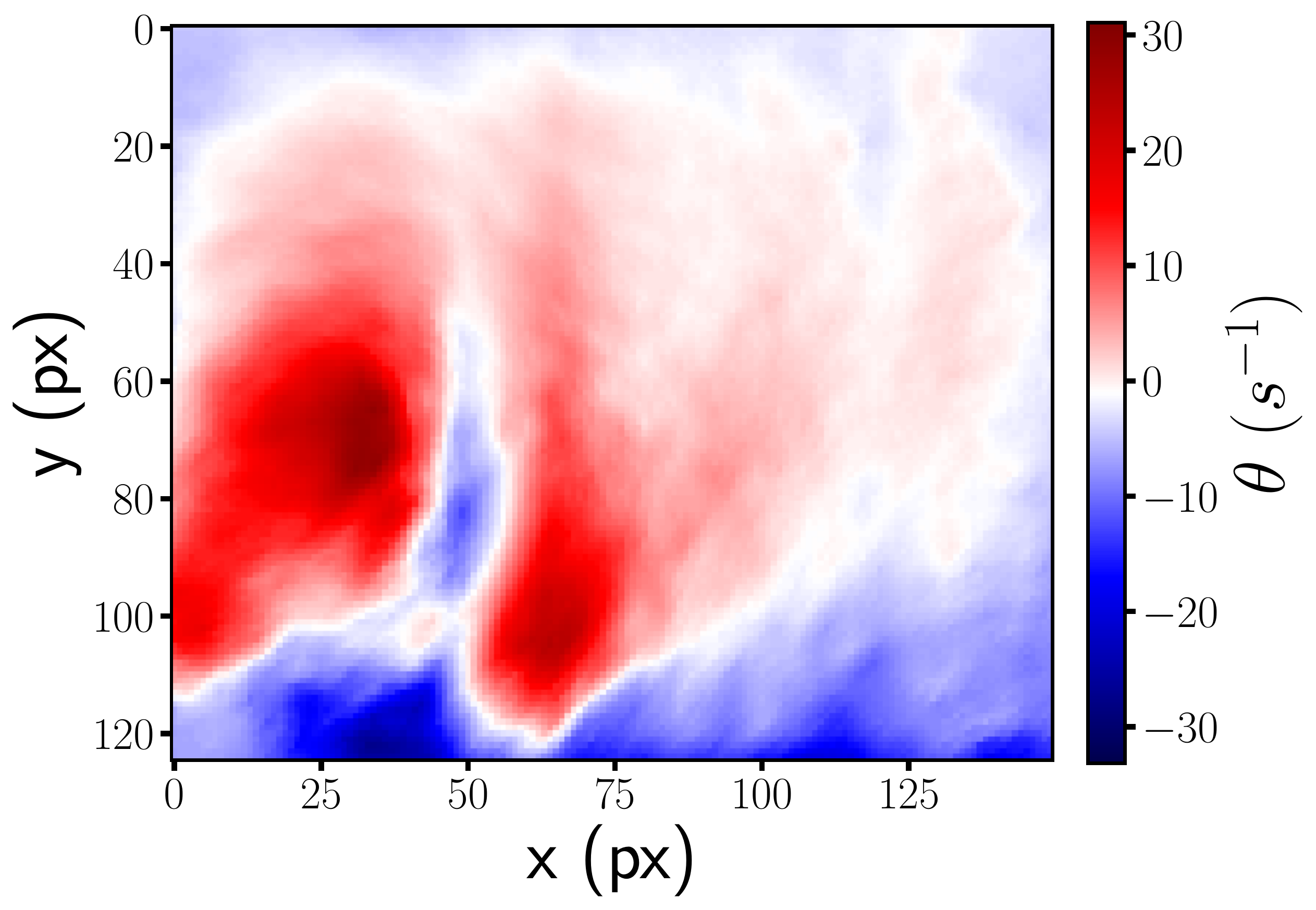}
		\caption{Pseudovorticity map at $\tilde{t}$=2.53.}
		\label{fig:psv7}
	\end{subfigure}
	\hfill
	\begin{subfigure}[b]{0.42\textwidth}
		\centering
		\includegraphics[scale=0.3]{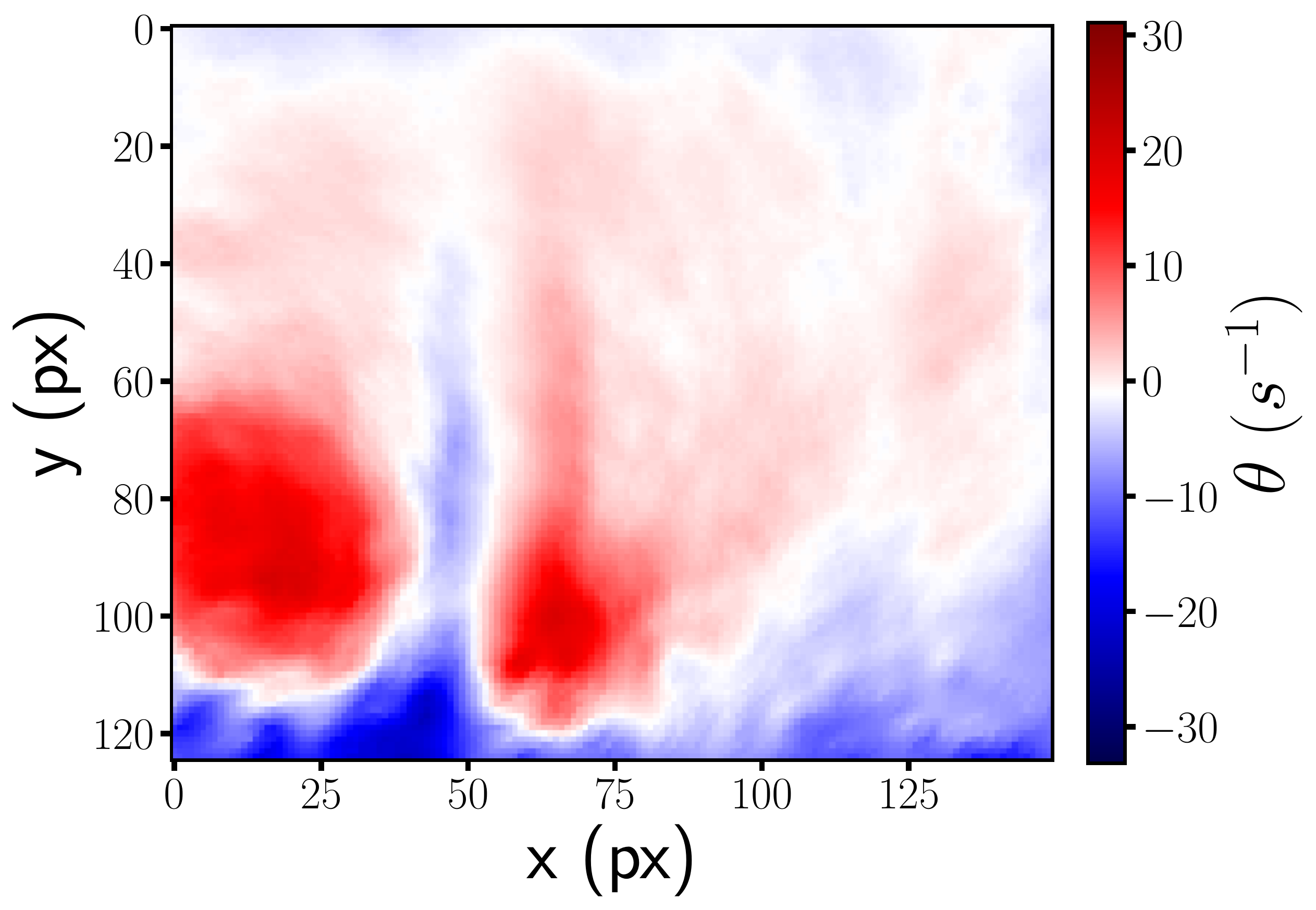}
		\caption{Pseudovorticity map at $\tilde{t}$=3.18.}
		\label{fig:psv8}
	\end{subfigure}
	\hfill
	\begin{subfigure}[b]{0.42\textwidth}
		\centering
		\includegraphics[scale=0.3]{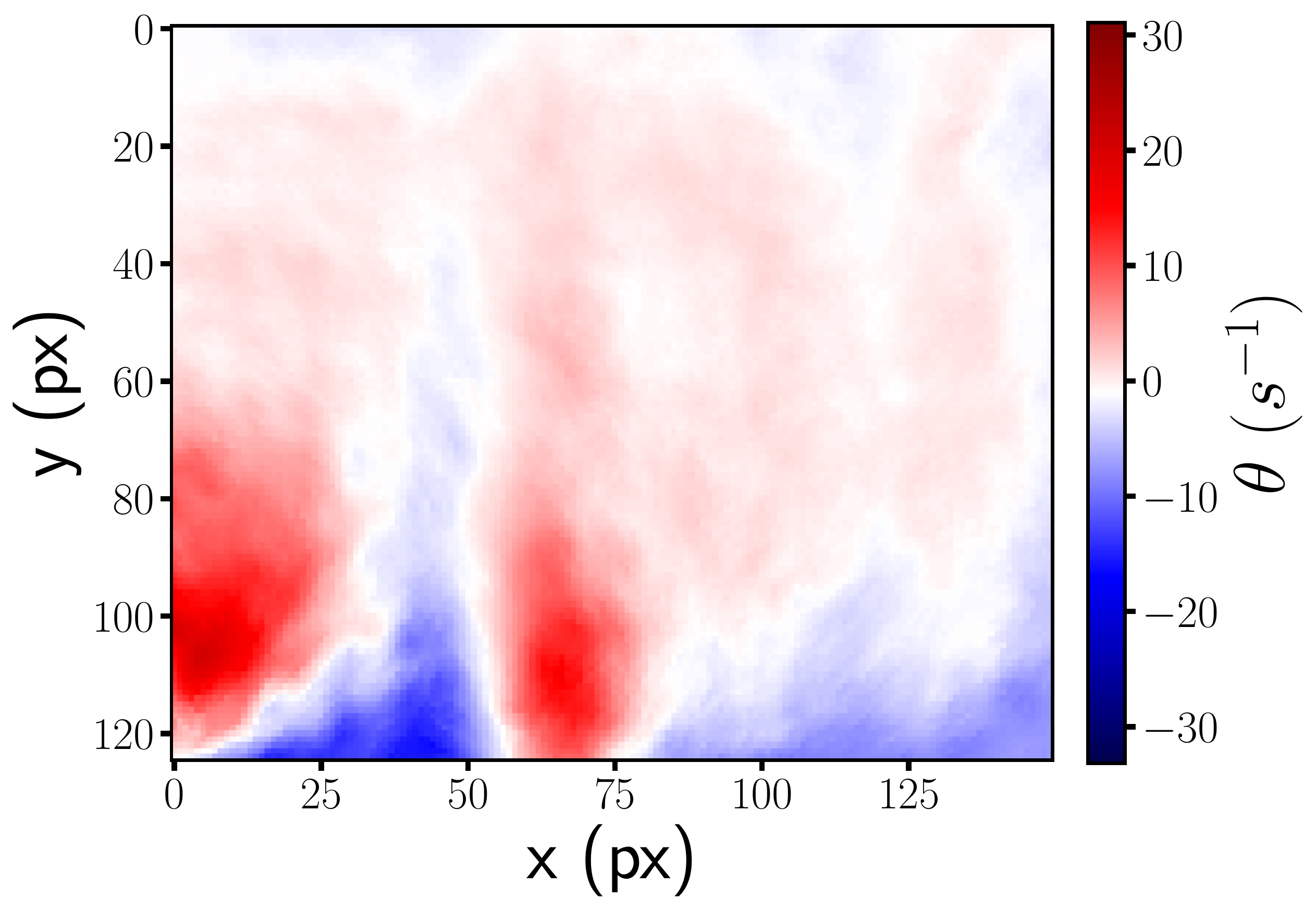}
		\caption{Pseudovorticity map at $\tilde{t}$=3.83.}
		\label{fig:psv9}
	\end{subfigure}
	\hfill
	\begin{subfigure}[b]{0.42\textwidth}
		\centering
		\includegraphics[scale=0.3]{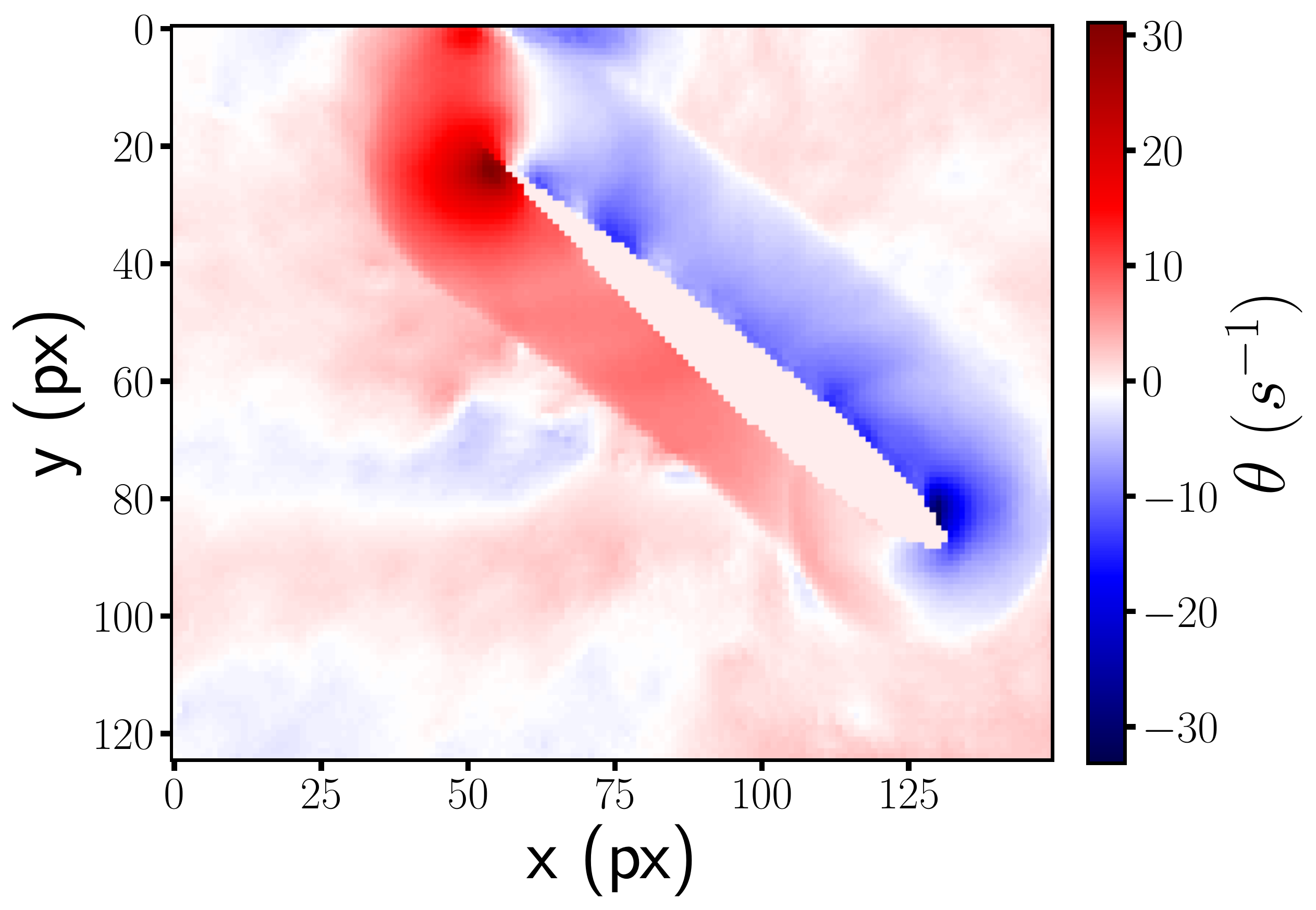}
		\caption{Pseudovorticity map at $\tilde{t}$=0.45 for the experiment Da.}
		\label{fig:psvAutre}
	\end{subfigure}
	\caption{Pseudovorticity maps at different time $\tilde{t}$ for the experiment Aa.}
	\label{fig:psvbis}
\end{figure}

For all the cases, the pseudovorticity map is very similar. An example of the pseudovorticity map at different times $\tilde{t}$ for the case Aa are given figures \ref{fig:psv} and \ref{fig:psvbis}. Soon after the airfoil started we can see the formation of a vortex at the trailing edge, which intensity grows with time (see figures \ref{fig:psv1} to \ref{fig:psv4}) while accompanying the airfoil and detaching itself from it (see figures \ref{fig:psv2} to \ref{fig:psv4}). Once the vortex is no longer attached to the trailing edge, it deviates with a straight trajectory towards the exterior (the left on the images, see figures \ref{fig:psv4}, \ref{fig:psv5}, \ref{fig:psv6} and \ref{fig:psv7}). An interesting feature is the creation of a secondary vortex at the trailing edge once the first one is separated from the airfoil (see figures \ref{fig:psv5} and \ref{fig:psv6}). This secondary vortex is going to merge with the first one (see figures \ref{fig:psv7} and \ref{fig:psv8}). Another interesting observation is the creation of a third vortex from the first one while it is drifting on the left (see figure \ref{fig:psv7} to \ref{fig:psv9}). This third vortex is created roughly at the same x-absissa as the creation of the first one. In some cases (see figure \ref{fig:psvAutre}) we observe a region of high pseudovorticity below the tip of the trailing edge. Interestingly, this phenomenon is only observed for the experiments Da to Ec. This could be linked to the fact that these experiments were realised some days after the experiments Aa to Cc, and therefore trace with fewer and bigger particles. However, no particular differences were visible in the region below the tip of the trailing edge looking at the trajectories. No changes have been observed when reducing the size of the mask or $R_{\textrm{max}}$ when computing the pseudovorticity. We still can pinpoint the lack of particles below the airfoil and especially below the tip of the airfoil (see figure \ref{fig:Traj}). This inhomogeneity is caused by the airfoil that prevent the particles which settles between the injections of helium gas and the starting of the airfoil to go below it. We also see less particles because of the shadding from the airfoil. Moreover, this vortex was not observed in previous experiments (see \cite{MasterThesis}), so we assume that this vortex is an artifact from the data processing. But this vortex is stronger than our vortex of interest, explaining why we took $p$=40 which is a rather low value. A last observation is the change of direction of the vortex. Some change of direction was already observed in the simulations made by Xu\cite{Xu2014} or in the experiments of Pullin\cite{Pullin_Perry_1980}. However, first their results do not go far enough in time to draw any similarities between our observations and theirs. Then, in their case, when the deviation happen, the vortex is still attached to the airfoil contrary to our case. So it is unlikely that these observations are directly comparable. This change of direction is not found in the various theories. An explanation would be that the deviation is a result of 3D-effects. Indeed, the aspect ratio of the wing is $\Sigma = \frac{L}{b}=0.6$, so criticism could be made to considering two-dimensional flows. Furthermore, the support of the airfoil and the metallic plate can have some effects.

\subsection{The vortex}

\begin{figure}[h!]
	\centering
	\input{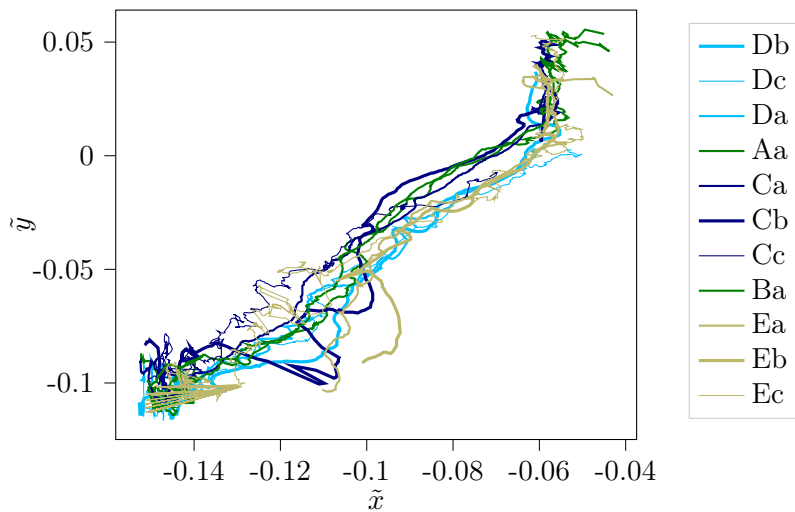}
	\caption{Trajectories of the starting vortex for all cases in the laboratory reference frame.}
	\label{fig:Trajxy}
\end{figure}

\begin{figure}[h!]
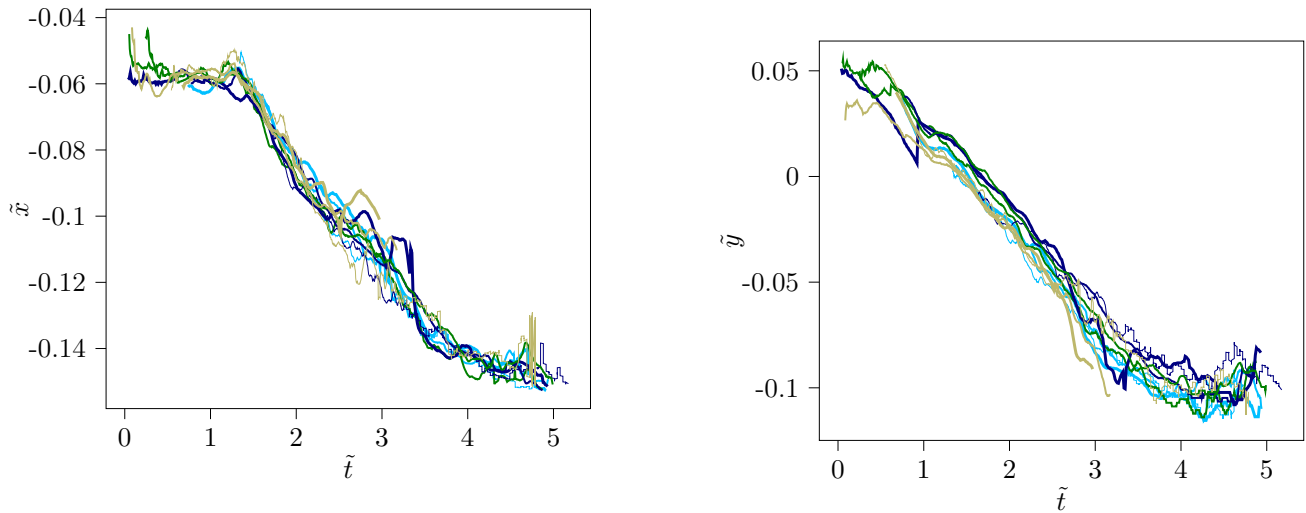

	\centering
	\begin{subfigure}[b]{0.45\textwidth}
		\centering
		\input{Imagesxt}
		\caption{Position of the starting vortex on the $\tilde{x}$-coordinate over the adimensional time $\tilde{t}$. The legend is the same as in the figure \ref{fig:Trajxy}. The section from $\tilde{t}=$1.4 to 3.9 can be fitted as $\tilde{x}=-0.031\tilde{t}-0.022$.}
		\label{fig:Trajxt}
	\end{subfigure}
	\hfill
	\begin{subfigure}[b]{0.45\textwidth}
		\centering
		\input{Imagesyt}
		\caption{Position of the starting vortex on the $\tilde{y}$-coordinate over the adimensional time $\tilde{t}$. The legend is the same as in the figure \ref{fig:Trajxy}.The section from $\tilde{t}=$1.4 to 3 can be fitted as $\tilde{y}=-0.049\tilde{t}+0.076$.}
		\label{fig:Trajyt}
	\end{subfigure}
	\caption{Trajectories over the time $\tilde{t}$ of the starting vortex for all cases in the laboratory reference frame.}
\end{figure}

To have a better understanding of the dynamics of the vortex and more quantitative data, we can plot the trajectories of the vortices. In the following the origin will be located in the middle of the airfoil of the first frame. As for the pseudovorticity map with a qualitative outlook, the trajectories are similar between each experiment and we find the elements qualitatively describes above. When the vorticies grow, the trajectories are surprisingly straight (see figure \ref{fig:Trajxy} from $\tilde{y}$=0.05 to 0.01 and the plateau until $\tilde{t}$= 1.05 in the figure \ref{fig:Trajxt}). Some abrupt variations along the $\tilde{x}$-absisca exist at the beginning of some trajectories because their vortex is ill detected at small times, creating jumps in the trajectory. The full separation of the vortex with the aifoil results in a plateau on the $\tilde{y}$ axis at $\tilde{y}\approx 0.01$ and an increase on the $\tilde{x}$ axis from $\tilde{t}$= 1.05 to 1.3. After that, the vortex deviates towards the exterior in a straight trajectory with an angle of 49.1° with the horizontal, which corresponds to the angle of attack of the airfoil. More experiments should be run at different angles to confirm any relation. We can pin point that the velocity is constant after $\tilde{t}<$1.4 (linear relation between $\tilde{x}/\tilde{y}$ and $\tilde{t}$), and neither the trajectory nor the velocity of the vortex is influenced by the movement of the airfoil once it has deviated, the vortex became independent. So we can assume that the airfoil gave some momentum to the vortex, allowing it to keep moving. At later times, the merging of the two vortices result in an ill definition of the vortex core, creating some large deviations between $\tilde{x}$=-0.1 and -0.14. Then the vortices are leaving the frames so their detected position is not accurate anymore.

Luchini and Tognaccini proposed \cite{LUCHINI_TOGNACCINI_2002}
in 2002 four stages for the evolution of the starting vortex. There is a first stage named the \textit{Rayleigh stage} where the flow is potential except, from a thin viscous layer with a constant thickness around the airfoil. Then the convective terms in the Navier-Stokes equation become comparable to the viscous terms. In this \textit{viscous stage} the vortex structure appears. The convective terms finally become dominant in the \textit{self-similar stage}, but the vortex remains independent of the geometry as it is small enough. At last, the vortex split up with the plate as the initial recirculation bubble opens up. However, the resolution in our data processing is not high enough to accurately observe and differentiate the three first stages. 

Xu and Nitsche \cite{Xu2014} verified a self similar theory developed by Pullin \cite{Pullin1978} for the third stage for a flat plate in a two-dimensional flow for the starting vortex. However, this theory is for a flow perpendicular to the plate, ie with an angle of attack of 90°, when we have a non-zero thickness and an angle of attack of 49.1°. We will nonetheless compare our result with it. In the following, all the quantities will be nondimensionalized, following the equation \ref{eq:adimensionnement}. The development with dimensionless parameters is only valid at early times, when the size of the starting vortex remains small compare to the plate length. In this case, the flow behaves locally independently of any external length scale. This is correct in the development of this theory as it is supposed to describe the self-similar stage. Xu uses the distance $d$ by which the plate has been displaced at time $t$ instead of the time. The displacement is then given by $d=t^2/2$ and its dimensionless version is $\tilde{d}=\tilde{t}^2/2$. The vortex sheet is assumed to separate at the edge of a semi-infinite plate and the background flow is accelerating and described by $\hat{\psi}_\infty = A\hat{t}\sqrt{\hat{r}}\cos (\theta/2)$ with $\hat{r}$, $\theta$ the polar coordinates centered at the plate tip, $A$ a dimensional constant equal to $\sqrt{L}a$ and $\psi$ the potential flow (defined as $(u,v)=\left( \pdv{\psi}{y},-\pdv{\psi}{x} \right)$). The rotation center and the circulation are given by 
\begin{equation}
	\tilde{x_c} + i\tilde{y_c} \approx \omega_0\left(\frac{A\tilde{t}^2}{2}\right)^{2/3},\quad \tilde{\Gamma}\approx J\frac{A^{4/3}\tilde{t}^{5/3}}{2^{1/3}},
\end{equation}
with $\omega_0$ and $J$ complex numbers. These equations give relations between the coordinates or the circulation and the vortex displacement which satisfy
\begin{equation}
	(x_c-0.5)+iy_c \propto d^{2/3}, \quad \Gamma \propto d^{5/6}. 	
\end{equation}

However, as already mentioned, this theory doesn't take into account the angle of the airfoil. A more complex theory was created by Pullin \cite{Pullin_Perry_1980}. The flow is assumed to be two-dimensional and inviscid. The limit $\nu\rightarrow 0$ can be modelled by separating the edges of the plate $x=\pm L/2$ in the form of a pair of unsteady vortex sheets. The flow is irrotational except along the vortex sheets. The main steps of the development of the theory can be found in the appendix \ref{AppThPullin}. The total shed circulation $\Gamma_+(\tilde{t})$ at time $\tilde{t}$ and the position of the vortex sheet at the trailing edge are given by:

\begin{subequations}
	\begin{equation}
		\Gamma_+(t) = K^{1/2}B^{4/3}L^{2/3}t^{5/3}(J_0 + J_1K^{1/2}L^{-1/3}B^{1/3}\sin\alpha^{1/3}t^{2/3} + \cdot\cdot\cdot),
	\end{equation}
	\begin{equation}
		\tilde{Z}_+(0,t) = KB^{2/3}L^{1/3}\sin\alpha^{2/3}t^{4/3}(\omega_0(1) + \omega_1(1)K^{1/2}L^{-1/3}B^{1/3}\sin\alpha^{1/3}t^{2/3} + \cdot\cdot\cdot),
	\end{equation}
\end{subequations}
with $K = (3/8)^{2/3}$, $B$ a constant, $\lambda$ a dimensionless circulation parameter, $J_0$ and $J_1$ dimensionless constants and $\omega_0(\lambda)$ and $\omega_1(\lambda)$ complex shape functions.\newline

\begin{figure}[h!]
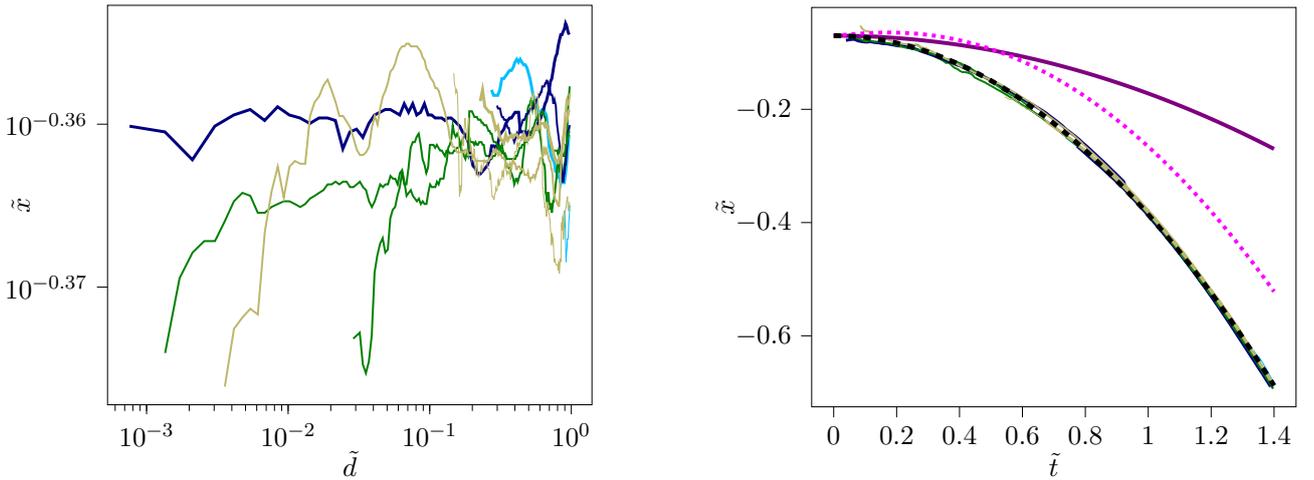

	\centering
	\begin{subfigure}[b]{0.45\textwidth}
		\centering
		\input{Imagesxu_x}
		\caption{Trajectories in Xu's reference frame over the displacement of the airfoil.}
		\label{fig:xu_x}
	\end{subfigure}
	\hfill
	\begin{subfigure}[b]{0.45\textwidth}
		\centering
		\input{ImagesPullin_x}
		\caption{Trajectories in Pullin's reference frame over the adimentional time.}
		\label{fig:pullin_x}
	\end{subfigure}
	\caption{Trajectories on the x-abscissa of the different cases over the displacement of the airfoil in the referent frames of the respective theories. Purple thick line: theoretical trajectory; pink dotted line: result obtained in \cite{MasterThesis}; black dashed line: fitting of the present work. The color of the different trajectories are those of the figure \ref{fig:Trajxy}.}
	\label{fig:th_x}
\end{figure}

As the theories describe the creation of the vortex at early times, the comparisons will only be made until $\tilde{t}$=1.4, when the vortex is completely detached from influence of the airfoil and changes direction, and therefore the theories are no longer valid. They were also developed in different reference frames, so changes had to be made to allow any comparison. More details about it can be found in the appendix \ref{AppThRefFrame}. All the fitting parameters that are calculated are the mean of the fitting parameters for each experiment. We can see figure \ref{fig:xu_x} that the trajectories do not exhibit any particular slope, much less $\tilde{x}\propto\tilde{d}^{2/3}$. A first explanation to this lack of similarity between the theory and the experimental results is the fact that this theory doesn't take into account the angle of attack of the airfoil. However, even with taking into account this angle of attack (see figure \ref{fig:pullin_x}), our results do not seem close to it. In the figure is also displayed previous results with the same airfoil and approximately the same angle of attack. We can see that already in that case, differences with the theory were observed. This might be due to the fact that the theoretical expression of the constant (see table \ref{tab:results}) were obtained with rather strong hypothesis of the vortex sheet being concentrated into a point. But our results are also quite different from the previous experiments. We can see that $\tilde{x}$ decreased sooner. And after the beginning of the decrease, the slope is similar. We attribute this difference at small times to stronger 3D effects. Indeed, in our experiments, the laser sheet was much closer to the support of the airfoil than previously. The diameter $d$ of our vortices vary between 2.2 and 2.7 mm and thus if we take $d=2.5$ mm, $\frac{d}{b/2}\sim 0.2$. \newline

\begin{figure}[h!]
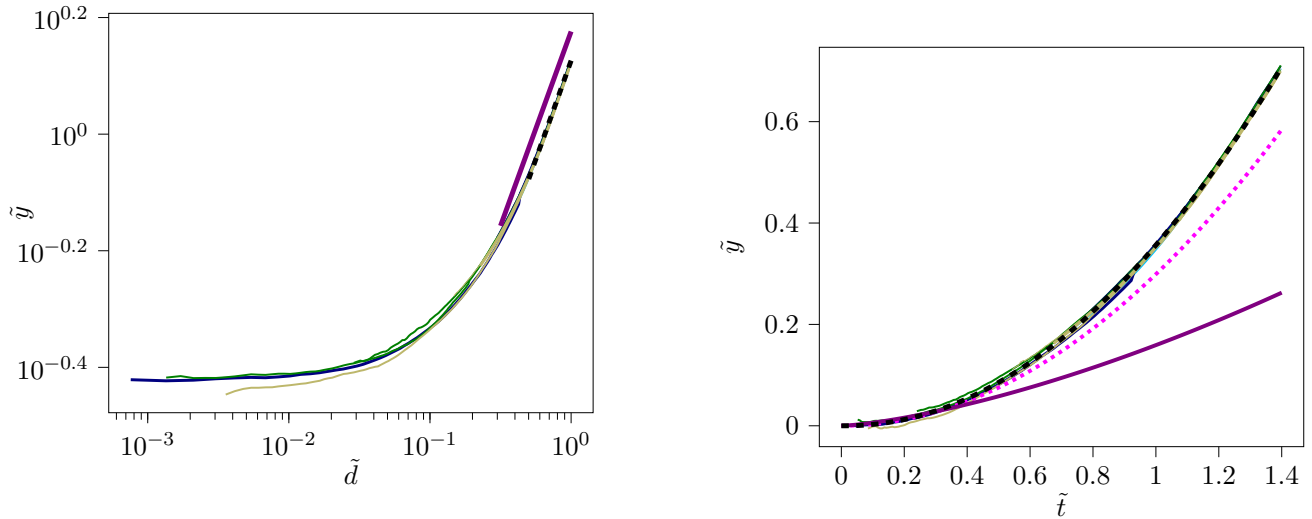

	\centering
	\begin{subfigure}[b]{0.45\textwidth}
		\centering
		\input{Imagesxu_y}
		\caption{Trajectories in Xu's reference frame over the displacement of the airfoil. A shift is added for the theoretical curve to ease the comparison.}
		\label{fig:xu_y}
	\end{subfigure}
	\hfill
	\begin{subfigure}[b]{0.45\textwidth}
		\centering
		\input{ImagesPullin_y}
		\caption{Trajectories in Pullin's reference frame over the adimentional time.}
		\label{fig:pullin_y}
	\end{subfigure}
	\caption{Trajectories of the different cases on the y-abscissa over the displacement of the airfoil or the time $\tilde{t}$ in the referent frames of the respective theories. Purple thick line: theoretical trajectory; pink dotted line: result obtained in \cite{MasterThesis}; black dashed line: fitting of the present work. The color of the different trajectories are those of the figure \ref{fig:Trajxy}.}
	\label{fig:th_y}
\end{figure}

For the trajectory on the $\tilde{y}$ axis, curiously our results are closer to the expectations of Xu's expression than the more complex one of Pullin. However, it is important to indicate that this scaling law, for this type of movement (constant acceleration) is only verified for $\tilde{d}>10^{-1}$, so we do not expect the trajectories in the figures \ref{fig:xu_x} and \ref{fig:xu_y} to verify the theoretical expression of Xu's theory for  $\tilde{d}<10^{-1}$. The exponent found in our experiment (figure \ref{fig:xu_y}) is close to the theoretical one (see tab \ref{tab:results}). For pullin's theory (see figure \ref{fig:pullin_y}), we see that it matches with the theory at early time $\tilde{t}<0.4$, before diverging from the theoretical expression. Once again, the deviation at later times could be an effect of the assumption to obtain the theoretical constants. Indeed, Pullin already observed in their simulations, with a viscous flow and an airfoil with a non-zero thickness, that the vortex was located further away from the airfoil than the theoretical expectations. So this is coherent with our observation of the distance between the airfoil and the vortex core increases faster in our experiments than in Pullin's theory. Nevertheless, these results for the trajectory on the $\tilde{y}$ abscissa is comforting to say that the vortex behaves as predicted. This is not the case for the $\tilde{x}$ abscissa, as in theory the center of the vortex core usually deviates toward the middle of the airfoil\cite{Pullin_Perry_1980}. But we can observe on the figures \ref{fig:Trajxy} and \ref{fig:Trajxt} that it is not the case, explaining the lack of agreement with the theories.\newline

\begin{figure}[h!]
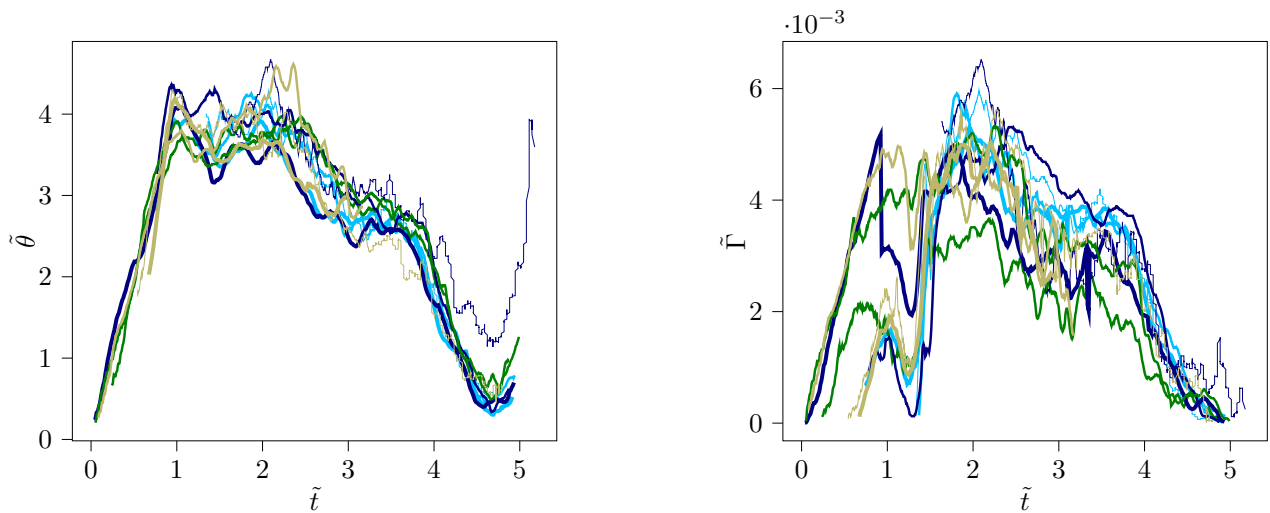

	\centering
	\begin{subfigure}[b]{0.45\textwidth}
		\centering
		\input{ImagesTheta}
		\caption{Dimensionless pseudovorticity over the dimensionless time $\tilde{t}$. The legend is the same as in the figure \ref{fig:Trajxy}.}
		\label{fig:theta}
	\end{subfigure}
	\hfill
	\begin{subfigure}[b]{0.45\textwidth}
		\centering
		\input{ImagesGamma}
		\caption{Dimensionless circulation over the dimensionless time $\tilde{t}$. The legend is the same as in the figure \ref{fig:Trajxy}.}
		\label{fig:gamma}
	\end{subfigure}
	\caption{Dimensionless circulation and pseudovorticity over the dimensionless time $\tilde{t}$.}
\end{figure}

To complete the analysis, we plot the strength and the circulation (see figures \ref{fig:theta} and \ref{fig:gamma}). The circulation is defined as:
\begin{equation}
	\Gamma = \oint_C\vb{v}\cdot d\vb{l}.
\end{equation}
In our case, it is calculated multiplying the strength of the vortex with its size. And the strength is the average value of $\theta$ inside the vortex. The strength is quite similar from $\tilde{t}$=0 to $\tilde{t}$=1 (see figure \ref{fig:theta}). They do not seem to have significant differences between the experiments. However, the lack of data at those times might cause this observation. We then notice a first maximum at $\tilde{t}_0\approx 1$, and just after, a minimum at $\tilde{t}_1=1.4$. This minimum can be observed by looking at the detected area being a vortex in the pseudovorticity maps $\mathcal{F}$. Indeed, we saw a reduction (which is also noticeable in the figure \ref{fig:gamma}) at $\tilde{t}=$ 1.4. Interestingly, the first maximum (resp. the plateau) corresponds to the beginning of the increase (resp. the plateau) in the figure \ref{fig:Trajxt} (resp. \ref{fig:Trajyt}), and the first minimum is found little after the beginning of the decrease in the same figure, which means that the vortex started to deviate through the left. So this reflects the detachment of the vortex from the influence of the airfoil. However, the cases Aa and Ca do not feature this decrease in the strength of the vortex. The strength then increases to reach another maximum around $\tilde{t}_2=$2.2. It corresponds to the moment when the vortices begin to merge. We wish to underline that for some cases the first minimum (Ea) or the second maximum (Cb, Ec, Eb, Dc) is not very pronounced. So we suggest that these variations might not be entirely physical, meaning that the strength of the vortex does not vary that much between the separation of the airfoil and the merging with the second vortex. And the minimum in the circulation can be highly criticized as it comes from a multiplication of the size of the vortex, which is flawed by the area of high pseudovorticity at the tip of the airfoil. Then the strength decreases. It is not due to the creation of the third vortex as it is created earlier (see figure \ref{fig:psv6}). The plateau from $\tilde{t}_3$=3.1 to 3.9 is not physical as it corresponds to the moment when the vortex begins to leave the observable area, and the following abrupt decrease can be explained by the fact that the code detects the outer region of the vortex. A last observation on those figures would be that the strength of the vortices at $f$=2 Hz is slightly lower than for the other frequencies. The circulation (see figure \ref{fig:gamma}) is coherent with the previous observations as the variations are concomitant with the ones in the strength. We also do not observe any particular trend in the time derivative of the pseudovorticity that could be interesting\cite{Lian1989}.

To compare with Pullin's theory, the fitting of the parameters is done by only taking into account the data that begins at the earliest times. In this case, our fitting curve is close to the theoretical expectations. However, we can see, in particular with later data, that the curve inflections are completely different. Similarly, we observe a coherence with the theoretical expectations and the experimental data with xu's comparison. But the lack of data at these times when experimental data and theory seem to match do not allow us to draw any real conclusions. \newline

\begin{figure}[h!]
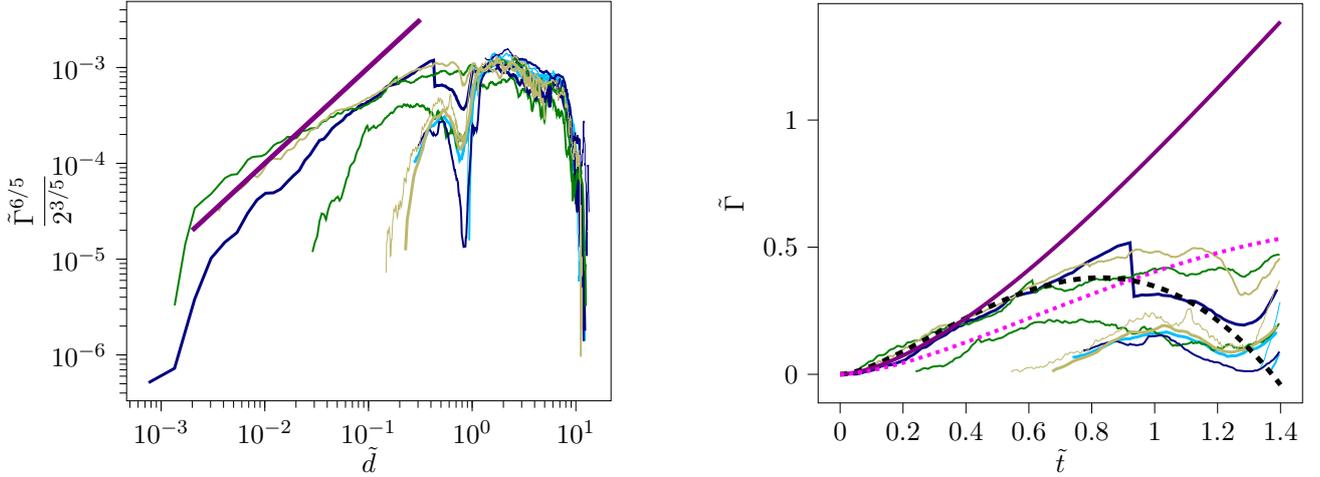

	\centering
	\begin{subfigure}[b]{0.45\textwidth}
		\centering
		\input{Imagesxu_g}
		\caption{Circulation of the vortices over the displacement of the airfoil compared with Xu's theory.}
		\label{fig:xu_g}
	\end{subfigure}
	\hfill
	\begin{subfigure}[b]{0.45\textwidth}
		\centering
		\input{ImagesPullin_g}
		\caption{Circulation of the vortices over the adimensional time compared with Pullin's theory.}
		\label{fig:pullin_g}
	\end{subfigure}
	\caption{Comparison of the circulation with the different theories. Purple thick line: theoretical trajectory; pink dotted line: result obtained in \cite{MasterThesis}; black dashed line: fitting of the present work. The color of the different trajectories are those of the figure \ref{fig:Trajxy}.}
	\label{fig:g}
\end{figure}

\begin{table}[h!]
	\centering
	\begin{tabular}{c|c|c|c|c}
		Theory & Scaling law/ & Value of the & This work & Previous work \\ 
		& parameter & parameter &  & \cite{MasterThesis}\\ \hline
		Xu & $\tilde{y}\propto\tilde{d}^{\gamma}$ & 2/3 & 0.68 & 1.07\\
		Xu & $\tilde{x}\propto\tilde{d}^{\gamma}$ & 2/3 &  & 0.61\\
		Xu & $\frac{\tilde{\Gamma}^{6/5}}{2^{3/5}}\propto\tilde{d}^{\gamma}$ & 1 &  & 0.58\\
		Pullin & $\Omega_0$ & 0.29i & 0.0064(0.005)-0.056i(0.005) & -0.4+0.01i\\
		Pullin & $\Omega_1$ & 0.36+0.12i & 1.1(0.007)+1.3i(0.007) & 1.3+1.04i\\
		Pullin & $J_0$ & 2.4 & 3.6 & 1.6\\
		Pullin & $J_1$ & -0.95 & -4.5 & -1.2
	\end{tabular}
	\caption{Table of the fitting parameters of the different theoretical expressions.  The fitting parameters for Pullin's theory were calculated with the function \texttt{curve$\_$fit} and the exponent for Xu's theory with \texttt{polyfit}, both from the library \texttt{scipy}.}
	\label{tab:results}
\end{table}
The discrepancies with the theory could be explained, as we already mentioned with the 3D effects, but also with the viscosity. Indeed the theories have been developed for inviscid fluids. The simulations made by Xu\cite{Xu2014} (at Re=500) to check his theory verified to some extent the expectations. However we already saw this theory doesn't take into account the angle of attack. Pullin's simulations at Re=800 however, only verify the expectations of its theory at early times $\tilde{t}<0.5$. But Pullin focuses on the forces, so we cannot say with any certainty if the verification is also valid for the trajectories, at least in the $(x,y)$ plane as the vortex move away from the airfoil faster than predicted. The shape of the tip can also influence the starting vortex, however it seems unlikely as the main observations are about the stability of the vortex sheet\cite{Schneider2014}. A last possibility would be that the vortices from the trailing edge are interacting. However quantitative measurements on the distance separating them could not be made as the second vortex is ill defined at small times because it is dominated by the vortex below the plate, and at larger time because its strength is not high enough to be accuratly detected by the code. The vortex below the plate could not be an artifact of the data processing as Pullin in 1978 \cite{Pullin_Perry_1980} observed this secondary vortex. However this secondary vortex only appeared at later times contrary to us, holding doubts on its physical meaning for our observation. Furthermore, we are unable to explain why we only observe it at some temperature in HeII. It is also unlikely that the walls of the inner cryostat affect the trajectory\cite{Auerbach_1987}. Furthermore, we do not observe any particular difference between the data in HeI and in HeII. We can explain that through the probe scale being larger or at the same order of the mean intervortex distance.

	\section{Conclusion}
	
The aim of this internship was to study the shedding of the starting vortex in $^4$He. The starting vortex was created by an airfoil with a significant angle of attack moving at constant acceleration. Three different accelerations in HeII (at temperature going from 1.3K to 1.95K), and one case in HeI were studied. To do so, a visualisation technique has been used. Deuterium particles seeded the liquid. They allowed us to run the PTV technique from which we calculated the pseudovorticity field. The pseudovorticity being proportional to the vorticity, we then detected and track the starting vortex. As it is proportional, with the exact proportionality not known, we can only make relative comparisons. To compare the experiments with different acceleration between them, all the quantities have been beforehand non-dimentionalized. We found no difference between the trajectories in HeII and HeI. We attribute this to the fact that the probe scale is larger or similar to the mean intervortex distance, inhibiting us from observing any quantum behaviour. The trajectories obtained poorly match the theory. However, some limitations can be drawn for the comparisons between the experiments and the theory. First, our plate is an airfoil with a non-zero thickness, contrary to the theory. Then, we have 3D effects due to the support of the airfoil and the metallic plate connecting the aifoil to the main shaft. And the theories were developed under the inviscid assumption, which is not verified in our experiments. The main difference is that the trajectories are not slightly drifting towards the middle of the airfoil during the growing of the vortex. When the vortex is no longer influenced by the airfoil, it deviates towards the exterior at constant velocity in a straight trajectory, even though it should not.

To go further, additional experiments should be run with different angles of attack to verify if the angle in the trajectory of the vortex in the $(x,y)$  plane is influenced by it, and with different fluids to confirm or rule out the role of viscosity in our results. This has not been done during this internship, but the leading vortex could also be studied. Improvement in the detecting vortex code should be made in order to detect the vortex at earlier times, as for now, the ill definition of the pseudovorticity at the border with the airfoil and the small strength the vortices have at the beginning do not enable a proper detection, and though tracking. This would enrich our knowledge on the starting vortex and could enlighten our present results.

\textcolor{white}{
	\nocite{Auerbach_1987_3D}
	\nocite{ZDRAVKOVICH1996427}
	\nocite{Stagg_2015}
	\nocite{GHARIB_RAMBOD_SHARIFF_1998}
	\nocite{Outgassing}
	}

\newpage
\emergencystretch 2em 
\bibliography{biblio}

\appendix
\appendixpage
\addappheadtotoc

\section{The mathematical description of the two fluid model}\label{App:2fluid}

The mathematical description can be fulfilled by writing the conservation of mass law

\begin{equation}
	\pdv{\rho}{t} + \grad\cdot(\rho_s\vb{v}_s + \rho_n\vb{v}_n ) = 0,
\end{equation}

as well as the Navier Stokes equations for each component. For the superfluid component we have:

\begin{equation}
	\pdv{\vb{v}_s}{t} + \rho_s(\vb{v}_s\cdot\grad)\vb{v}_s = -\frac{\rho_s}{\rho}\grad p + \rho_sS\grad T + \frac{\rho_n\rho_s}{\rho}\grad(\vb{v}_n-\vb{v}_s)^2 - \vb{F}_{ns};
\end{equation}
and for the normal component:

\begin{equation} 
	\pdv{\vb{v}_n}{t} + \rho_n(\vb{v}_n\cdot\grad)\vb{v}_n = -\frac{\rho_n}{\rho}\grad p + \rho_nS\grad T - \frac{\rho_n\rho_s}{\rho}\grad(\vb{v}_n-\vb{v}_s)^2 + \vb{F}_{ns} + \eta\laplacian\vb{v}_n,
\end{equation}

with $\vb{F}_{ns}$ the mutual friction force, $\frac{\rho_n\rho_s}{\rho}\grad(\vb{v}_n-\vb{v}_s)^2$ the interaction of the counterflow. With the existence of the two velocity fields, various types of flow can exist. If $\vb{v_n}\approx\vb{v_s}$, the two flows move in the same direction, this is a \textit{cloflow}. If  $\vb{v_n}=0$ we talk about \textit{superflow}. Finally if the superfluid and the normal fluid move in opposite direction, the phenomenon is named \textit{counterflow}. On the other hand, the mutual friction describes the interaction between the components. It can be view as the scattering  of phonons and rotons.

\section{Relation between the pressure of helium and its temperature}\label{App:SatVap}

Because of this relation, the manometer can also serve as a thermometer. However, this relation is only true at saturated vapour. But this relation indicates that at saturated vapour the temperature can be controlled through the pressure, which is what we have done.

\begin{figure}[H]
	\centering
\begin{tikzpicture}

\definecolor{darkgray176}{RGB}{176,176,176}
\definecolor{purple}{RGB}{128,0,128}
\definecolor{steelblue31119180}{RGB}{31,119,180}

\begin{axis}[
log basis y={10},
tick align=outside,
tick pos=left,
x grid style={darkgray176},
xlabel={Temperature (K)},
xmajorgrids,
xmin=0.635, xmax=4.265,
xtick style={color=black},
y grid style={darkgray176},
ylabel={Saturated vapour pressure (Torr)},
ymajorgrids,
ymin=0.00637650797558168, ymax=1173.06207413253,
ymode=log,
ytick style={color=black}
]
\path [draw=purple, semithick, dash pattern=on 5.55pt off 2.4pt]
(axis cs:0.8,37.81)
--(axis cs:2.1768,37.81);

\path [draw=purple, semithick, dash pattern=on 5.55pt off 2.4pt]
(axis cs:2.1768,0.0110634145)
--(axis cs:2.1768,37.81);

\addplot [semithick, steelblue31119180]
table {%
0.8 0.0110634145
0.85 0.02185680668
0.9 0.04034583498
0.95 0.07034831498
1 0.1167846534
1.05 0.1858653636
1.1 0.28502356
1.15 0.423409999
1.2 0.6111505176
1.25 0.860321114
1.3 1.184347898
1.35 1.596881998
1.4 2.11517484
1.45 2.755727788
1.5 3.53654233
1.55 4.47787014
1.6 5.598462768
1.65 6.918571888
1.7 8.46069936
1.75 10.24584692
1.8 12.28601556
1.85 14.61870838
1.9 17.1764198
1.95 20.19166904
2 23.46943998
2.05 27.09223944
2.1 31.06006742
2.15 35.3654233
2.2 40.0158077
2.25 45.0412231
2.3 50.4791726
2.35 56.34465744
2.4 62.66017948
2.45 69.44073996
2.5 76.7313426
2.55 84.4569812
2.6 92.7826694
2.65 101.633401
2.7 111.0841822
2.75 121.0600068
2.8 131.635881
2.85 142.886811
2.9 154.7377906
2.95 167.1888198
3 180.389911
3.05 194.1910518
3.1 208.8172608
3.15 224.0435194
3.2 240.0948462
3.25 256.896235
3.3 274.4476858
3.35 292.8242048
3.4 311.9507858
3.45 331.9774412
3.5 352.8291648
3.55 374.5809628
3.6 397.2328352
3.65 420.7097758
3.7 445.161797
3.75 470.5888988
3.8 496.916075
3.85 524.2183318
3.9 552.4956692
3.95 581.8230934
4 612.2006044
4.05 643.6282022
4.1 676.1058868
};
\end{axis}

\end{tikzpicture}
	\caption{Saturated vapour line as a function of T. The dotted line indicates the temperature ($T_\lambda = 2.1768$ K) and the pressure ($P_\lambda = 37.81$ Torr) of the lambda transition. Data from \cite{TheObservedPropertiesofLiquidHeliumattheSaturatedVaporPressure}}
	\label{AppFig:SaturatedVapourLine}
\end{figure}
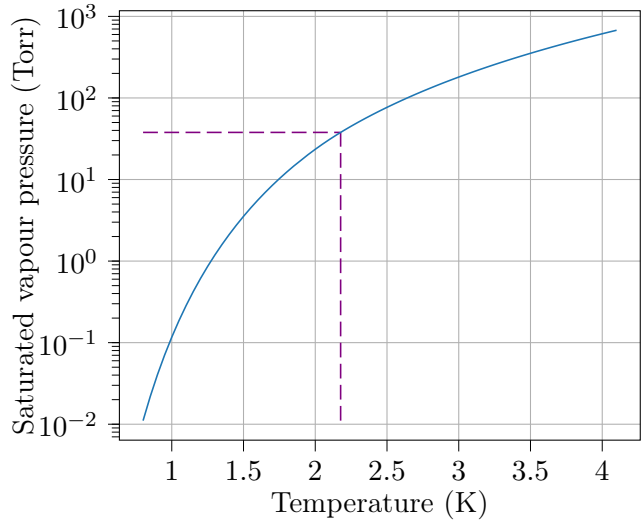

\section{Additional information on the sensors}\label{App:ForceTorque}

\begin{figure}[h!]
	\centering
	\includegraphics[scale=0.3]{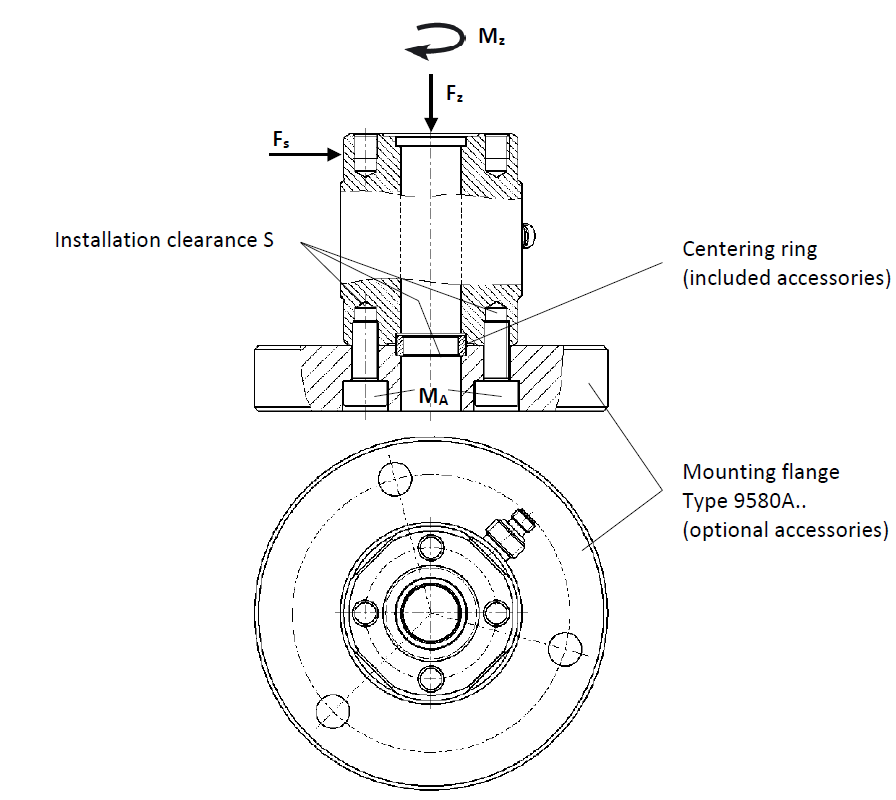}
	\caption{Drawing of the installation of the torque sensor with a mounting flange from the kistler manual. $M_z$ is the torque (N$\cdot$m), $F_z$ the compression force (N) and $F_s$ the transverse force (N).}
	\label{AppFig:Torque}    
\end{figure}

The torque sensor is mounted between two mounting flanges. A drawing of the torque mounted on one flange is displayed figure \ref{AppFig:Torque}. The torque sensor is centered on the mounting flange with a centering ring. It is then fixed to it with four screws per flange. The torque is applied to the sensor through the flanges.\newline

The force is introduced to the sensor with a coupling element. The mounting thread introduces coaxial forces.

\begin{figure}[h!]
	\centering
	\includegraphics[scale=0.3]{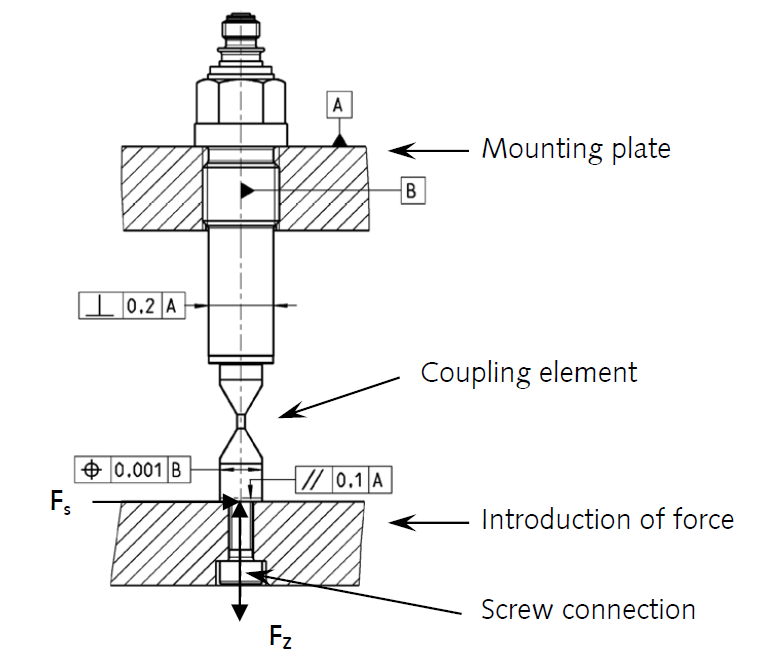}
	\caption{Drawing of the installation of the force sensor with a coupling element from the kistler manual. $F_z$ stands for the tensile/compressive force and $F_s$ for the lateral force.}
	\label{AppFig:Forque}    
\end{figure}

\section{Some results of prior experiments for the sensors}\label{App:sensors}

\begin{figure}[h!]
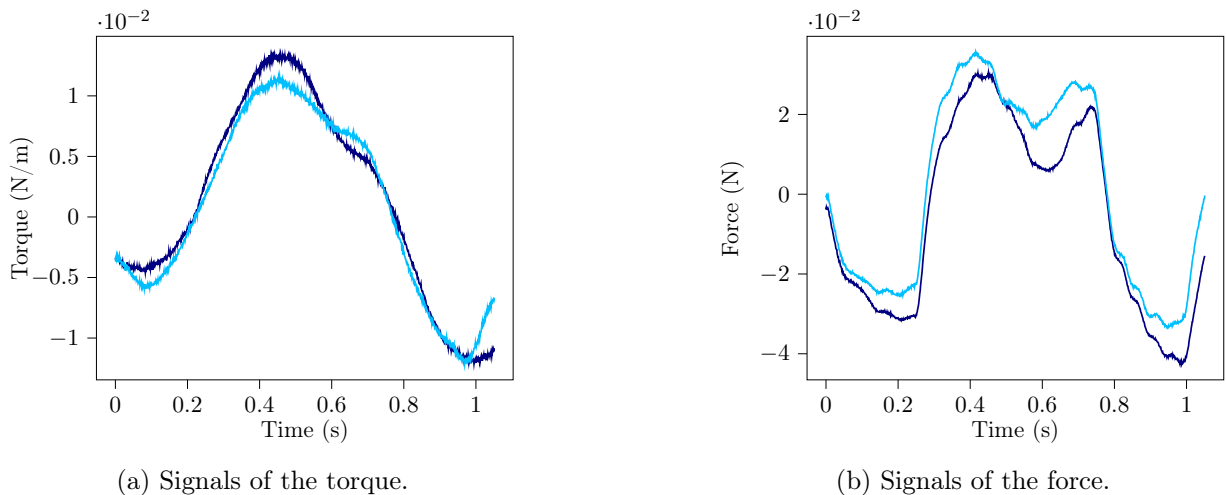

	\centering
	\begin{subfigure}[b]{0.45\textwidth}
		\centering
		\input{ImagesTorqueAdd}
		\caption{Signals of the torque.}
		\label{AppFig:TorqueSensors}
	\end{subfigure}  
	\hfill
	\begin{subfigure}[b]{0.45\textwidth}
		\centering
		\input{ImagesForceAdd}
		\caption{Signals of the force.}
		\label{AppFig:ForceSensors}
	\end{subfigure}
	\caption{Signals of the force and of the torque at room temperature and at $f$=1 Hz. This experiment was run prior to the main one. In dark blue: vacuum; in light blue: helium gas. The sensitivity of the torque sensor was 1 pC/N$\cdot$m, and the range was -2189 N$\cdot$m. The sensitivity of the force sensor was  -103.3 pC/N and the range was 5 N.}
	\label{AppFig:Sensors}
\end{figure}

The purpose of showing the figure \ref{AppFig:Sensors} is to indicate the pattern of the signal of the force that we would expect, and we did not have in most of our results. This pattern was consistent whatever the sensor, the wing, the angle of the wing or the connector between the wing and the main shaft used.  We can see a  temporal shift in this example. We also observed this phenomenon in experiments in water and air. This is due to the fact that helium (resp water) have to puts more resistance than vacuum (resp air). It is thus odd that we did not observe this between cold vacuum and helium (liquid or gas). On the other hand, the signal of the force is highly dependent of the various parameters. However, the results in the figure \ref{AppFig:Sensors} used the exact same apparatus. A remarkable feature is that we do also observe an asymmetry,
but this asymmetry is reversed between the two experiments.

\section{Pseudovorticity and vorticity}\label{App:pseudovorticity}

Švančara \textit{et al.}\cite{Svancara2020} and Outrata \textit{et al.}\cite{Outrata_Pavelka_Hron_LaMantia_Polanco_Krstulovic_2021} have demonstrated that a relation exists between the pseudovorticity and the vorticity. The equation \ref{eq:pseudovorticity} can be rewritten as 
\begin{equation}\label{eq:pseudovorticityBis}
	\theta(\vb{r}, \vb{R}) = \frac{1}{\pi \vb{R}^2}\int_{D(\vb{r},\vb{R})}f(\vb{r}')\frac{[(\vb{r}'-\vb{r})\times\vb{u}(\vb{r}')]_z}{|\vb{r}'-\vb{r}|^2}d^2\vb{r}',
\end{equation}
with $\vb{D}(\vb{r},{R})$ the considered circle centered at $\vb{r}$ with a radius $\vb{R}$. $f$ is the particle distribution function. The vorticity of two-dimensional vortex with a purely azimutal fluid velocity is defined as:
\begin{equation}
	\omega = \frac{1}{r}\pdv{rv_a}{r} = \frac{\Gamma}{\pi}g'(r^2),
\end{equation}
with $v_a$ the pseudovorticity of the flow, $\Gamma$ the constant fluid circulation and $g'(r^2)$ the derivation of $g(r^2)$ a smooth, non-negative function vanishing at the origin and constant at infinity. Thus the equation \ref{eq:pseudovorticityBis} can be rewritten as 
\begin{equation}
	\theta (x_0), \vb{R} =  \frac{1}{N}\int_{D(x_0,\vb{R})}\frac{\vb{r}-(x_0,0)}{\vb{r}-(x_0,0)}\times \lbrack\frac{\Gamma}{2\pi r}g(r^2)\vb{e}_a\rbrack d^2\vb{r}',
\end{equation}
with $\vb{e}_a$ the azimutal veocity direction and $(x_0,0)$ the considered inspection point. After some development at the fourth order\cite{Outrata_Pavelka_Hron_LaMantia_Polanco_Krstulovic_2021} we obtained 
\begin{equation}
	\theta(\vb{r},\vb{R}) = \frac{\omega(\vb{r})}{2} + \frac{\vb{R^2}}{32}\laplacian\omega(\vb{r}),
\end{equation}
a quadratic function of $R$ with $\laplacian$ the laplacian operator.

It was named pseudovorticity by \cite{Duda2017} because it was expected in the ideal condition (infinitely dense sampling) to converge to the flow vorticity.

\section{Exemple of the data processing}\label{AppDataProcessing}

\begin{figure}[h!]
	\centering
	\begin{subfigure}[b]{0.45\textwidth}
		\centering
		\includegraphics[scale=0.2]{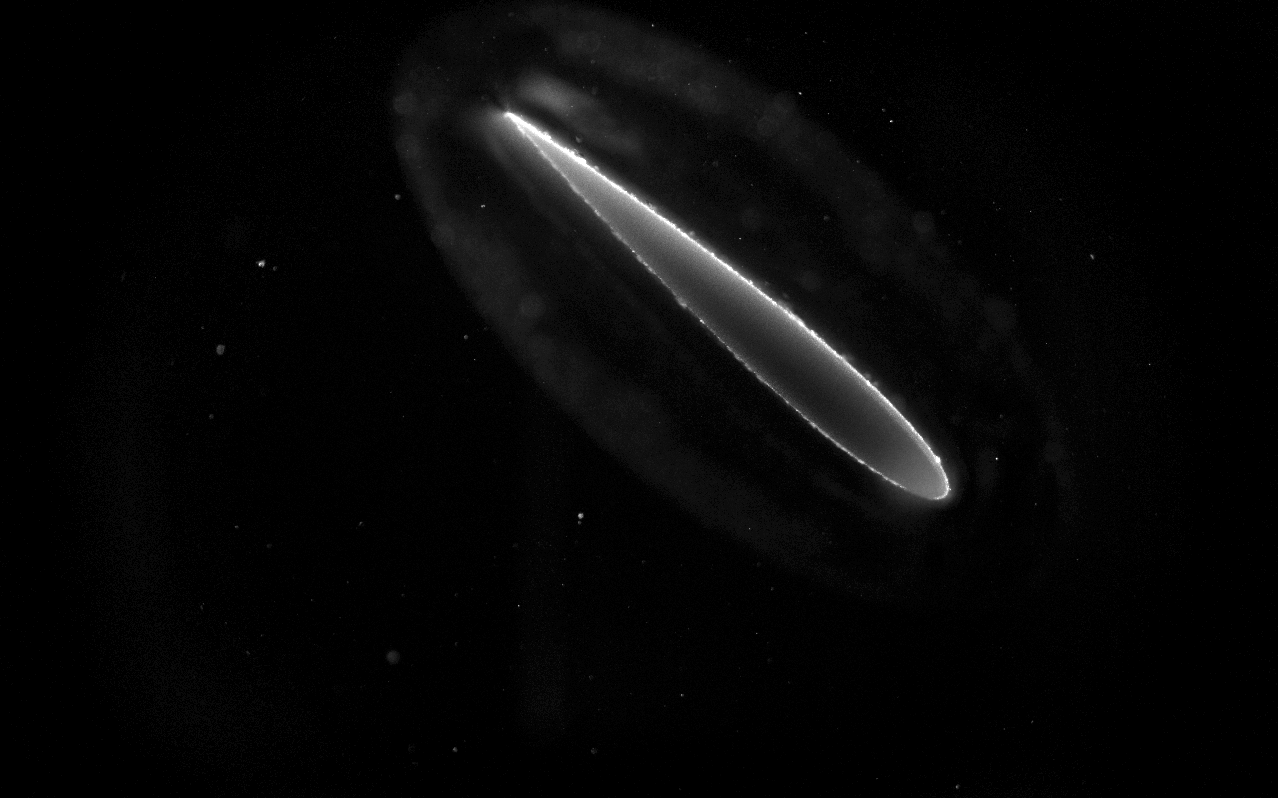}
		\caption{Raw image from the experiment Ec.}
		\label{AppFig:DataProcessing_raw}
	\end{subfigure}
	\hfill
	\begin{subfigure}[b]{0.45\textwidth}
		\centering
		\includegraphics[scale=0.2]{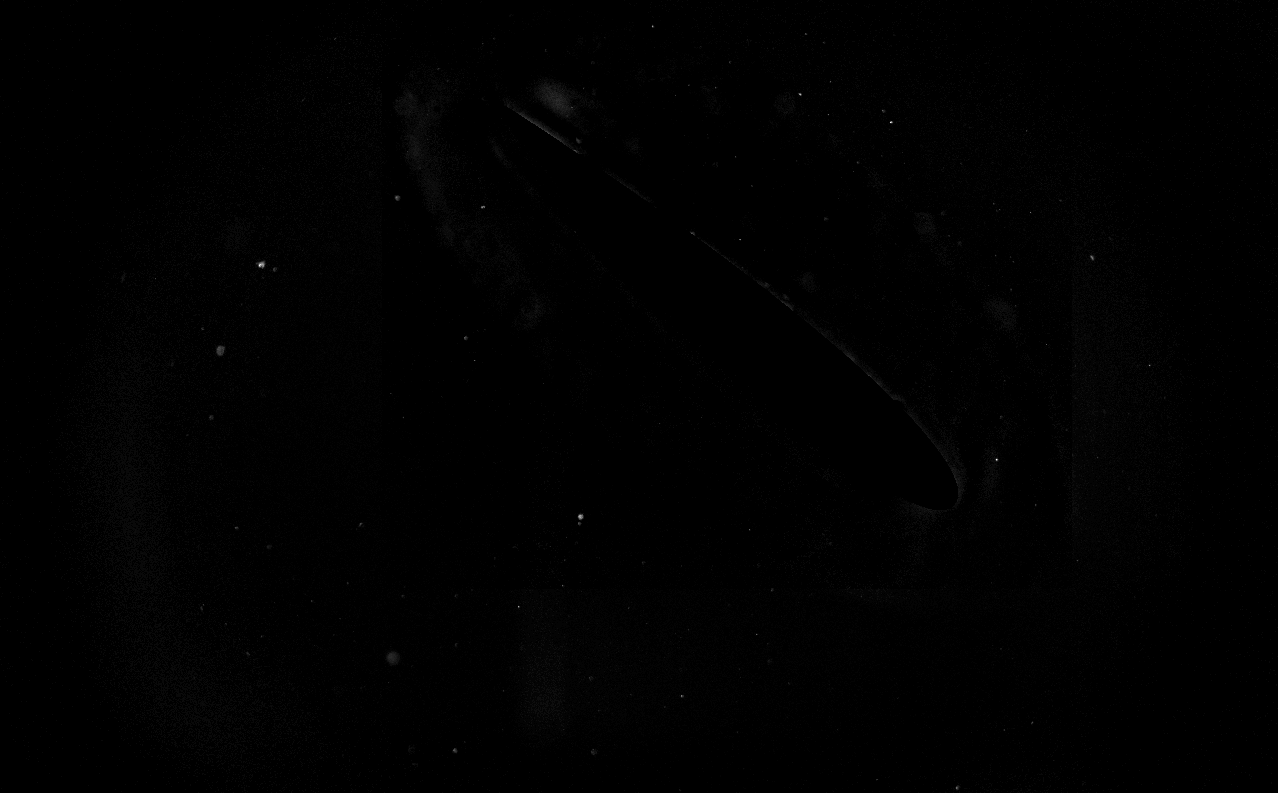}
		\caption{Image after the masking of the airfoil and the subtraction of the support}
		\label{AppFig:DataProcessing_masking}
	\end{subfigure}    
	\hfill
	\begin{subfigure}[b]{0.45\textwidth}
		\centering
		\includegraphics[scale=0.2]{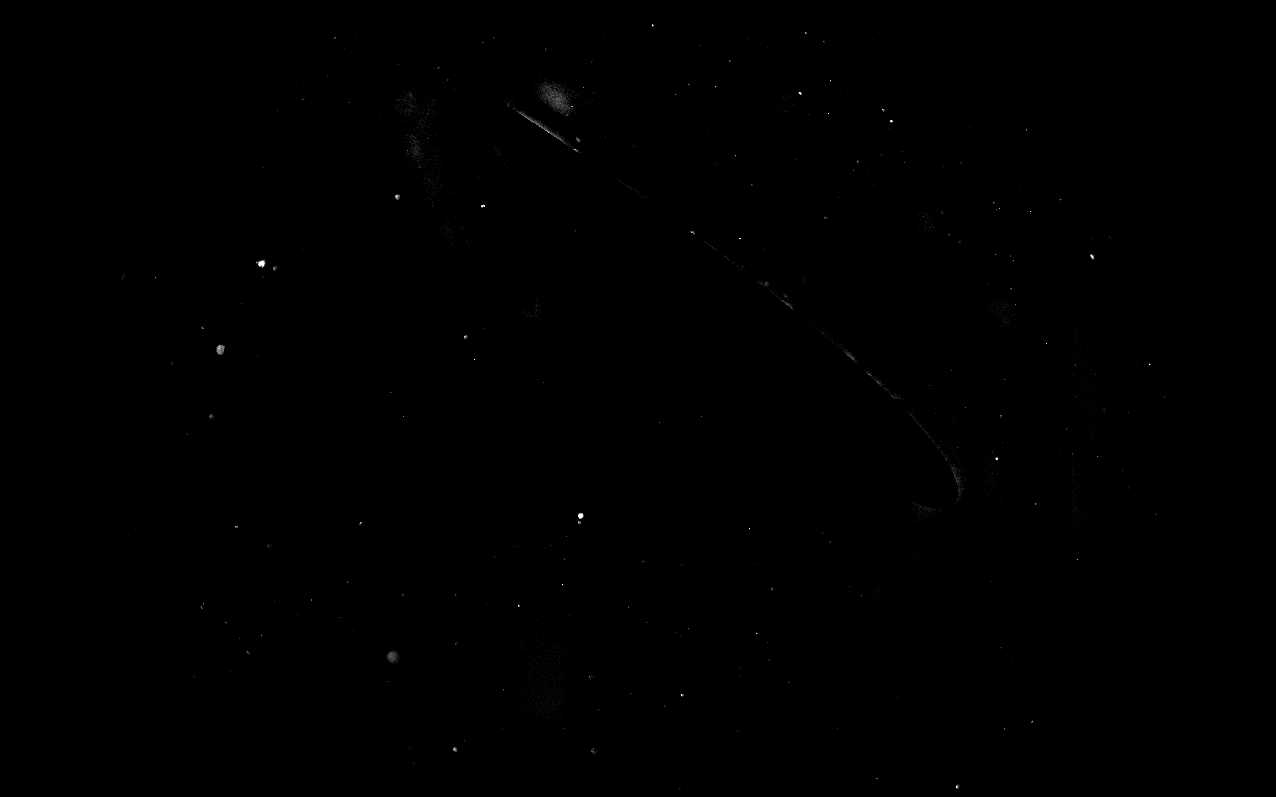}
		\caption{Image after enhancing the contrast.}
		\label{AppFig:DataProcessing_enhanced}
	\end{subfigure}
	\hfill
	\begin{subfigure}[b]{0.45\textwidth}
		\centering
		\includegraphics[scale=0.21]{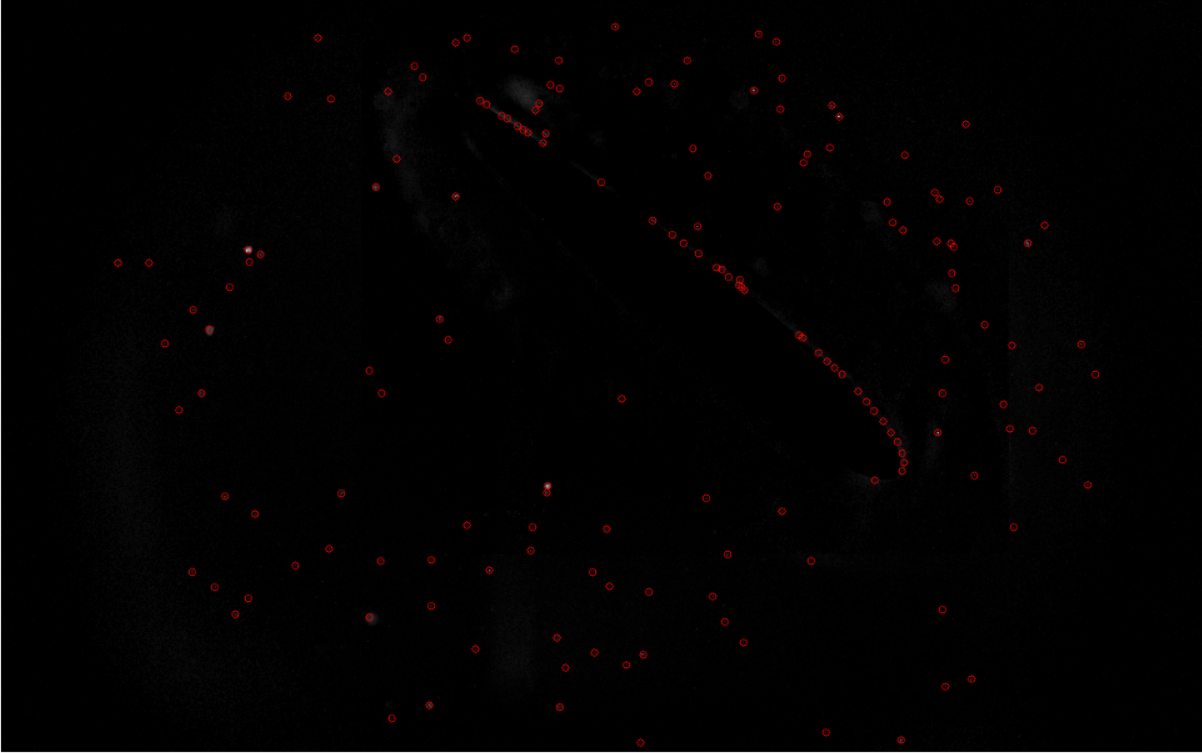}
		\caption{Image after detecting the particles.}
		\label{AppFig:DataProcessing_particles}
	\end{subfigure}  
	\hfill
	\begin{subfigure}[b]{0.45\textwidth}
		\centering
		\includegraphics[scale=0.21]{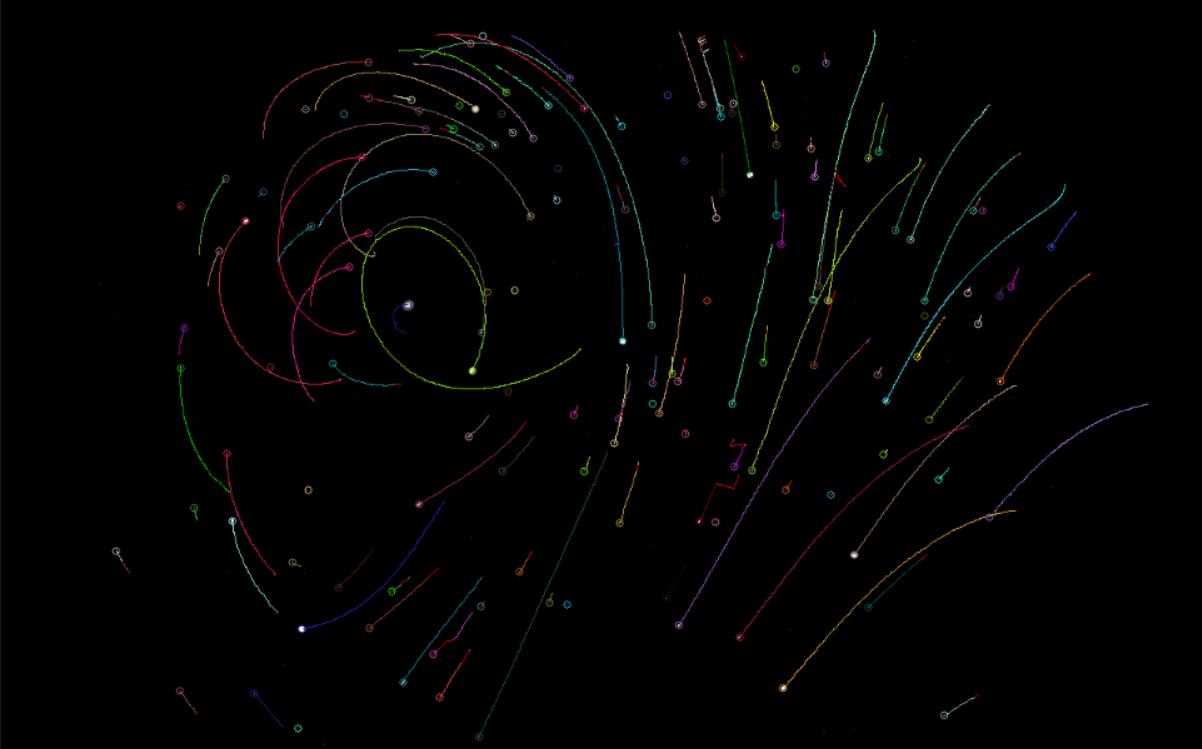}
		\caption{Image after the PTV.}
		\label{AppFig:DataProcessing_tracks}
	\end{subfigure}
	\caption{Exemple of the different steps of the data processing. The data are taken from the experiment Aa.}
	\label{AppFig:DataProcessing}
\end{figure}

\section{Additional information related to the theories used}\label{AppTh}

\subsection{Change of reference frame}\label{AppThRefFrame}

Our results were in the reference frame of the laboratory (see figures \ref{fig:Trajxy}, \ref{fig:Trajxt} and \ref{fig:Trajyt}), the origin being at the center of the airfoil in the first frame. However, the theories were not developped in the same reference frame. So to compare our results with it we needed to change it. \newline

Xu's theory uses the reference frame of the plate with its origin at the tip of the airfoil. As we study the starting vortex at the trailing edge in this work we will therefore choose this tip. The change of reference frame is then done following:
\begin{equation}
	\begin{pmatrix}
		x'(t) \\
		y'(t)
	\end{pmatrix}
	= 
	\begin{pmatrix}
		x(t) \\
		y(t)
	\end{pmatrix}
	+
	\begin{pmatrix}
		-L/2\sin\alpha \\
		L/2\cos\alpha
	\end{pmatrix}
	+
	\begin{pmatrix}
		0 \\
		1/2at^2
	\end{pmatrix}.
\end{equation}\newline

For his theory, Pullin also uses the reference frame of the plate, but its origin is at the center of the plate and the axis are rotated so the $x''$ axis is parallel the the background flow and $y''$ perpendicular. The change of reference frame is then done following:
\begin{equation}
	\begin{pmatrix}
		x''(t) \\
		y''(t)
	\end{pmatrix}
	= 
	\begin{pmatrix}
		\sin\alpha & -\cos\alpha\\
		\cos\alpha & \sin\alpha
	\end{pmatrix}
	\begin{pmatrix}
		x(t) \\
		y(t) + 1/2at^2
	\end{pmatrix}
\end{equation}

\begin{figure}[h!]
	\centering
	\includegraphics[scale=0.3]{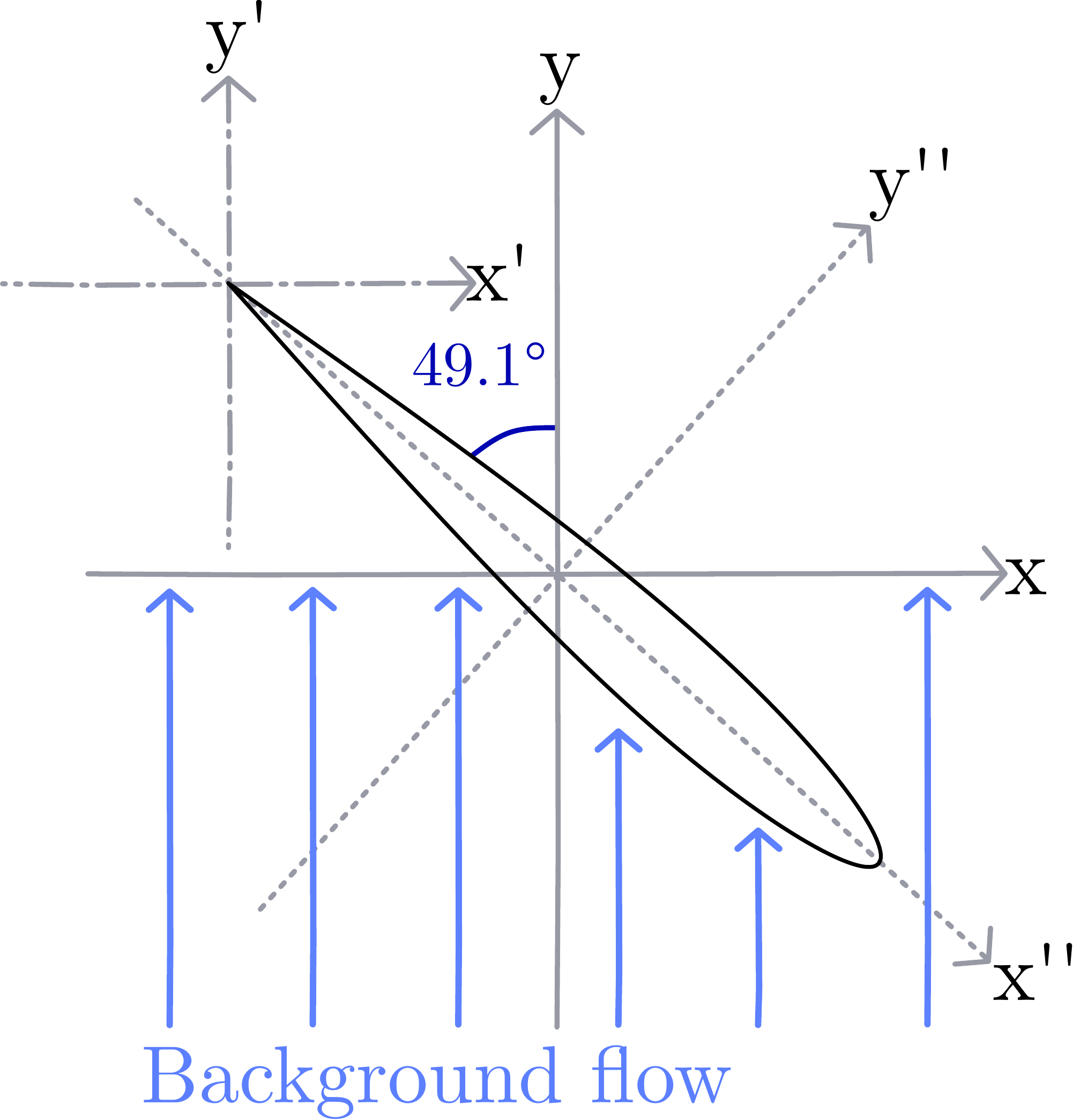}
	\caption{Drawing of the airfoil with the different reference frame at $t=0$. $(x,y)$ the initial reference frame of the laboratory, $(x',y')$ the reference frame used by Xu and $(x'',y'')$ the reference frame used by Pullin.}
	\label{AppFigRefFrame}
\end{figure}

\subsection{Main steps of Pullin's theory}\label{AppThPullin}

To explain the theory we need to introduce the complex potential 
\begin{equation}
	W(z,t) = \phi(x,y,t) + i\psi (x,y,t),
\end{equation}
with $\phi$ the velocity potential and $\psi$ the stream function, and the conformal mapping
\begin{equation}
	\zeta(z) + \frac{1}{2}(z+\sqrt{z^2-L^2/4})
\end{equation}
with $z = x+iy$. With this conformal mapping, the complex potential can be re-written
\begin{equation}
	W(z,t) = U(t)w_a(z) + W_u(\zeta(z),t),\quad w_a(z) = z\cos\alpha - i\sin\alpha\sqrt{z^2-L^2/4},
\end{equation}
with $W_u(z,t)$ the complex potential produced by separation at the plate edges. The boundary conditions $\Im [w_a]=0$ at $y''$=0, $-L/2<x''<L/2$.
They introduce $\Gamma_{\pm}(t)$, the total shed circulation at time $t$ on the sheet $\pm$, which is related to the trailing and leading edges at $z=+L/2$ and $z=-L/2$. Two vortex sheets $Z_+(\Gamma,t)$ and $Z_-(\Gamma,t)$ at the edges of the plate are introduced. They are functions for which $\Gamma$ is a conserved quantity moving with the average velocity of the two instantaneously adjacent fluid particles. They satify the Euler dynamics of the vortex-sheet evolution. $\Gamma$ is a lagrangrian parameter describing the total circulation on the sheet between a point and the sheet tip. Then, the vortex sheets mapped onto the $\zeta$-plane are
\begin{equation}
	\mathcal{L}_\pm (\Gamma,t) = \frac{1}{2}(Z_\pm + \sqrt{Z_\pm + L^2/4}),
\end{equation}
which gives 
\begin{equation}
	W_u(\zeta,t) = \frac{1}{2i\pi}\sum_\pm\int_0^{\Gamma_\pm (t)}\left(\log (\zeta - \mathcal{L}_\pm(\Gamma,t)) - \log\left(\zeta-\frac{L^2}{16\mathcal{L}^*_\pm(\Gamma,t)}\right)\right) \textrm{d}\Gamma,
\end{equation}
where $|\zeta|=L/4$ is used to satisfy the boundary condition $\psi_u = \Im [W_u]$. To obtain a solution for the dynamical evolution of $Z_{\pm}(\Gamma,t)$ and $\mathcal{L}_\pm (\Gamma,t)$, they solved the Birkhoff-Rott equation for the vortex sheet at $z = L/2$ is given by 
\begin{equation}\label{eq:BR}
	\begin{split}
		\pdv{Z^*_+(\Gamma,t)}{t} &= \dv{W}{z}(z=Z_+(\Gamma,t))\\
		&= Gt\left(\cos\alpha-\frac{iZ_+\sin\alpha}{\sqrt{Z_+^2-L^2/4}}\right) + \frac{1}{2}\left(1+\frac{Z_+}{\sqrt{Z_+^2-L^2/4}}\right)\\
		& \times\frac{1}{2i\pi}\int_0^{\Gamma (t)} \left(\frac{1}{\mathcal{L}_+ - \mathcal{L}_+'} - \frac{1}{\mathcal{L}_+ - L^2/(16\mathcal{L}_+^{*,})}\right) \textrm{d}\Gamma'.
	\end{split}
\end{equation}
with the influence from the other tip of the airfoil being omitted. As the equation \ref{eq:BR} is non linear, a similarity expansion is done to have an analytical expression. This expression will then only accurately describe the vortex dynamics at small times such that $|Z_+(\Gamma,t)-L/2|/L\ll1$. By putting $\tilde{Z}_+ = Z_+ - L/2$, an expansion in power $\tilde{Z}_+/L\ll 1$ gives
\begin{equation}
	\dv{W_a}{z}(z=Z_+)=-\frac{i}{2}L^{1/2}B\sin\alpha t \tilde{Z}_+^{-1/2}+\cdot\cdot\cdot,
\end{equation}
with $B$ a constant. After some developments it finally gives

\begin{subequations}
	\begin{equation}
		\Gamma_+(t) = K^{1/2}B^{4/3}L^{2/3}t^{5/3}(J_0 + J_1K^{1/2}L^{-1/3}B^{1/3}\sin\alpha^{1/3}t^{2/3} + \cdot\cdot\cdot),
	\end{equation}
	\begin{equation}
		\Gamma_-(t) = K^{1/2}B^{4/3}L^{2/3}t^{5/3}(-J_0 + J_1K^{1/2}L^{-1/3}B^{1/3}\sin\alpha^{1/3}t^{2/3} + \cdot\cdot\cdot),
	\end{equation}
	\begin{equation}
		\tilde{Z}_+(0,t) = KB^{2/3}L^{1/3}\sin\alpha^{2/3}t^{4/3}(\omega_0(1) + \omega_1(1)K^{1/2}L^{-1/3}B^{1/3}\sin\alpha^{1/3}t^{2/3} + \cdot\cdot\cdot),
	\end{equation}
	\begin{equation}
		\tilde{Z}_-(\Gamma,t) + \frac{L}{2} = KB^{2/3}L^{1/3}\sin\alpha^{2/3}t^{4/3}(-\omega_0^*(\lambda) + \omega_1^*(\lambda)K^{1/2}L^{-1/3}B^{1/3}\sin\alpha^{1/3}t^{2/3} + \cdot\cdot\cdot),
	\end{equation}
\end{subequations}
with $K = (3/8)^{2/3}$ a convinient scaling constant, $\lambda$ a dimensionless circulation parameter, $J_0$ and $J_1$ dimensionless constants and $\omega_0(\lambda)$ and $\omega_1(\lambda)$ complex shape functions.

Some theoretical expression of the constantes $J_0$, $J_1$ $\omega_0(\lambda)$ and $\omega_1(\lambda)$ can be found using the point-vortex model. In this model, the entire vortex sheet is concentrated into a single point. The previous steps to obtain the expressions of $\Gamma_+(t)$, $\Gamma_-(t)$, $\tilde{Z}_+(0,t)$ and $\tilde{Z}_-(\Gamma,t)$ can be done again with this approximation. The constant are at the end

\begin{subequations}
	\begin{equation}
		\Omega_0 = 0.29i,\quad \Omega_1 = 0.35 + 0.12i
	\end{equation}
	\begin{equation}
		J_0 = 2.4,\quad J_1=-0.95
	\end{equation}
\end{subequations}

\end{document}